\documentclass[notitlepage, nofootinbib, no reprint, amsmath, amssymb, aps, twocolumn, pra]{revtex4-2}

\usepackage{amssymb}
\usepackage{hyperref}
\usepackage{bm}
\usepackage{soul}
\usepackage{dsfont}
\usepackage{amsthm}
\usepackage{amsmath}
\usepackage{amssymb}
\usepackage{verbatim} 
\usepackage{qcircuit}
\usepackage{circuitikz}
\usepackage{tabularx}
\usepackage{color,soul}
\usepackage{esint}
\usepackage{physics}
\usepackage{braket}
\usepackage{mathtools}
\usepackage{bbold}
\usepackage{float}
\usepackage{placeins}
\usepackage{mathrsfs}
\usepackage[normalem]{ulem}

\usepackage{todonotes}

\def\S{\mathrm{S}}
\def\T{\mathrm{T}}
\def\R{\mathbb{R}}

\def\Id{\mathbb{1}}
\def\balpha{\boldsymbol{\alpha}}
\def\btheta{\boldsymbol{\theta}}
\def\rhoalpha{\rho_\alpha}

\def\wls{w_\mathrm{ls}}
\def\wvar{w_\mathrm{var}}

\newcommand{\forwardPi}{\overset{\rightarrow}{\Pi}}
\newcommand{\backwardPi}{\overset{\leftarrow}{\Pi}}

\begin{document}

\author{Yerassyl Balkybek}
\email{Yerassyl.Balkybek@skoltech.ru}
 \affiliation{Skolkovo Institute of Science and Technology, Bolshoy Boulevard 30, Moscow, 121205, Russia}

\author{Andrey Kardashin}
\altaffiliation{Former affiliation. Current address: Donostia International Physics Center, Donostia-San Sebastián, Spain.}
\affiliation{Skolkovo Institute of Science and Technology, Bolshoy Boulevard 30, Moscow, 121205, Russia}
 
\author{Konstantin Antipin}
\affiliation{Faculty of Physics, M.V. Lomonosov Moscow State University,\\
Leninskie gory, GSP-1, Moscow 119991,  Russia}
\affiliation{Skolkovo Institute of Science and Technology, Bolshoy Boulevard 30, Moscow, 121205, Russia}

\author{Vladimir V. Palyulin}
 \affiliation{Skolkovo Institute of Science and Technology, Bolshoy Boulevard 30, Moscow, 121205, Russia}

\begin{abstract}

    Quantum regression tasks for predicting properties of quantum states are commonly addressed using variational quantum algorithms. 
    While variational quantum circuits are highly expressive and allow to achieve reasonable accuracy, training these circuits may demand a considerable amount of time and resources. 
    In this work, we propose an approach of constructing problem-specific quantum regression models with encoding relevant symmetries and regularizing the variance. 
    The proposed method is based on finding the coefficients of the linear combination of suitably chosen observables.
    Although it requires the knowledge of the symmetries of the problem in question, the method does not involve parameterized quantum circuits, and the training is done efficiently once the observables are measured.
    We demonstrate this method on two examples: Prediction of the transverse field strength in the Ising model, and quantification of entanglement in bipartite qubit systems. 
    Our approach is accurate and less resource-intensive than conventional variational methods.
\end{abstract}

\title{Parametrized-circuit-free quantum regression with variance regularization}
\maketitle

\section{Introduction}
   
   Quantum machine learning (QML)\cite{schuld2015introduction,lloyd2018quantum, schuld2019quantum, biamonte2017quantum} has emerged as a rapidly developing research field in quantum computing and is particularly relevant in the current noisy intermediate-scale quantum (NISQ) era \cite{preskill2018quantum,cerezo2021variational, schuld2021machine, meyer2021fisher}. 
   QML methods enable the modeling of complex relationships between classical data points and quantum states through various tasks, such as classification \cite{park2020theory, banchi2021generalization} and regression \cite{schuld2016prediction, reddy2021hybrid, kard2025}. 
   While classification tasks in QML focus on determining the discrete class labels for the input states $\rho$, regression tasks map the inputs into real-valued continuous variables. 

    Variational quantum circuits (VQCs) \cite{cerezo2021variational,biamonte2021universal} have been the predominant tool for these tasks, forming the core of variational quantum algorithms (VQAs). In a standard VQA, a parameterized unitary $U(\boldsymbol{\theta})$ prepares a quantum state, and the expectation value of an observable $G$ is measured as $\langle G \rangle_{\boldsymbol{\theta}} = \Tr\left[ U(\boldsymbol{\theta}) \rho \, U^\dagger(\boldsymbol{\theta}) \, G \right]$.
    This expectation value serves as the model's output. A classical optimizer then adjusts the parameters $\boldsymbol{\theta}$ to minimize a cost function, creating a hybrid quantum-classical feedback loop. VQAs have shown promise for various problems, including linear algebra \cite{bravo2023variational, bravo2019variational, xu2021variational, peruzzo2014variational, lubasch2020variational}, Hamiltonian diagonalization \cite{zeng2021variational, larose2019variational} and finding ground states of many-body Hamiltonians \cite{kokail2019self, grimsley2019adaptive}.

    However, VQAs face significant challenges limiting their practical utility. A primary concern is the barren plateaus \cite{larocca2025barren, uvarov2021barren, wang2021noise}, where the cost function gradients vanish exponentially with the number of qubits, making training intractable \cite{mcclean2018barren}. Furthermore, the expressivity of a given VQC ansatz could mismatch the problem structure, potentially ignoring relevant symmetries or requiring excessive circuit depth for adequate approximation \cite{larocca2023theory,holmes2022connecting,holmes2022connecting, cerezo2022challenges}.

    To circumvent these limitations, alternative approaches can be used \cite{huang2023post, greiter2018method, huang2023learning}. One can construct a fixed, problem-inspired ansatz consisting of $M$ predefined observables $\mathcal{G}=\{G_1, G_2, \dots, G_M\}$. A variational model is then built not by tuning a quantum circuit, but by forming a \textit{classical} linear combination of their expectation values:  
    \begin{equation}
        f_{\boldsymbol{\theta}}(\rho) = \sum_{j=1}^{M} \theta_j \, \Tr\left[ \rho \, G_j \right],
    \label{eq:postvar_model}
    \end{equation}
    where $\boldsymbol{\theta} = (\theta_1, \dots, \theta_M)$ are trainable parameters. Crucially, the quantum circuit used to measure the expectation values $\Tr[\rho G_j]$ is independent of $\boldsymbol{\theta}$.
    All variational optimization is therefore performed as a classical post-processing step after the quantum data collection, effectively decoupling the parameter training from the quantum execution.
    
    In this work, we leverage the approach given by Eq.~(\ref{eq:postvar_model}) to solve two specific quantum regression problems. 
    First, we predict the relative strength of an external field in a transverse-field Ising model from its ground state. 
    Second, we quantify the entanglement negativity for bipartite qubit systems.
    For each task, we design a problem-specific ansatz $\mathcal{G}$ and employ variance-based regularization to control the variance of the observable and enhance generalization.
    Our results show that our approach, when combined with problem-symmetry focused ans\"atze, provides an effective and trainable framework for predicting targeted quantum properties, overcoming the key limitations of variational methods. 
    In particular, for entanglement negativity prediction, we propose an ansatz $\mathcal{G}$ inspired by entanglement symmetries that outperforms variational hardware-efficient and $k$-local Pauli-based ansätze. 
    Moreover, our ansatz requires fewer training samples and can be trained on simpler datasets (such as pure and isotropic states) to predict random mixed states, where hardware-efficient and $k$-local Pauli ansätze tend to overfit. 
    For the transverse-field Ising model, we consider the task of predicting the relative strength of an external field by employing a $3$-local Pauli-string ansatz constrained to respect the spin-flip and time-reversal symmetries. 
    For an 8-qubit Hamiltonian, this results in an ansatz comprising 90 Pauli strings for an 8-qubit system, yielding high prediction accuracy with low variance for the training set of only 10 states.

    \section{Problem statement}

Consider the following set:
\begin{equation}
    \label{eq:training_set}
    \mathcal{T} = \big\{(\rho_{i}, \alpha_i)\big\}_{i=1}^{T},
\end{equation}
where $\rho_{i} \equiv \rho(\alpha_i)$ are data comprised of $n$-qubit quantum states with labels $\alpha_i \in \R$.
From a training set $\mathcal{T}$ we want to learn a model that predicts the label $\alpha$ for an unseen state $\rhoalpha$.

This is a regression problem with the peculiarity that the data points are quantum states $\rhoalpha$. 
The label $\alpha$ of a state $\rhoalpha$ is then the expectation value of an observable $H$, i.e., $\Tr H\rhoalpha$. 
Such an observable can be found as
\begin{equation} 
    \label{eq:regression-H}
    H^* \in \arg\min_{H} \left[\wls f_\mathrm{ls} + \wvar  f_\mathrm{var}\right],
\end{equation}
where
\begin{align}
    \label{eq:cost}
    \begin{split}
        f_\mathrm{ls}&=\sum_{i=1}^T \Big( \Tr H \rho_i - \alpha_i \Big)^2, \\
        f_\mathrm{var}&=\sum_{i=1}^T \left(\Tr H^2 \rho_i - \Tr^2 H \rho_i\right).
    \end{split}
\end{align}
Thus, we simultaneously minimize (i) the weighted least-squares error between the predictions $\Tr H\rho_i$ and the true labels $\alpha_i$, as well as (ii) the weighted variance of $H$ in the states $\rho_i$. The weights $\wls$ and $\wvar$ control the trade-off between these two objectives.

\section{Methods}
\label{sec:methods}
    This section details our approach. 
    We formalize the observable parametrization as a linear combination of Hermitian operators for variance-regularized regression, derive the optimal parameters by solving a linear system of equations, and highlight the advantage of the proposed method in extending the solution to new datasets.

    \subsection{Parametrization of the observable}
        \label{sec:method_param}

        Commonly used variational quantum algorithms (VQAs) rely on iterative parameter updates in parametrized quantum circuits (or ans\"atze). Both by us \cite{kard2025} and others 
        \cite{kreplin2024reduction, zhang2022variational, thanasilp2023subtleties} VQAs were shown to effectively solve regression problems with variance regularization. One key advantage of parametrized ans\"atze is their ability to address data-agnostic problems. However, the conventional VQAs require parameter updates on a quantum computer while optimizing a cost function on a classical computer.

        Inspired by Refs.~\cite{greiter2018method, huang2023post}, we take a different approach: Instead of iteratively updating parameters on the quantum computer, we predefine a 
        parametrized linear combination of observables, measure them, and perform all parameter updates classically afterwards. 
        We represent the observable $H$ as 
        \begin{equation}
            \label{eq:ham-par}
            H_{\btheta} = \sum_{j=1}^M \theta_j G_j,
        \end{equation}
        where $\btheta = ( \theta_i )_{i=1}^M \subset \R$ are variational parameters, and $G_i$ are Hermitian operators that are preset and can be problem specific.
        In this work, we shall call the set $\mathcal{G} = \{ G_j \}_{j=1}^M$ as \textit{an ansatz}.

        An obvious disadvantage of such parametrization of an observable is a need of a \textit{guess} for $\mathcal{G}$ for a given training set $\mathcal{T}$ (hence, ``ansatz'').
        Alternatively, as commonly done in the literature \cite{huang2023learning, huang2020predicting, zhou2020single}, one can set $G_j$ to be Pauli strings of a bounded locality.

         An important advantage of the parametrization \eqref{eq:ham-par} is that the expectations required for evaluating the cost function in \eqref{eq:regression-H} need to be computed \textit{only once}.
         Indeed, given a training set as in \eqref{eq:training_set} and defining the arrays 
        \begin{gather}
            L = \big[\Tr \rho_i G_j\big]_{i,j=1}^{T,M}, \label{eq:L}\\
            S = \big[ \frac{1}{2} \Tr \rho_i (G_j G_k+G_k G_j)  \big]_{i,j,k=1}^{T,M,M}, \label{eq:S}
        \end{gather}
        we can write
        \begin{gather}
            \Tr H_{\btheta} \rho_i   = L_i  \btheta, \label{eq:L-trace} \\
            \Tr H_{\btheta}^2 \rho_i = \btheta^T S_i \btheta, \label{eq:S-trace}
        \end{gather}
        where $S_i$ is the $i$th (square symmetric)  matrix from the array \eqref{eq:S}.
        
        Let us denote the vector of labels $\balpha \equiv (\alpha_i)_{i=1}^T$ and $S_\Sigma \equiv \sum_{i=1}^{T} S_i$, in which $S_\Sigma$ appears to be a symmetric matrix.
        Putting \eqref{eq:L-trace} and \eqref{eq:S-trace} into \eqref{eq:cost}, we arrive to an optimization problem
        \begin{equation}\label{eq:theta_argmin}
            \btheta^* \in \arg\min_{\btheta} f(\btheta)
        \end{equation}
        with the cost function 
        \begin{multline}
            \label{eq:num_cost}
            f(\btheta) = \wls\left(\balpha^\T \balpha - 2 \balpha^\T L \btheta \right) \\ +  \wvar \btheta^\T S_\Sigma \btheta + (\wls - \wvar) \btheta^\T L^\T L \btheta.
        \end{multline}
        Differentiation with respect to $\btheta$ \cite{petersen2008matrix} (denoted $\partial_{\btheta}$) reads
        \begin{gather}
             \partial_{\btheta} f(\btheta)  = -2\wls L^\T\balpha + 2 \wvar S_\Sigma\btheta + 2(\wls - \wvar) L^\T L \btheta, \label{eq:cost_1der} \\
             \partial_{\btheta} \big( \partial_{\btheta} f(\btheta) \big)  = 2\wvar S_\Sigma + 2(\wls - \wvar) L^\T L. \label{eq:cost_2der} 
        \end{gather}
        
        If $\wls>\wvar$, then the second derivative in Eq.~\eqref{eq:cost_2der} is non-negative, since $L^TL$ and $S_\Sigma$ are both positive semidefinite (see Appendix \ref{app:sec:hessian}). Thus, $\partial_{\btheta}(\partial_{\btheta} f(\btheta))$ is a positive semi-definite Hessian matrix of $f(\btheta)$.
        Therefore, we can equate the gradient \eqref{eq:cost_1der} to zero and solve the equation for $\btheta$.
        Denoting $k = \wls/\wvar$, one obtains a solution of \eqref{eq:theta_argmin} in the form $\btheta^* = A^+b$, where
        \begin{gather}
           \label{eq:matrix_A_num}
            A = (k -1) L^\mathrm{T} L +  S_\Sigma, \\
            b = k L^\mathrm{T} \balpha,
        \label{eq:vector_b_num}
        \end{gather}
        and $A^+$ is a pseudoinverse of $A$.
        When the analytical form of $\rho_\alpha$ is known, one can derive exact expressions for $A$ and $b$, see Appendix~\ref{app:sec:theory_Ab}.

        An extra advantage of the proposed method comes from the fact that it allows combining non-overlapping datasets $\{\mathcal{T}_l\}_l$ through summation of their corresponding arrays $A_l$,$b_l$ (for the full description, see Appendix~\ref{app:sec:comb_datasets}).

\subsection{Prediction and variance}
\label{sec:estim_var_crb}
Once the parametrized observable in Eq.~(\ref{eq:ham-par}) has been learned from the training set~(\ref{eq:training_set}) by minimizing the total cost~(\ref{eq:cost}), we define a prediction for the unknown label $\alpha$ of a state $\rho_\alpha$ as
\begin{equation}
    \label{eq:estimator}
    \mathsf{a}(\btheta, \rho_\alpha) \equiv \Tr\left(H_{\btheta} \rho_\alpha\right) = \sum_{j=1}^M \theta_j \Tr\left(G_j \rho_\alpha\right)
\end{equation}
which returns an expectation value $\langle H_{\btheta} \rangle_{\rho_\alpha} \equiv \Tr(H_{\btheta} \rho_\alpha)$. Using the variance of $H_{\btheta}$ in the state $\rho_\alpha$, $\Delta_{\rho_\alpha}^2 H_{\btheta} \equiv \langle H_{\btheta}^2 \rangle_{\rho_\alpha} - \langle H_{\btheta} \rangle_{\rho_\alpha}^2$, we define the reduced variance
\begin{equation}
    \label{eq:estimation_variance}
    \operatorname{rVar}(H_{\btheta}) 
    = \frac{\Delta_{\rho_\alpha}^2 H_{\btheta}}{\big| \partial_\alpha \langle H_{\btheta} \rangle_{\rho_\alpha} \big|^2}.
\end{equation}
This quantity is lower bounded by the classical Cramer-Rao bound (cCRB) \cite{sidhu2020geometric}:
\begin{equation}
    \label{eq:cramer-rao_classical}
    \operatorname{rVar}(H_{\btheta}) \geqslant \frac{1}{I_c(\boldsymbol{\Pi}, \rho_\alpha)},
\end{equation}
where $I_c(\boldsymbol{\Pi}, \rho_\alpha)$ is the classical Fisher information (cFI). In our setting, the projectors $\boldsymbol{\Pi}=\{\Pi_i\}_i$ are the eigenprojectors of $H_{\btheta}$.
The cFI can be written as \cite{meyer2021variational}
\begin{equation}
    \label{eq:fisher_classical}
    I_c(\boldsymbol{\Pi}, \rho_\alpha) = \sum_i \frac{1}{p_i(\alpha)} \left( \frac{\partial p_i(\alpha)}{\partial \alpha} \right)^2,
\end{equation}
where $p_i(\alpha) = \Tr[\Pi_i \rho_\alpha]$.

The classical Fisher information is bounded from above by the quantum Fisher information (qFI) $I_q(\rho_\alpha)$, which can be expressed via the fidelity $F(\rho, \tau) = \Tr\sqrt{\sqrt{\rho}\,\tau\,\sqrt{\rho}}$ of two states $\rho$ and $\tau$ as
\begin{equation}
    \label{eq:qfi-fidelity}
    I_q(\rho_\alpha) = \lim_{\mathrm{d}\alpha\to0} 8\frac{1 - F\!\left(\rho_\alpha, \rho_{\alpha+\mathrm{d}\alpha}\right)}{\mathrm{d}\alpha^2}.
\end{equation}

Combining the two bounds together, we obtain the chain of inequalities
\begin{equation}
    \label{eq:var-qcrb}
    \frac{\Delta_{\rho_\alpha}^2 H_{\btheta}}{\big| \partial_\alpha \langle H_{\btheta} \rangle_{\rho_\alpha} \big|^2} 
    \geqslant \frac{1}{I_c(\boldsymbol{\Pi}, \rho_\alpha)} 
    \geqslant \frac{1}{I_q(\rho_\alpha)},
\end{equation}
where the leftmost quantity is the reduced variance of the observable used for prediction, the middle is the classical CRB, and the rightmost is the quantum CRB (qCRB). 

\section{Relation to the literature}
\label{sec:literature}

A closely related approach to our method, though applied to a different purpose, is the work by Greiter \textit{et al.}~\cite{greiter2018method}. To construct a parent Hamiltonian $H$ for a given trial wavefunction  $\ket{\psi_0}$, similarly to \eqref{eq:ham-par}, Greiter \textit{et al.} propose expressing $H$ as a linear combination of local Hermitian operators $\{H_i\}_i$:
\begin{equation}
    \label{eq:herms_i}
    H = \sum_{i=1}^L a_i H_i,
\end{equation}
where $\{a_i\}_i$ are variable coefficients to determine. Defining $\ket{\psi_i} = H_i \ket{\psi_0}$ (with $H_0 \equiv \Id $) and $M_{ji} = \braket{\psi_j | \psi_i}$, they derive a system of linear equations:
\begin{equation}
\sum_{i=0}^L M_{ji} a_i = 0 \quad \text{for} \quad j = 0,1,\dots,L.    
\end{equation}
This system yields an exact parent Hamiltonian $H$ for $\ket{\psi_0}$ if a non-trivial solution exists for the chosen $\{H_i\}_i$. However, if $\det(M_{ji}) \neq 0$, only the trivial solution $a_i = 0$ is possible. The authors argue that when no exact solution exists for the chosen $\{H_i\}_i$, an \textit{approximate} parent Hamiltonian can often be obtained instead.

Alternatively Huang \textit{et al.}~\cite{huang2023learning} introduce a provably efficient classical algorithm for predicting expectation values of an unknown quantum process $\mathcal{E}$ acting on $n$ qubits. The goal is to learn $f(\rho,O) = \Tr\!\left[ O\,\mathcal{E}(\rho) \right]$, for input states $\rho$ drawn from a distribution $\mathcal{D}$. If $\mathcal{D}$ is invariant under single-qubit Clifford transformations and the observable $O$ has bounded (constant) spectral norm, then the observable after the process
can be approximated by a truncated Pauli expansion,
\begin{equation}
    \label{eq:k-local_Huang}
    O^{(k)} = \sum_{\substack{P \in \{I,X,Y,Z\}^{\otimes n}:\,  |P| \le k}} \alpha_P P,
\end{equation}
where $|P|$ is the Pauli weight (the number of qubits on which $P$ acts non-trivially) and the mean-squared prediction error decays exponentially with $k$.
By combining classical shadow tomography with low-degree approximations, the method achieves polynomial sample and computational complexity for constant error and quasipolynomial complexity for $\epsilon=1/\operatorname{poly}(n)$.

Furthermore, the work \cite{huang2023post} proposes post-variational strategies for a data-agnostic quantum learning. A variational observable $\mathscr{D}:=U^\dagger(\boldsymbol{\theta}) O U(\boldsymbol{\theta})$ is decomposed as
$\mathscr{D}=\sum_{j=1}^M \mathcal{F}_j(\boldsymbol{\theta})\,\mathscr{D}_j,$
where $\mathcal{F}_j$ are classical functions and $\mathscr{D}_j$ are fixed Hermitian operators, effectively shifting optimization to classical post-processing. This decomposition is again similar to \eqref{eq:ham-par}, and can be achieved either by expanding the ansatz around a suitable initialization and truncating the resulting Taylor series, or by constructing observables from $k$-local Pauli operators combined with classical shadow measurements. $p$ expanded ans\"atze and $q$ constructed observables yield $p\times q$ fixed operators, and numerical experiments show performance comparable to standard variational circuits while avoiding iterative quantum optimization.

A widely used approach to quantum state measurement and tomography is based on randomized measurements and can be viewed as a form of classical post-processing. An $N$-qubit quantum state $\rho$ is locally rotated by random unitaries $U_i$ drawn from a tomographically (or informationally) complete ensemble, such as Clifford or single-qubit Pauli rotations~\cite{elben2023randomized}. 
For the latter, each unitary is constructed as a tensor product $U_i=\bigotimes_{n=1}^N U_{n,i}$, with independently chosen single-qubit unitaries $U_{n,i}$. The rotated states $U_i\rho U_i^\dagger$ are measured $K$ times in the computational basis. Collecting data for $M$ such unitaries yields a total of $KM$ measurement outcomes. The protocol is efficient for estimating nonlinear properties such as purities, fidelities, entanglement entropies, and expectation values of complex many-body observables, without requiring full state reconstruction.

Similar approaches can be observed in specific tasks in combination with classical machine learning techniques.
For instance, randomized measurements have been widely applied to the detection and quantification of entanglement in multipartite qubit and qudit systems~\cite{elben2020mixed, elben2023randomized, elben2023randomized,rath2021importance, huang2020predicting, van2012measuring, carrasco2024entanglement, elben2019statistical, brydges2019probing,ketterer2019characterizing, neven2021symmetry, yu2021optimal}. One common strategy estimates higher-order partial-transpose (PT) moments of a quantum state, motivated by the positive partial transposition (PPT) criterion. The PPT criterion, introduced by Peres~\cite{peres1996separability}, states that if a density matrix $\rho = \rho_{AB}$ is separable, then its partial transpose $\rho^{T_B}$ must have non-negative eigenvalues, where $T_B$ denotes transposition with respect to subsystem $B$. The contrapositivity certifies entanglement between subsystems $A$ and $B$.
In Ref.~\cite{gray2018machine}, the authors consider moments of partially transposed bipartite states $\rho_{AB}$ of the form
    $\mu_c = \Tr\!\left[\left(\bigotimes_{q=1}^c \rho_{A_q B_q}^{T_{B_q}}\right)\mathbb{P}^c\right]
          = \Tr\!\left[\left(\bigotimes_{q=1}^c \rho_{A_q B_q}\right)(\mathbb{P}^c)^{T_B}\right],$
where $\mathbb{P}^c$ is a linear combination of cyclic permutation operators of order $c$, and we consider $c$ copies as $\rho^{\otimes c}\equiv\rho_{AB}^{\otimes c}\equiv\bigotimes_{q=1}^c \rho_{A_q B_q}$. 
These permutation operators can be decomposed into products of swap operators $\S_{A_q A_r}$ and $\S_{B_q B_r}$, exchanging the states of subsystems $A$ (resp.\ $B$) across different copies of $\rho_{AB}$ indexed by $q$ and $r$. 
For example, for $c=3$ one has $\mathbb{P}^3 = \S_{A_2 A_3}\S_{A_1 A_2}\S_{B_2 B_3}\S_{B_1 B_2}$, while $(\mathbb{P}^3)^{T_B} = \S_{A_2 A_3}\S_{A_1 A_2}\S_{B_1 B_2}\S_{B_2 B_3}$. 
Reference~\cite{gray2018machine} further employs machine learning techniques to relate these moments to entanglement measures.
The first two moments, $\mu_1 = \Tr[\rho] = 1$ and $\mu_2 = \Tr[\rho^2]$, quantify state normalization and purity, respectively. The third moment yields a sufficient condition for entanglement: if $\mu_3 < \mu_2^2$, then $\rho_{AB}$ violates the PPT criterion and is therefore entangled~\cite{elben2020mixed}.

Combined with classical shadows and U-statistics, the local randomized measurements allow efficient estimation of the third PT moment~\cite{elben2020mixed}. Similarly, Ref.~\cite{zhou2020single} shows that the third-order PT moments can be post-processed using classical shadows, constructing Bell-basis observables to quantify entanglement negativity. Building on this approach, Ref.~\cite{li2024directly} analyzes the measurement efficiency and robustness of such schemes under noise and experimental imperfections. Beyond the bipartite setting, higher-order PT moments together with local randomized measurements were used for multipartite entanglement detection: Ref.~\cite{coffman2024local} employs local randomized measurements forming projective 2-designs to construct unbiased estimators for concentratable entanglement measures.
Beyond the PT moments, various nonlinear permutation moments have been introduced to compute entanglement measures such as concurrence, R\'enyi entropies \cite{elben2019statistical}, and realignment-based criteria \cite{chen2002matrix, wang2024moments}. 

In this work, for the entanglement quantification task, we construct a tailored ansatz based on a selected set of permutation operators that excludes PT moments but includes permutation moments associated with second-order realignment and concurrence. For the Ising-Hamiltonian problem, we instead choose $k$-local Pauli operators as the basis of the ansatz.

\section{Predicting the transverse field in the Ising Hamiltonian}
    \label{sec:ising}
    We consider an open chain of $n$ spins described by the transverse field Ising Hamiltonian
    \begin{equation}
        \label{eq:ising}
        \mathcal{H}_h = -\sum\limits_{i=1}^{n-1} J_{i,i+1} Z_i Z_{i+1} + h \sum\limits_{i=1}^{n} X_i,
    \end{equation}
    where we set $n=8$, and the coefficients $J_{i,i+1}$ are sampled randomly and uniformly from $[0.1, 2]$.
    We aim to predict the value of the transverse field $h$ given the ground state $\ket{\psi_h}$ of $\mathcal{H}_h$.
    In the ansatz $H_{\boldsymbol{\theta}} = \sum_{j=1}^M \theta_j G_j$, we consider $G_j$ to be Pauli strings as in \eqref{eq:k-local_Huang}. 

      \begin{figure*}[tbh]
        \centering
        \includegraphics[width=.445\textwidth]{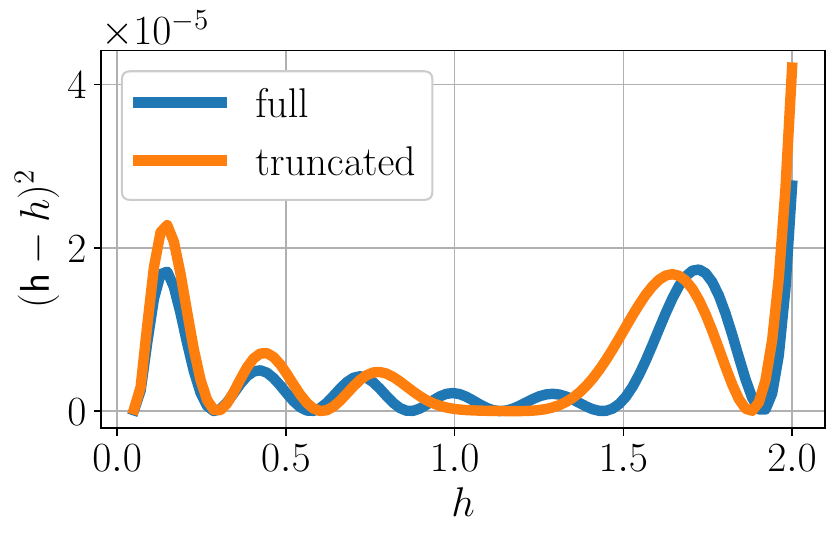}
        \hspace{0.25cm}
        \includegraphics[width=.4125\textwidth]{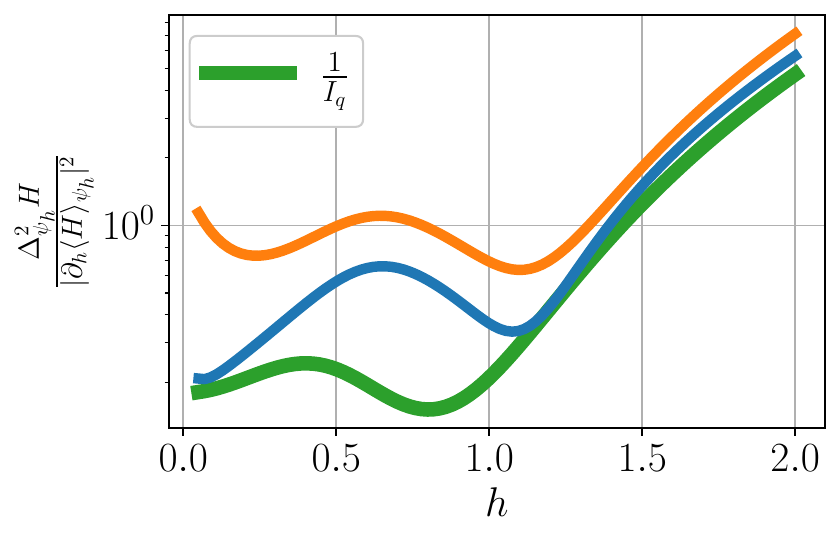}
        \caption{
            Error between the predicted $\mathsf{h} = \langle H_{\boldsymbol{\theta}^*} \rangle_{\psi_h}$ and the true $h$ transverse field of the Ising Hamiltonian (left), and the reduced variance \eqref{eq:estimation_variance} of $H_{\boldsymbol{\theta}^*}$ (right).
            The ``full'' (blue) and ``truncated'' (orange) lines correspond to the ans\"atze described in the main text.
            The solid green line indicates the bound \eqref{eq:var-qcrb}.
        }
        \label{fig:ising-chain}
    \end{figure*}
    
    By inspecting Eq.\eqref{eq:ising}, one can reduce the number of ansatz terms $M$.
    First, since $\mathcal{H}_h$ contains only nearest-neighbor interactions, we use only the Pauli strings of relatively low locality $k = 3$.
    Second, by the same reasoning, we keep the Pauli strings acting only on the neighbor qubits \textit{without} periodic boundary conditions (i.e., the strings such as $X_1 X_2 X_4$ or $X_1 X_7 X_8$ are excluded).
    Third, since $\mathcal{H}_h$ commutes with $X^{\otimes n}$ (spin-flip symmetry) and is invariant to the complex conjugation (time-reversal symmetry), we discard the terms not preserved under these two operations (i.e., remove the strings with odd number of $Y$ and $Z$).
    Thus, we introduce a \textit{truncated} ansatz $H_{\boldsymbol{\theta}}$ made of $M = 90$ Pauli strings.

    In Figure~\ref{fig:ising-chain}, we show the results of predicting the transverse field $h$ as $\mathsf{h} = \langle H_{\boldsymbol{\theta}^*} \rangle_{\psi_h}$. The observable $H_{\boldsymbol{\theta}^*}$ was trained on a set $\mathcal{T} = \big\{ (\ket{\psi_{h_i}}, h_i) \big\}_{i=1}^{10}$ with $h_i$ picked equidistantly from $[0.05, 2]$.
    Recall that the optimal parameters are obtained as $\boldsymbol{\theta}^*=A^{+}b$ with $A$ and $b$ defined in \eqref{eq:matrix_A_num} and \eqref{eq:vector_b_num}.
    As can be seen, our observable accurately predicts the transverse field with the variance relatively close to qCRB \eqref{eq:var-qcrb} for $h \gtrsim 1$.

    For comparison, in Figure~\ref{fig:ising-chain} we also plot the results obtained with the full ansatz having all $M = 1789$ neighboring 3-local Pauli strings with periodic boundary conditions.
    This full ansatz is significantly bigger, but performs nearly the same as the truncated one.


\section{Entanglement quantification}
\label{sec:entanglement}

    In this section, we construct an observable for quantifying the entanglement of a random bipartite state $\rho$ using the method introduced in Sec.~\ref{sec:methods}.
    For this, we consider multiple copies $c$ of a two-qubit state $\rho$, i.e,  $\rho^{\otimes c}$, in Eqs.~\eqref{eq:L} and~\eqref{eq:S}, since a single copy is insufficient to find an observable that predicts the entanglement of arbitrary states (see Refs.~\cite{sancho2000measuring, lu2016tomography, liu2022fundamental, larocca2022group, banchi2025statistical} and Appendix~\ref{app:sec:ent_sym_intu}).

    As an entanglement measure for a bipartite state $\rho$, we consider the negativity~\cite{vidal2002computable}
    \begin{equation}
        \label{eq:negativity} 
        N(\rho)= ||\rho^{T_B}||_1-1,
    \end{equation}
    where $\|\cdot\|_1$ denotes the trace norm, and $\rho^{T_B} \equiv (\Id \otimes T)[\rho]$ is the partial transpose with respect to subsystem $B$. 
    This measure derives from the PPT criterion and quantifies the violation of the PPT condition.

    In Sec.~\ref{sec:Pauli}, we show that Pauli-based ansätze used for predicting the negativity require large training datasets and tend to overfit when data are limited. In Sec.~\ref{sec:anz_choosing}, we analyze the construction of suitable observables using two copies within a regression framework with variance regularization. Finally, in Sec.~\ref{sec:ent_neg} and beyond, we apply this approach to predict entanglement negativity using multiple copies.

\subsection{Training on random mixed states using $k$-local Pauli operators}
\label{sec:Pauli}

We train the observable (\ref{eq:ham-par}) to predict the negativity \eqref{eq:negativity} of two-qubit states $\rho$ using a Pauli-string ansatz with locality $k$. Similar to \eqref{eq:k-local_Huang}, we construct the Hamiltonian
\begin{equation}
    \label{eq:Pauli_ansatz}
    H=\sum_{P\in \{\mathbb{I},X,Y,Z\}^{\otimes 2c}:\,|P|\le k} \theta_P P.
\end{equation}
For $c=4$ copies of $\rho$ (an effective 8-qubit system) a locality of $k=4$ yields a total of 1537 Pauli strings. From extensive numerical experiments 
we find that the combination $c=4$ and $k=4$ achieves the highest prediction quality when trained on 1000 random mixed states.

\begin{figure*}
        \centering
        \includegraphics[width=.495\textwidth]{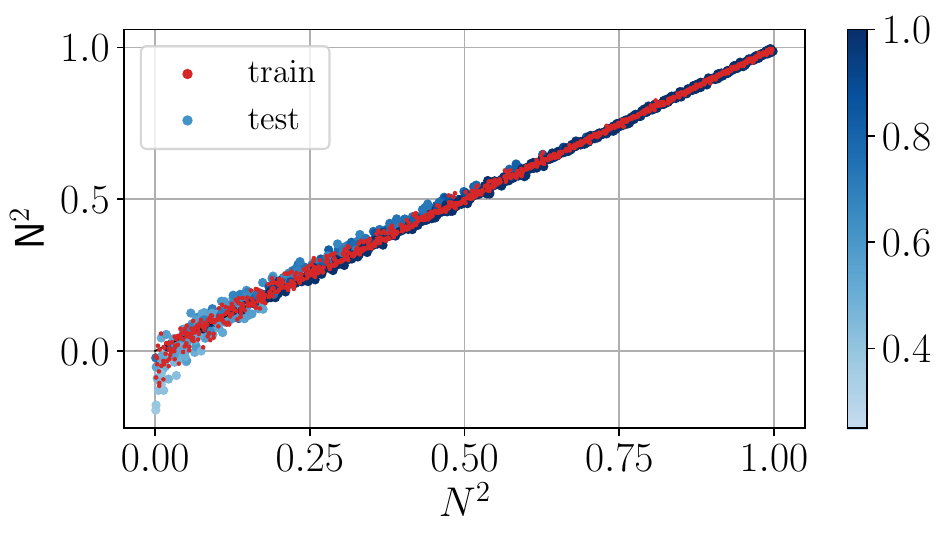}
        \includegraphics[width=.495\textwidth]{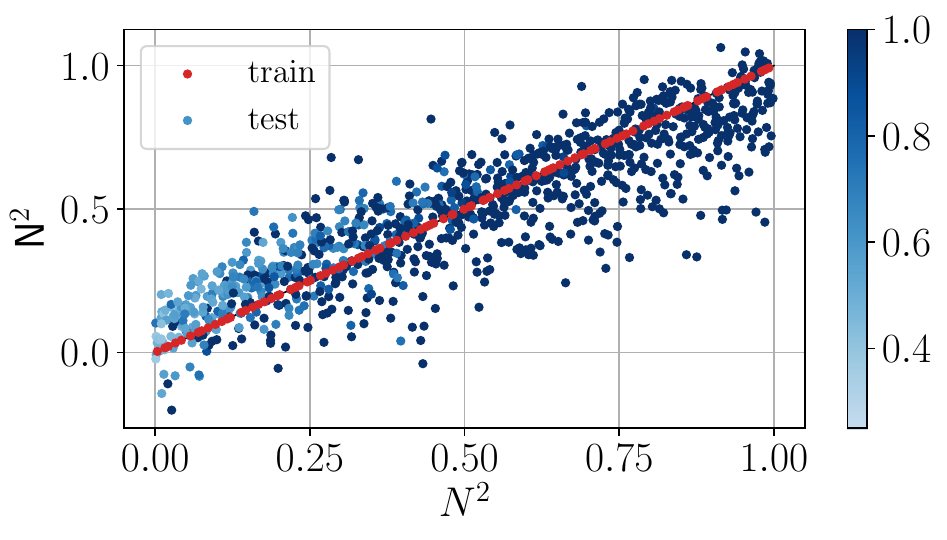}
        \caption{
        Predicted versus true squared negativity $N^2$ for 1000 random mixed states, using 4-local Pauli strings as an ansatz. The model was trained on 1000 (left) and 100  (right) random mixed states, with each training state represented as $\rho_j^{\otimes 4}$ and labeled by its true negativity squared $N_j^2$. The color bars represents the purity of the states.
        }
        \label{fig:Pauli_entanglement}
    \end{figure*}

In Fig.~\ref{fig:Pauli_entanglement}, the left panel shows the prediction performance of the Pauli-string ansatz evaluated on 1000 random test states. Despite its substantially larger parameter space, this ansatz achieves at best comparable prediction quality to the variational model in~\cite{kard2025} which is trained on 1000 random mixed states using 256 parameters and 4 copies of the input state. Hence, an increase of ansatz complexity alone does not necessarily improve entanglement prediction accuracy.

Moreover, when the training dataset is reduced to 100 random mixed states, the Pauli-string ansatz exhibits a pronounced overfitting (right panel of Fig.~\ref{fig:Pauli_entanglement}). Similarly to generic variational circuits, the behavior stems from the high expressivity of the $k$-local Pauli ansatz and fails to exploit the intrinsic structure and symmetries of the entanglement estimation problem. Motivated by this limitation, we avoid such general-purpose ans\"atze and instead design problem-specific constructions. In particular, for entanglement quantification, we propose ans\"atze based on structured combinations of swap operators.

        \subsection{Choosing an ansatz for entanglement quantification problems}
        \label{sec:anz_choosing}

Here we present theoretical insights into the suitability of ans\"atze consisting of swap operators and their products for quantifying entanglement on several copies of random pure and mixed states. For simplicity, we demonstrate the entanglement prediction on $c=2$ copies of a two-qubit state $\rho$. 
Formally, our task is to predict a quantity $\alpha \in [a,b]$ via a function $\xi(\alpha)$ on a fixed family of states $\rho_\alpha$. 

Following the quantum regression method described in Section~\ref{sec:method_param}, one seeks to find an observable $H_0$ such that
\begin{equation}
    H_0\in \arg\min_{H} F[H]
\end{equation}
with
\begin{align}
    &F[H] = \int dU_A\, dU_B \int_a^b d\alpha\, \Bigl[ \wls\left(\xi(\alpha) - \Tr[H \rho_{\scriptscriptstyle \alpha, U_A, U_B}^{\otimes 2}]\right)^2 \nonumber \\
    &\quad+ \wvar\left(\Tr[H^2 \rho_{\scriptscriptstyle \alpha, U_A, U_B}^{\otimes 2}] - \Tr^2[H \rho_{\scriptscriptstyle \alpha, U_A, U_B}^{\otimes 2}]\right) \Bigr], \label{eq:haar_var_reg}
\end{align}
where
\begin{equation}
    \rho_{\scriptscriptstyle \alpha, U_A, U_B} = (U_A \otimes U_B)\, \rho_\alpha\, (U_A \otimes U_B)^{\dagger}. \label{eq:state_rot}
\end{equation}
The state~\eqref{eq:state_rot} is defined over the full range $\alpha \in [a, b]$, and the integration in~\eqref{eq:haar_var_reg} with respect to the Haar measures $dU_A$ and $dU_B$ over unitaries acting on subsystems $A$ and $B$ implements the randomization of states. The functional $F[H]$ approximates the cost function~\eqref{eq:cost} provided that the training labels $\alpha$ are sufficiently densely and uniformly distributed.

The optimal observable $H_0$ can be found by plugging in ~(\ref{eq:haar_var_reg}) a weakly perturbed operator $H=H_0+ \epsilon Y$, where $\epsilon Y$ is an arbitrary Hermitian operator. The derivation procedure follows Ref.~\cite{kard2025}, the only difference is additional integration with respect to Haar measure which can be postponed to the final stage. Setting the terms in front of $\epsilon$ to zero yields the necessary condition of extrema, the equation for the optimal observable is
\begin{multline}
\label{eq:opt_obs_reg}
\frac12\left(\Tilde{\rho} H_0+H_0\Tilde{\rho}\right) - k\int dU_AdU_B\int_a^b \xi(\alpha)\rho_{\scriptscriptstyle \alpha,U_A,U_B}^{\otimes2}d\alpha \\
    + (k-1)\int dU_AdU_B\int_a^b\mathrm{Tr}\{H_0\rho_{\scriptscriptstyle \alpha,U_A,U_B}^{\otimes2}\}\rho_{\scriptscriptstyle \alpha,U_A,U_B}^{\otimes2}d\alpha = 0, 
\end{multline}
where $k = \wls/\wvar$ and
\begin{equation}\label{eq:rho_global}
    \Tilde{\rho} = \int\,dU_A\,dU_B\,\int_a^b\,\rho_{\scriptscriptstyle \alpha,U_A,U_B}^{\otimes2}\,d\alpha.
\end{equation}
The limit $k \to \infty$ in Eq.~\eqref{eq:opt_obs_reg} corresponds to neglecting variance regularization, i.e., $\wls = 1$, $\wvar = 0$.
Moreover, a solution can be sought in a specific form motivated by symmetry:
\begin{equation}\label{eq:haar_sol_form}
    H_0 = \theta_1\Id + \theta_2 \S_{A_0A_1} + \theta_3 \S_{B_0B_1} + \theta_4 \S_{A_0A_1}\S_{B_0B_1},
\end{equation}
where $A_0, A_1$ and $B_0, B_1$ are copies of subsystems $A$ and $B$ of $\rho$, respectively, and $\S$ is the swap operator exchanging the states of the indicated subsystems. 

This ansatz is natural because the identity and swap operators generate the algebra of operators that commute with arbitrary local unitaries of the form \( U^{\otimes n} \).\footnote{More generally, by Schur-Weyl duality, any operator commuting with all unitaries \( U^{\otimes n} \) is a linear combination of permutation operators.} Since many entanglement quantifiers, witnesses, and purity checks are constructed from the permutation operators, this form is particularly suitable for entanglement problems.

Crucially, due to this symmetry, the expectation value \( \mathrm{Tr}(H_0\,\rho_{\alpha,U_A,U_B}^{\otimes 2}) \) is independent of the local unitaries \( U_A \),\( U_B \). 
Therefore, the Haar integration in Eq.~(\ref{eq:opt_obs_reg}) acts non-trivially only on \( \rho_{\alpha,U_A,U_B}^{\otimes 2} \). 
This integration yields a linear combination of at most 4 linearly independent operators (see Appendix~\ref{app:est_theor} for details) which implies that the number of constraints on the coefficients \( \theta_1, \theta_2, \theta_3, \theta_4 \) does not exceed 4. Consequently, a solution to Eq.~(\ref{eq:opt_obs_reg}) in the form of Eq.~(\ref{eq:haar_sol_form}) is guaranteed to exist.

In view of the above, to solve the optimization problem~(\ref{eq:regression-H}) for entanglement quantification numerically, as an ansatz for $H_{\btheta}$ in \eqref{eq:ham-par} we take   
\begin{equation}    
        \mathcal{G} = \{\Id_{AB}, \S_{A_0 A_1}, \S_{B_0B_1}, \S_{A_0 A_1}\S_{B_0B_1}\}.
        \label{eq:ansatz_2copies}
    \end{equation}
While we obtain the observable in Eq.~\eqref{eq:haar_sol_form} analytically by solving Eq.~\eqref{eq:opt_obs_reg} for structured   bipartite states $\rho_\alpha$, we use the ansatz in~\eqref{eq:ansatz_2copies} to numerically implement the method introduced in Sec.~\ref{sec:methods}. We then compare the theoretical and numerical observables and their performance on a test set.

\subsection{Entanglement negativity quantification}
\label{sec:ent_neg}

Above we derived Eq.~\eqref{eq:opt_obs_reg} for the optimal observable. 
For $c=2$ copies of $\rho_\alpha$, we can seek a solution of the form of Eq.~\eqref{eq:haar_sol_form}. In this section, we apply our method to predict entanglement, focusing on the squared negativity, $\alpha = N^2$, with $\xi(\alpha) = \alpha$.
Accordingly, the training data are sampled uniformly from $N^2$.

We aim at the squared negativity $N^2$ rather than $N$, since the expectation value $\Tr[H\, \rho^{\otimes c}]$ is polynomial in $\rho$, while $N$ is not, see \eqref{eq:negativity}. The squared negativity $N^2$ admits a much better polynomial approximation, making it a better target. A further insight is that the squared negativity $N^2$ is upper-bounded by a two-copy observable, whose expectation value is polynomial in $\rho$ \cite{NegUpBound08}. Moreover, for two-qubit pure states, the negativity coincides with the concurrence $C$ \cite{NegConcRel15} which admits a polynomial lower bound from a suitable observable~\cite{MintBuch07}. For two-qubit pure states, the squared negativity $N^2$ can be predicted exactly by a linear combination of permutation operators~\cite{NegUpBound08}.

\subsubsection{Pure and isotropic states with for $c=2$ copies}
\label{sec:pure_iso_states}

\begin{figure*}[tbh]
            \centering
            \includegraphics[width=.495\textwidth]{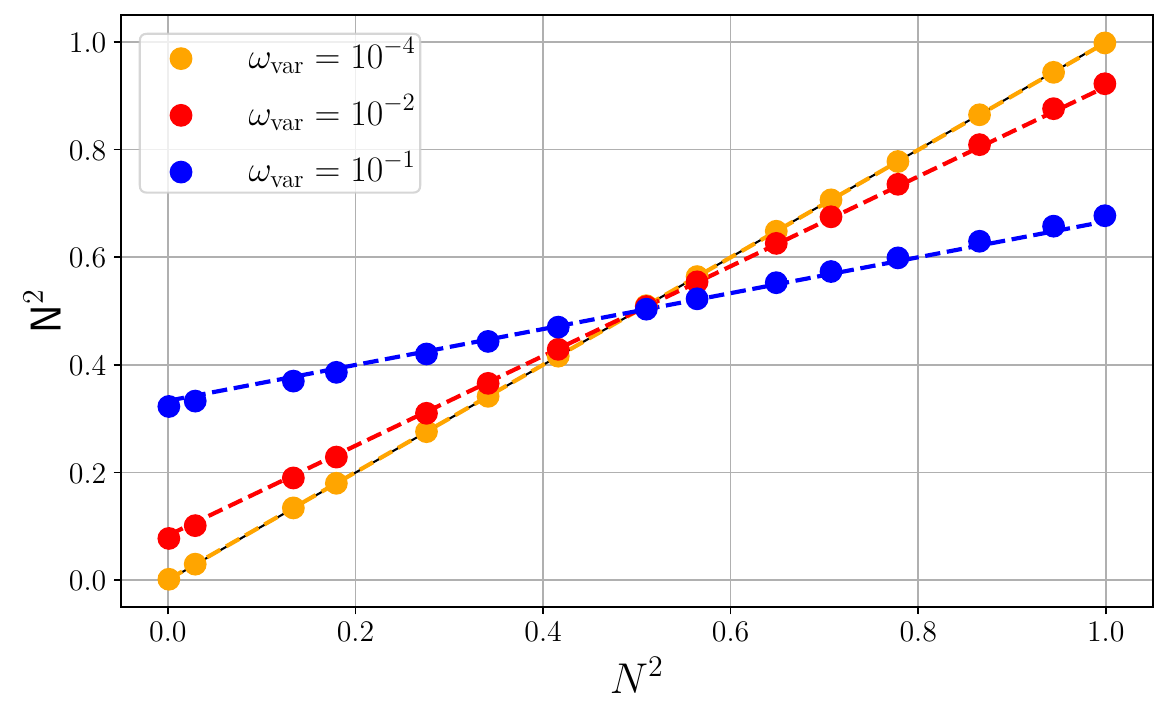}
            \includegraphics[width=.495\textwidth]{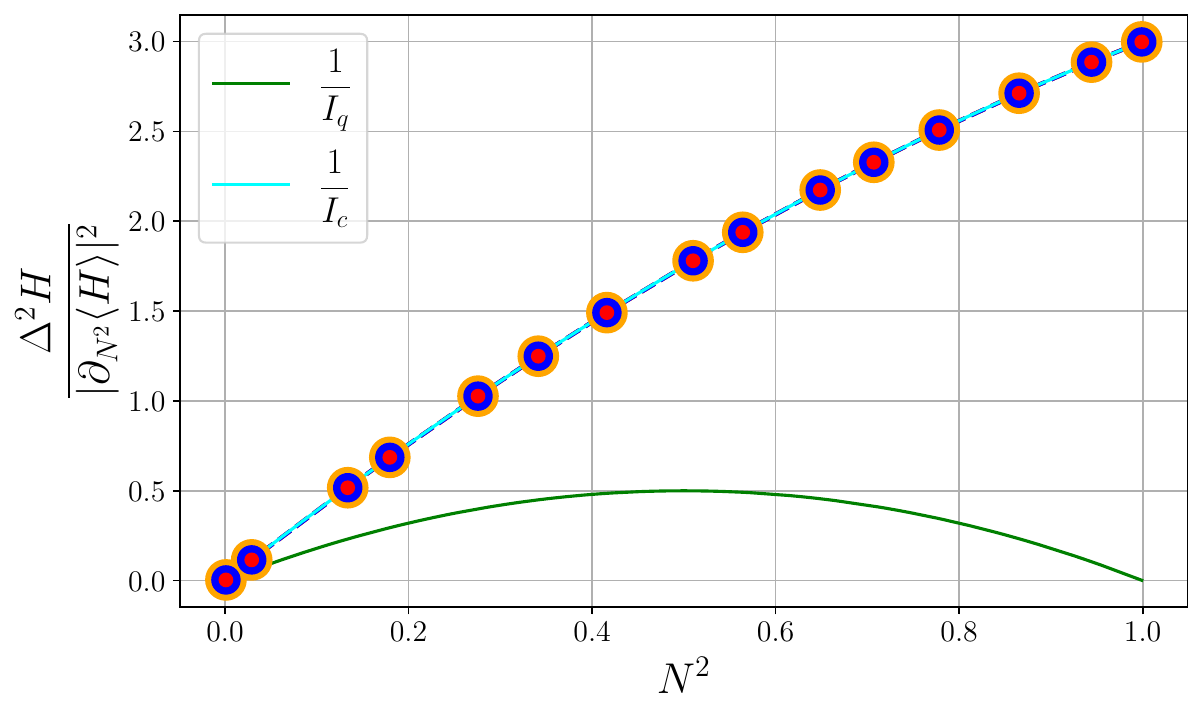}
            \caption{
                Left: Predicted squared negativity $\mathsf{N}^2 = \langle H \rangle_{\psi^{\otimes 2}}$ of random pure states $\ket{\psi}$ vs. the true squared negativity $N^2$ for $c=2$ copies. The black line connecting $(0,0)$ and $(1,1)$ is the ground truth. 
                Right: Reduced variance \eqref{eq:estimation_variance} of the trained observable $H$. 
                The dashed lines correspond to the analytical results obtained by solving \eqref{eq:opt_obs_reg} (see Appendix~\ref{sec:results_neg_pure}).
                The solid lines show the classical and quantum Cramer-Rao bounds \eqref{eq:var-qcrb}.
                The the observable $H$ has the form \eqref{eq:haar_sol_form} and its coefficients are trained on a set $\mathcal{T} = \big\{(\ket{\psi_j}^{\otimes 2}, N_j^2 )\big\}_{j=1}^{20}$ with random $\ket{\psi_j}$ and $N_j^2$ evenly distributed on $[0, 1]$. 
            }
            \label{fig:pre_neg_sq_pure}
        \end{figure*}

\begin{figure*}[tbh]
            \centering
            \includegraphics[width=.495\textwidth]{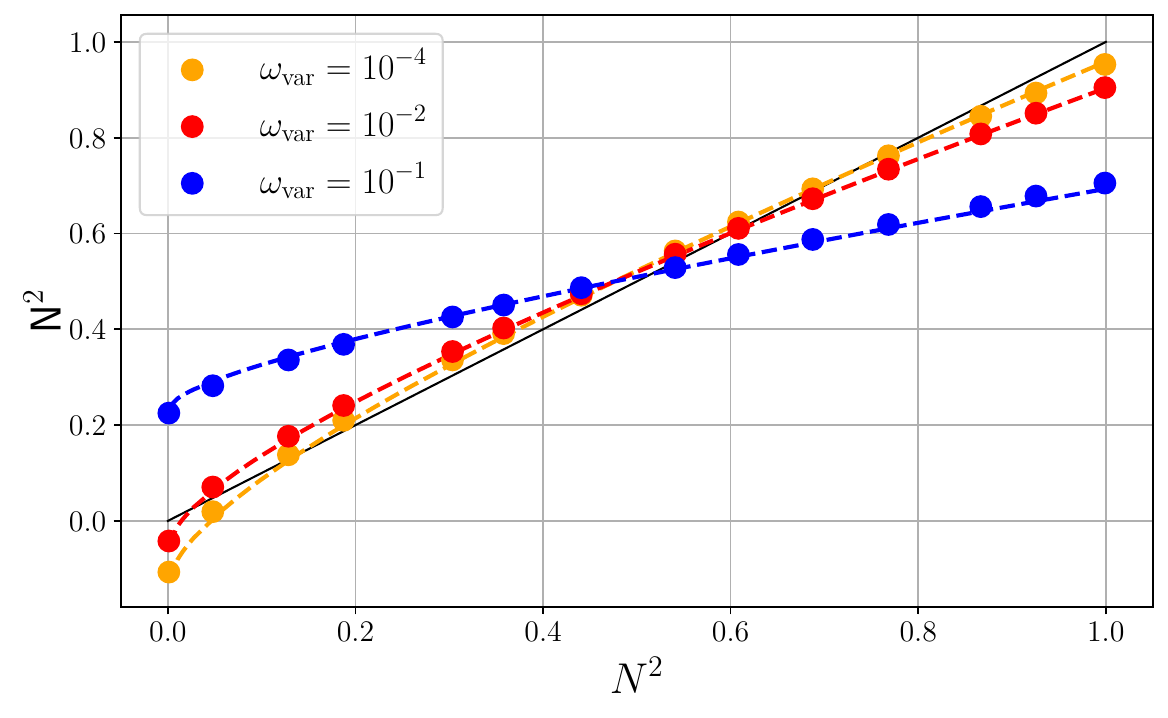}
            \includegraphics[width=.495\textwidth]{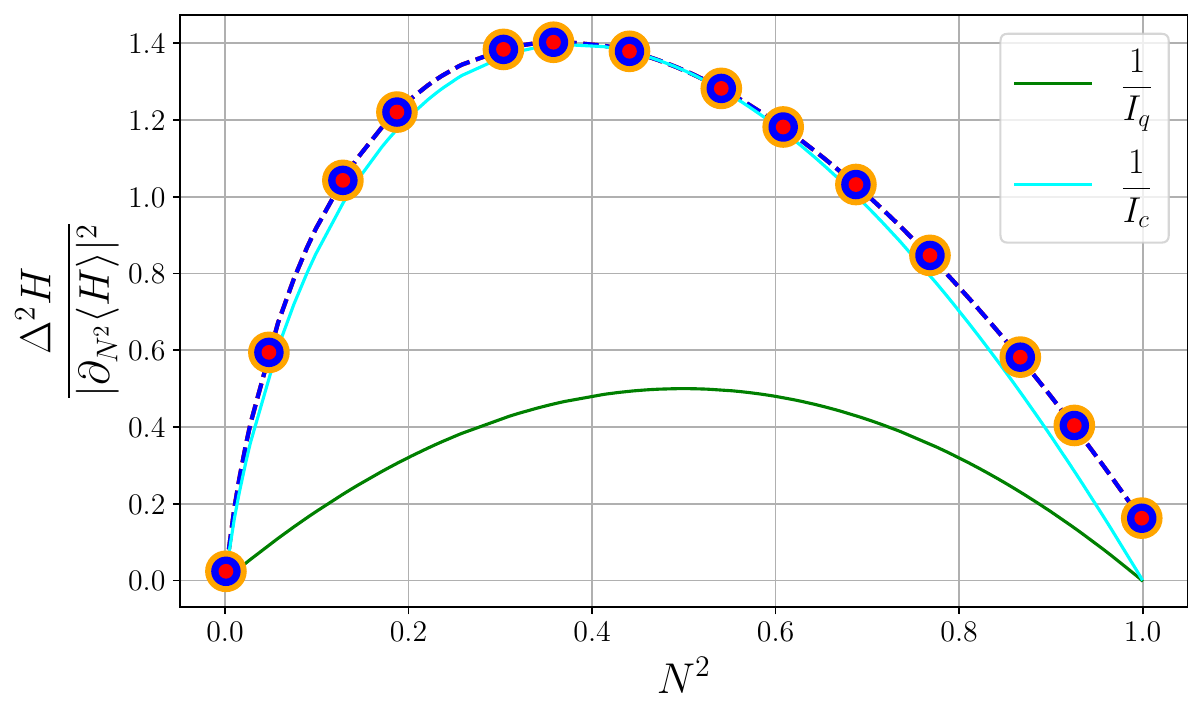}
            \caption{
                Left: Predicted squared negativity $\mathsf{N}^2 = \Tr H \rho_q^{\otimes 2}$ of random isotropic states $\rho_q$ vs. the true squared negativity $N^2$ for $c=2$ copies. 
                The black line connecting $(0,0)$ and $(1,1)$ is the ground truth. 
                Right: Reduced variance \eqref{eq:estimation_variance} of the trained observable $H$.
                The dashed lines correspond to the analytical results obtained by solving \eqref{eq:opt_obs_reg} (see Appendix~\ref{sec:results_neg_iso}).
                The solid lines show the classical and quantum Cramer-Rao bounds \eqref{eq:var-qcrb}.
                The the observable $H$ has the form \eqref{eq:haar_sol_form} and its coefficients are trained on a set $\mathcal{T} = \big\{(\rho_j^{\otimes 2}, N_j^2 )\big\}_{j=1}^{20}$ with random $\rho_j$ and $N_j^2$ evenly distributed on $[0, 1]$.
            }
            \label{fig:pre_neg_sq_iso}
        \end{figure*}

We predict the entanglement of bipartite pure and isotropic states of $n=2$ qubits. We consider the pure states
\begin{equation}
    \label{eq:pure_qubit}
    \ket{\psi}_{\scriptscriptstyle U_A, U_B} = \left(U_A\otimes U_B\right)\sum_{i=1}^2\sqrt{\lambda_i}\ket{i_A i_B},
\end{equation}
and the two-qubit isotropic states
\begin{equation}
    \label{eq:iso_qubit}
    \rho_q = q\ketbra{\Phi} + \frac{1-q}{4}\Id,
\end{equation}
where $\ket{\Phi} = \frac{1}{\sqrt{2}}(\ket{00}+\ket{11})$ is the Bell state.
Recall that finding an observable $H$ for prediction is done by obtaining the optimal coefficients for $H$ as  $\boldsymbol{\theta}^*=A^{+}b$ with $A$ and $b$ defined in \eqref{eq:matrix_A_num} and \eqref{eq:vector_b_num}, and with the observable $H$ having the form \eqref{eq:haar_sol_form}.
The prediction results for the optimized observable are shown in Fig.~\ref{fig:pre_neg_sq_pure} for pure states and in Fig.~\ref{fig:pre_neg_sq_iso} for isotropic states, for regularization weights $\wls=1$ and $\wvar \in \{10^{-4}, 10^{-2}, 10^{-1}\}$. 
The theoretical values are obtained by solving Eq.~\eqref{eq:opt_obs_reg} with a Hamiltonian of the form~\eqref{eq:haar_sol_form} for each family of states separately. For the analytical expressions, see Appendix~\ref{sec:results_neg_pure} (pure states) and Appendix~\ref{sec:results_neg_iso} (isotropic states).

As can be seen in the figure, the best prediction accuracy is achieved for $\wls=1$ and $\wvar=10^{-4}$ in both cases.
While the resulting observable saturates the classical Cramer--Rao bound in \eqref{eq:var-qcrb}, it does not saturate its quantum counterpart for the considered values of $\wvar$.
Analogous analytical and numerical results for the entanglement tangle, using two copies of bipartite qubit and qutrit states, are presented in Appendix~\ref{app:est_theor}.

\subsubsection{Permutation operators as an ansatz for $c>2$ copies}

So far we have considered only $c = 2$ copies. We now turn to $c > 2$ to improve the prediction quality and to identify an optimal set of operators $G_j$ for the observable \eqref{eq:ham-par}. We begin with the general formulation.

Let $S_c$ denote the symmetric group on $c$ elements. For $\pi \in S_c$, the permutation operator $G_d(\pi)$ on $\mathcal{H}^{\otimes c}$ is defined by its action on product states:
\begin{equation}
    G_d(\pi)\,\ket{\psi_1} \otimes \dots \otimes \ket{\psi_c} = \ket{\psi_{\pi^{-1}(1)}} \otimes \dots \otimes \ket{\psi_{\pi^{-1}(c)}},
\end{equation}
for all $\ket{\psi_1},\dots,\ket{\psi_c}\in\mathbb{C}^d$. For a bipartite system with subsystems $A$ and $B$, and permutations $\pi_A, \pi_B \in S_c$, we define the bipartite permutation operator
\begin{equation}
    G(\pi_A, \pi_B) = G_{d_A}(\pi_A) \otimes G_{d_B}(\pi_B),
\end{equation}
acting on $(\mathcal{H}_A \otimes \mathcal{H}_B)^{\otimes c}$. From here on we consider only the qubit systems, i.e., $d_A = d_B = 2$, and, thus, drop the indices $d_A$ and $d_B$. The operators $G(\pi_A)$ and $G(\pi_B)$ act only on their respective subsystems. We emphasize that the qubits are not physically permuted; rather, we use these operators to perform measurements, constructing our ansatz elements as $G_j = G(\pi_A) \otimes G(\pi_B) \equiv G_A \otimes G_B$.
\begin{figure}[!t]
            \centering
            \includegraphics[width=.495\textwidth]{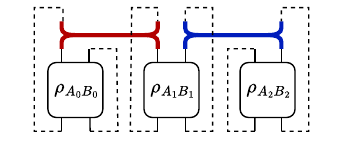}
            \caption{
                Representation of $\operatorname{Tr}(\mathrm{S}_{A_0A_1}\mathrm{S}_{B_1B_2}\rho_{AB}^{\otimes 3})$. Dashed lines indicate trace operation, while the red and blue lines represent the swap operators $\mathrm{S}_{A_0A_1}$ and $\mathrm{S}_{B_1B_2}$, respectively.
            }
            \label{fig:sw_example}
        \end{figure}

Since any permutation in $S_c$ can be decomposed into a product of transpositions, each operator $G_j$ can be expressed as a product of elementary swap operators $\S_{A_i A_k}$ and $\S_{B_i B_k}$, which exchange the $i$-th and $k$-th copies within subsystems $A$ and $B$, respectively. 
In Fig.~\ref{fig:sw_example}, we show, as an example, a diagrammatic representation of $\operatorname{Tr}(\mathrm{S}_{A_0 A_1}\mathrm{S}_{B_1 B_2}\, \rho_{AB}^{\otimes 3})$, in which the observable $\mathrm{S}_{A_0 A_1}\mathrm{S}_{B_1 B_2}$ is measured in the state $\rho_{AB}^{\otimes 3}$. 
As in Eq.~\eqref{eq:L}, we employ the operators $G_j$ to perform measurements on the $c$-copy state $\rho_{AB}^{\otimes c}$. 
As noted in Sec.~\ref{sec:anz_choosing} and Appendix~\ref{app:sec:ent_sym_intu}, permutation operators commute with local unitaries, making them natural candidates for entanglement-related observables. Our goal is to identify observables $H$ expressible as linear combinations of permutation operators.

\begin{table*}[!t]
\centering
\caption{Number of permutation operators for different copy counts $c$ in $\rho_{AB}^{\otimes c}$.}
\label{tab:c_counts}  
\begin{tabular}{|l|c|c|c|c|}
\hline
\textbf{Case} & \textbf{Overall Terms} & \textbf{Independent} & \textbf{Independent \& Hermitized}  & \textbf{Independent \& Hermitian} \\
\hline
$c=2$ & $2_A!\times2_B!=4$ & $2_A\times2_B=4$ & $2_A\times2_B=4$ &$2_A\times2_B=4$ \\
\hline
$c=3$ & $3!\times3!=36$& $5_A\times5_B=25$ &$17$& $4_A\times 4_B=16$ \\
\hline
$c=4$ & $4!\times4!=576$ & $14_A\times14_B=196$ &$116$ & $10_A\times10_B= 100$ \\
\hline
\end{tabular}
\end{table*}

\begin{table}[!t]
\centering
\caption{Ten classes of permutation operators for $c=4$ copies of $\rho\equiv\rho_{AB}$. $\rho_{A,B}$ denotes the reduced state $\Tr_{B,A} \rho$}.
\label{tab:10-class_short}  
\begin{tabular}{|l|c|c|c|c|}
        \hline
        Class & Number of operators & Formula  \\
        \hline
        1& 3 & $(\Tr \big[\rho^2\big])^2$ \\
        \hline
        2& 6 & $\Tr \big[\rho^2\big]$\\
        \hline
        3&6 & $\Tr\big[(R^\dagger R)^2\big]$ \\
        \hline
        4A&6 & $\Tr \big[\rho_A^2\big]\Tr \big[\rho^2\big]$\\
        4B&6 & $\Tr \big[\rho_B^2\big]\Tr \big[\rho^2\big]$\\
        \hline
        5A & 12& $\Tr\big[\Tr_A\big[((\rho_A\otimes\Id)\rho)^2\big]\big]$\\
        5B & 12 &$\Tr\big[\Tr_B\big[((\Id\otimes \rho_B)\rho)^2\big]\big]$\\
        \hline
        6AB & 6 and 6 & $\Tr \big[\rho_A^2\big]$ and $\Tr \big[\rho_B^2]$ \\
        \hline
        7 &24 & $\Tr \big[\rho(\rho_A\otimes \rho_B)\big]$ \\
        \hline
        8 & 6 & $\Tr\big[\rho_A^2\big] \Tr\big[\rho_B^2]$\\
        \hline
        9AB & 3 and 3 & $(\Tr\big[\rho_A^2\big])^2$ and $(\Tr\big[\rho_B^2])^2$ \\
        \hline
        10 & 1 & 1 \\
        \hline
\end{tabular}
\end{table}

For $c > 2$ copies of $\rho_{AB}$, arbitrary permutations $\pi_A, \pi_B$ may yield non-Hermitian operators $G_j$ which are unphysical as observables. 
Specifically, these non-Hermitian permutations occur when $\pi_A$ or $\pi_B$ contains at least one cycle of three or more elements.\footnote{For example, $G_A = \S_{A_0 A_1}\S_{A_1 A_2}\S_{A_2 A_3}$ implements the $4$-cycle $\pi_A = (0\,1\,2\,3)$ and is therefore non-Hermitian, while $G_A = \S_{A_0 A_2}\S_{A_1 A_3}$ implements the disjoint transpositions $\pi_A = (0\,2)(1\,3)$ and is Hermitian.}
To address this limitation, we can construct physical observables by \textit{hermitizing} the ansatz elements $G_j$ through the transformation:
\begin{equation}
    G_j\rightarrow \frac{1}{2}(G_j + G_j^\dagger).
    \label{eq:hermitizing}
\end{equation}
This procedure ensures that the resulting operators are Hermitian and therefore represent valid physical observables while retaining the ansatz permutation structure.

For the case of $c=2$ copies of $\rho_{AB}$, we have only four operators, forming a candidate observable as in Eq.~\ref{eq:haar_sol_form}. For $c>2$, the number of permutation operators grows as $c! \times c!$, creating an excessive overhead for quantum devices. For instance, with $c=4$, this yields $576$ potential terms, accounting for all combinations of permutations $\pi_A$ and $\pi_B$ acting on subsystems $A$ and $B$, respectively.
To achieve a practical implementation, we restrict the ansatz to only linearly independent permutation operators $G_A \otimes G_B$ generated from $\pi_A$ and $\pi_B$. For $c=4$, this reduces the number of terms to $14$ operators per subsystem, yielding $14^2 = 196$ combined operators $G_A \otimes G_B$. 
Within these $14$ operators per subsystem, $10$ can be selected to be Hermitian, directly providing Hermitian permutation operators $G_A^{\text{herm}}$, $G_B^{\text{herm}}$. The tensor products $G_A^{\text{herm}} \otimes G_B^{\text{herm}}$ account for $10^2 = 100$ elements. 
The remaining $96$ operators require hermitization as in Eq.~\eqref{eq:hermitizing}. The hermitization process reduces the independent 196 operators to 116 linearly independent operators. Table~\ref{tab:c_counts} summarizes these counts for $c=2,3,4$.

With numerical tests, we find that 100 independent and inherently Hermitian permutation operators for predicting the entanglement of $\rho_{AB}$ yield nearly the same performance as 116 independent and hermitized permutation operators.
Therefore, we focus our subsequent analysis on the performance and the structure of these 100 (inherently) Hermitian permutation operators $G_A^{\text{herm}} \otimes G_B^{\text{herm}}$.

\subsubsection{Hermitian permutation operators}
\label{sec:herm}

As established previously, for $c=4$ copies of $\rho_{AB}$ there exist $100$ independent Hermitian permutation operators. 
We used these operators as an ansatz for the observable $H_{\btheta}$ in \eqref{eq:ham-par} and trained it on an set of $N=1000$ random mixed states, and observed that the optimized coefficients $\btheta^*$ for these Hermitian permutation operators cluster into distinct groups with nearly identical values of $\btheta^*$. 
By analyzing the operations within these groups, we found that each group is characterized by either a unique mathematical operation or a pair of operations symmetric across subsystems $A$ and $B$. This analysis led us to classify the operators into ten distinct classes summarized in Table~\ref{tab:10-class_short}. Full details of these 100 operators of 10 classes are presented in Appendix \ref{sec:app:10-class_derivations}. 

The operators within each class yield nearly identical optimized parameters $\btheta^*$ and can be combined into a single Hermitian operator per class. Formally, considering 10 classes, we define the class representatives as:
\begin{equation}
    \label{eq:Hj_ansatz}
    H_j = \sum_{G_A^{\text{herm}} \otimes G_B^{\text{herm}} \in \text{Class j}} G_A^{\text{herm}} \otimes G_B^{\text{herm}},
\end{equation}
which allows us to express the observable \eqref{eq:ham-par} as
\begin{equation}
    \label{eq:H10_ansatz}
    H = \sum_{j=1}^{10} \theta_j H_j.
\end{equation}

Notably, as can be seen in Table~\ref{tab:10-class_short}, this ansatz naturally incorporates fundamental entanglement-theoretic concepts: Class~2 encodes purity measurements, Class~3 corresponds to an operation $R(\rho_{AB})_{i,j;k,l} = (\rho_{AB})_{i,k;j,l}$ which is referred to as realignment \cite{chen2002matrix}, and Class~6 captures the linear entropy entanglement.
Additionally, it encompasses various polynomial operations involving $\rho$ and its reduced states $\rho_{A} \equiv \Tr_{B} \rho$ and $\rho_{B} \equiv \Tr_{A} \rho$. 
These classes are discussed in more detail in Appendix \ref{sec:app:10-class_derivations}

\begin{figure*}[tbh]
        \centering
        \includegraphics[width=.495\textwidth]{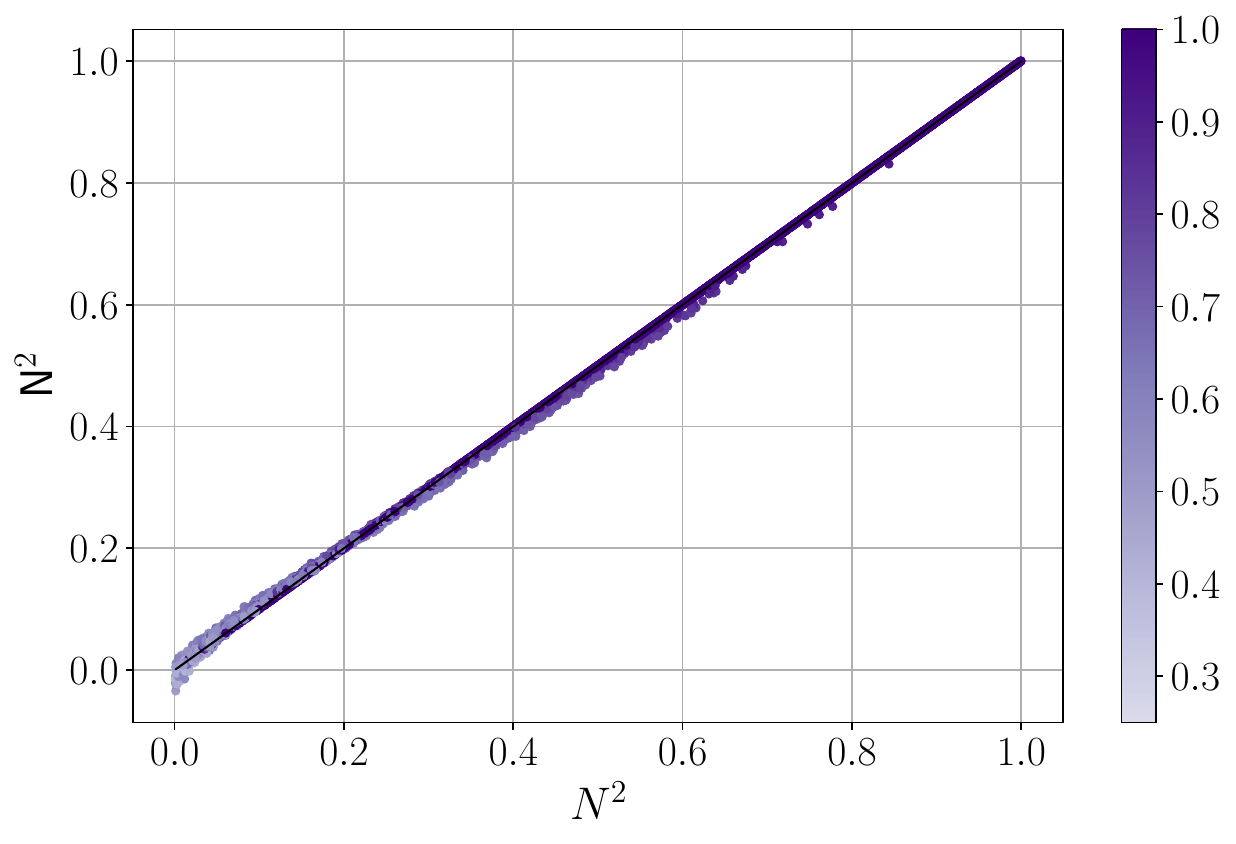}
        \includegraphics[width=.495\textwidth]{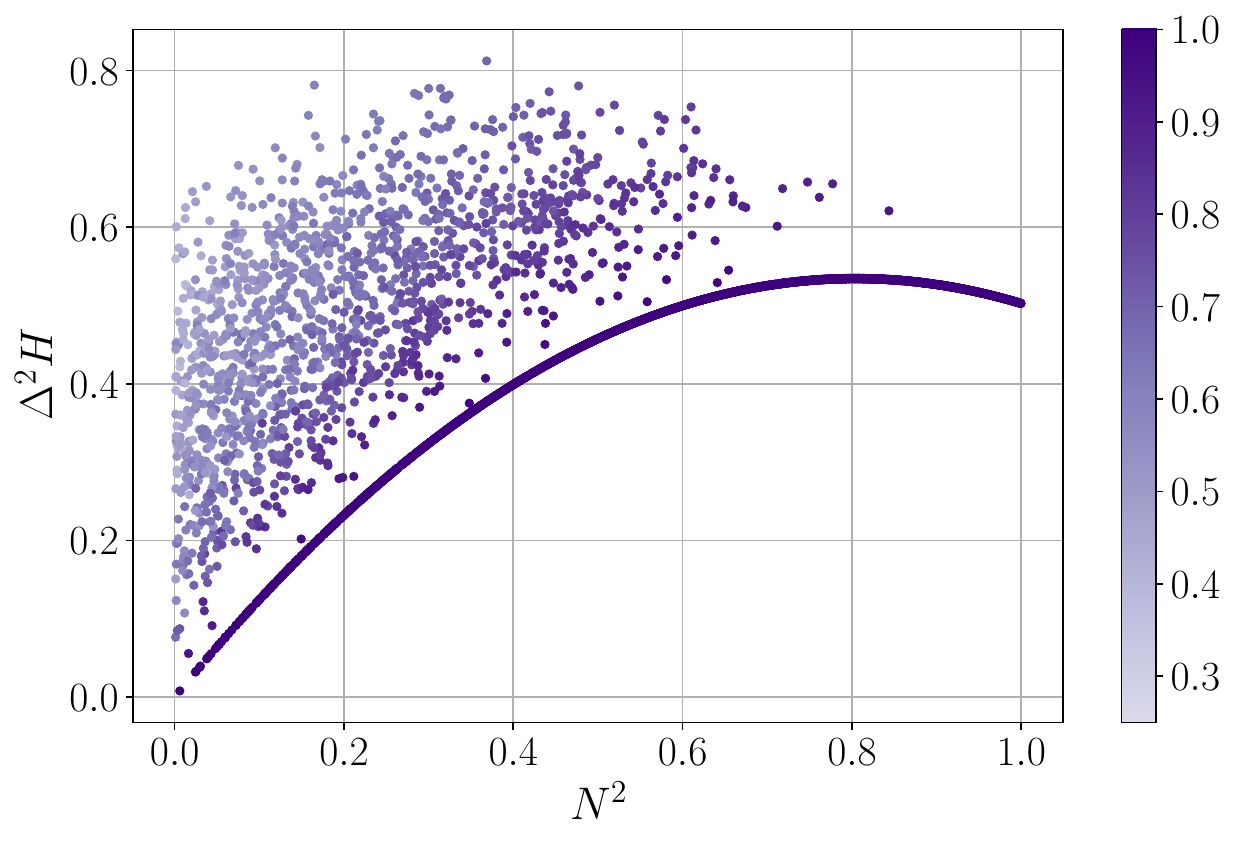}
        \caption{
             Results of training and testing on \textit{mixed} states. Left: Predicted squared negativity $\mathsf{N}^2 = \Tr H \rho^{\otimes 4}$ of $5000$ random mixed states $\rho$ vs. the true negativity $N^2$ for $c=4$ copies. The black line connecting $(0,0)$ and $(1,1)$ is the ground truth. 
             Right: Variance of the trained observable $H$.
             The color of each point represents the purity of the corresponding state. The observable is trained using the 10-class Hermitian ansatz in Eq.~\eqref{eq:H10_ansatz} on a dataset of 100 mixed states uniformly distributed on $N^2\in [0,1]$.
        }
        \label{fig:neg_square10mixedXmixed}
    \end{figure*}

In Fig.~\ref{fig:neg_square10mixedXmixed}, we show the prediction of the squared negativity $\mathsf{N}^2 =\Tr\big[\rho_{N^2}H\big]$ and the variance\footnote{Note that we cannot compute the reduced variance $\operatorname{rVar}(H_{\btheta})$ as in Eq.~\eqref{eq:estimation_variance}, since the random sampling of $\rho_\alpha$ does not yield a differentiable dependence on $\alpha$.}
$\Delta^2H=\Tr\big[\rho_{N^2}H^2\big]-(\Tr\big[\rho_{N^2}H\big])^2$ of the observable $H$ in Eq.~\eqref{eq:H10_ansatz}, with individual operators defined in Eq.~\eqref{eq:Hj_ansatz} for a test dataset $\mathcal{T}_{\mathrm{test}}=\{(\rho_i,N^2_i)\}_{i=1}^{5000}$ with random two-qubit states $\rho_i$.
The prediction maintains a good accuracy even when trained on 100 mixed states sampled randomly. 

One may ask how to measure the observables \eqref{eq:Hj_ansatz} in practice.
In Appendix~\ref{app:sec:measurements}, we propose an experimental procedure that, using only 9 parallel‑measurement setups, fully captures the outcomes for all 100 Hermitian operators.
These outcomes enable the evaluation of the expected values of 10‑Hermitian‑operator ansatz in~\eqref{eq:H10_ansatz}.

\subsection{Train on easy states, test on mixed states}
\label{sec:easy_states}

In the previous section, we considered training and testing an observable on random mixed states.
Here, we study the performance of observables trained on ``easy'' states, i.e., states with analytically tractable entanglement measures and experimental feasibility.
Among such states are the pure states~\eqref{eq:pure_qubit} and isotropic states~\eqref{eq:iso_qubit}.
We focus on the case of $c=4$ copies of $\rho$, employing the 10 class ansatz in Eq.~\eqref{eq:H10_ansatz}.
For comparison, the case of $c=2$ as input with the ansatz Eq.~\eqref{eq:ansatz_2copies} is presented in Appendix~\ref{app:sec:easy_2copies}.

As discussed in Sec.~\ref{sec:methods} and Appendix~\ref{app:sec:comb_datasets}, datasets for training can be processed independently. 
That is, for pure and isotropic states (see Section~\ref{sec:pure_iso_states}) one computes
\begin{align}
    A_{\mathrm{easy}} &= A_{\mathrm{pure}} + A_{\mathrm{iso}}, \nonumber\\
    b_{\mathrm{easy}} &= b_{\mathrm{pure}} + b_{\mathrm{iso}}.
\end{align}
Here, the matrices $A_{\mathrm{pure,\, iso}}$ and vectors $b_{\mathrm{pure,\, iso}}$ are constructed using Eqs.~\eqref{eq:matrix_A_num} and \eqref{eq:vector_b_num} with evaluating the arrays $L$ and $S$ in \eqref{eq:L} and \eqref{eq:S} for pure and isotropic states using the ansatz \eqref{eq:Hj_ansatz}.
The optimal coefficients of the ansatz operators can be obtained as $\boldsymbol{\theta}^* = A_{\mathrm{easy}}^+ b_{\mathrm{easy}}$. If analytical expressions for the states are available, these quantities can alternatively be computed directly, as discussed in Appendix~\ref{app:sec:theory_Ab}.

In Fig.~\ref{fig:neg_square10classisopureXmixed}, we show the performance of the resulting observable.
As can be seen, the ansatz~\eqref{eq:Hj_ansatz} yields an observable with strong generalization capabilities.
That is, trained on easy (i.e., pure and isotropic) states, it accurately predicts the entanglement of random mixed states.
This is in sharp contrast to the results in Sec.~\ref{sec:Pauli}, where we showed that an ansatz of $k$-local Pauli strings tends to overfit.

    \begin{figure*}
        \centering
        \includegraphics[width=.495\textwidth]{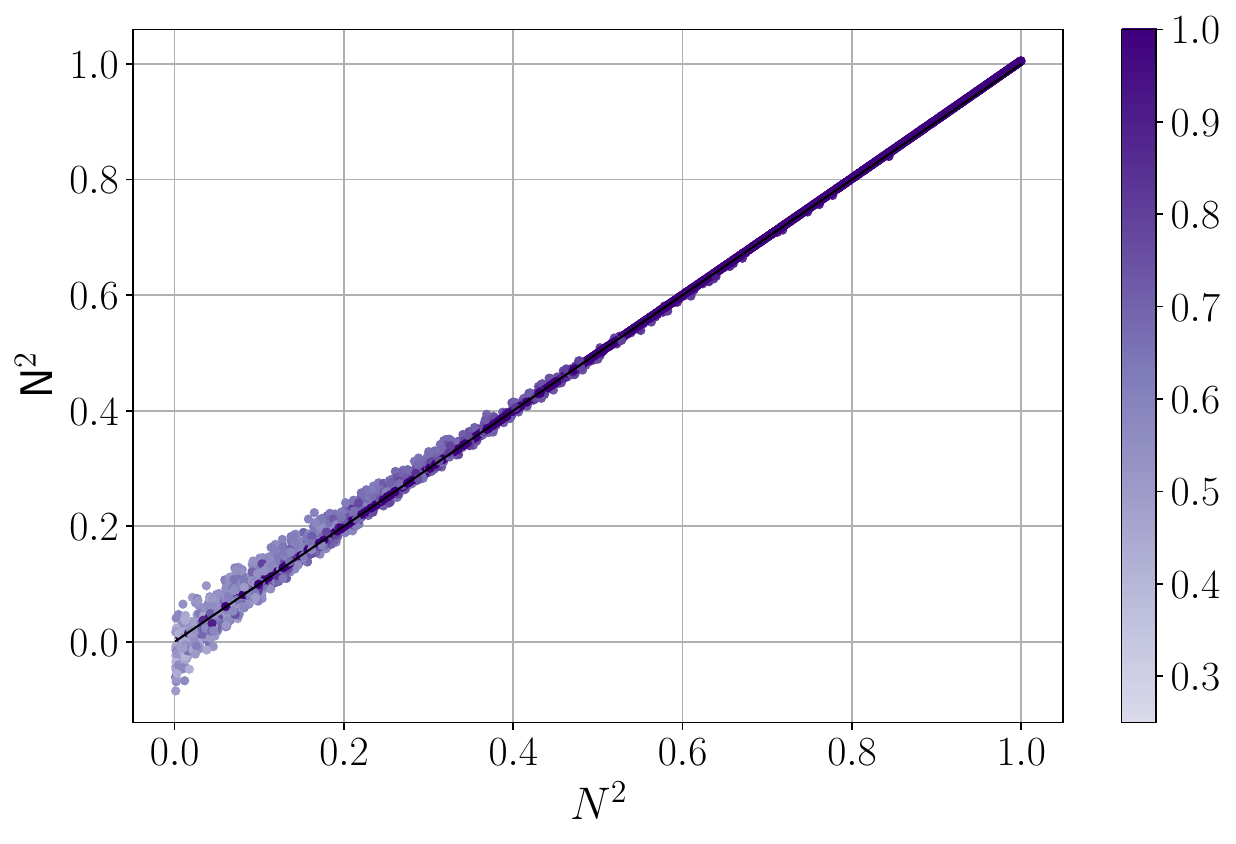}
        \includegraphics[width=.495\textwidth]{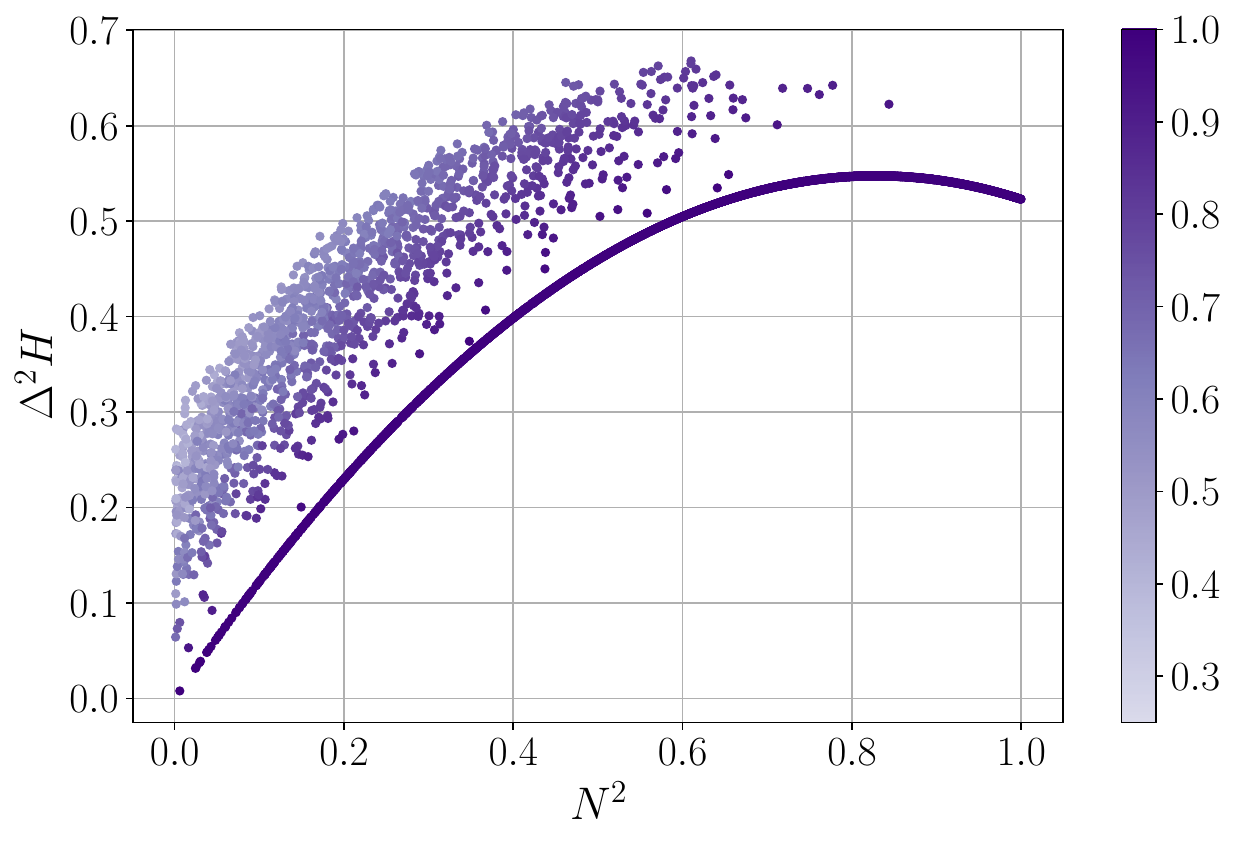}
        \caption{
             Results of training on \textit{easy} (pure and isotropic) states and testing on mixed states. Left: Predicted squared negativity $\mathsf{N}^2 = \Tr H \rho^{\otimes 4}$ of $5000$ random mixed states $\rho$ vs. the true negativity $N^2$ for $c=4$ copies. The black line connecting $(0,0)$ and $(1,1)$ is the ground truth. 
             Right: Variance of the trained observable $H$.
             The color of each point represents the purity of the corresponding state. The model is trained using the 10-class Hermitian ansatz in Eq.~\eqref{eq:H10_ansatz} on a dataset of 100 uniformly distributed \textit{easy} states.
        }
        \label{fig:neg_square10classisopureXmixed}
    \end{figure*}

\section{Discussion}
    
    In this section, we discuss the key results of our work, highlighting their main strengths and limitations. 
    We also outline future research directions, both for the specific problems considered and for our general regression method with variance regularization.

    \subsection{General method}
    In Sec.~\ref{sec:methods}, we described the construction of the regression model with variance regularization to predict the intrinsic parameter $\alpha$ of a quantum state $\rho_\alpha$. To this end, we constructed an observable $H$ as a linear combination of Hermitian operators, as in~\eqref{eq:ham-par}, subject to the simultaneous optimization of least-squares and variance-regularization terms, as given in~\eqref{eq:regression-H}. Once the optimal observable $H$ is found using a training set of the form~\eqref{eq:training_set}, we predict the parameter of an unknown state $\rho_\alpha$ as the expectation value $\Tr(H\rho_\alpha)$ and (where applicable) compare the variance of the observable $H$ against the Cramer--Rao bounds \eqref{eq:var-qcrb}, including the ground state of the transverse-field Ising model and entanglement in pure and isotropic states. 
    
    A key advantage of our method is that, once a symmetry-inspired ansatz is identified, the tensors $L$ and $S$ (as described in Eqs.~\eqref{eq:L} and \eqref{eq:S}) can be constructed by measuring the given dataset. Then the training is performed entirely classically. 
    Moreover, the method naturally supports incremental learning, as additional datasets can be incorporated by computing and adding their corresponding $L$ and $S$ tensors without recomputing the full dataset, as demonstrated for pure and isotropic states in Sec.~\ref{sec:easy_states}. 
    Compared to using variational ans\"atze such as in \cite{kard2025}, this provides a significant computational advantage. 
    
    The method, however, is not data-agnostic and requires identifying suitable symmetry-inspired ansatz operators. We demonstrated the approach on two tasks: 
    Prediction of the relative strength of an external field in a transverse-field Ising model, and quantification of entanglement represented by the squared negativity. 

    \subsection{Predicting transverse field Ising Hamiltonian}
    
    In Section~\ref{sec:ising}, for the prediction of transverse field $h$ of the Ising Hamiltonian $\mathcal{H}_h$ in Eq.~\eqref{eq:ising} under open boundary conditions, we applied physical reasoning to construct an ansatz that consists of 3-local Pauli strings and respects the relevant symmetries (spin-flip and time-reversal) with an amenable number of terms. 
    Our results demonstrate high prediction accuracy with a relatively low variance, even approaching the quantum Cramer--Rao bound \eqref{eq:var-qcrb} for $h>1.0$.
    However, one might get an inspiration for the ansatz from other considerations.
    For instance, one can regard the thermal state $\rho_{h} \equiv \frac{1}{\mathcal{Z}}e^{-\beta \mathcal{H}_h} = \sum_{k=0}^{\infty}\frac{(-\beta \mathcal{H}_h)^k}{k!}$ with inverse temperature $\beta$ and $\mathcal{Z} = \Tr e^{-\beta \mathcal{H}_h}$, and take the operators for the ansatz $H_{\boldsymbol{\theta}}$ by truncating this sum.
    If the Hamiltonian has symmetries $S$ (such as the flip-symmetry $S = X^{\otimes n}$), for this ansatz the condition $[H_{\boldsymbol{\theta}}, S] = 0$ will also hold.
    This, however, would not be generally true for the individual Pauli strings of $H_{\boldsymbol{\theta}}$.
    In this case, one could consider symmetrizing the individual operators via twirling \cite{nguyen2024theory,meyer2023exploiting}.

    \subsection{Predicting entanglement in bipartite systems.}
        For entanglement prediction, in Section \ref{sec:entanglement}, we argued that a suitable ansatz can be constructed as a linear combination of permutation operators. This choice is motivated by the local-unitary invariance of bipartite entanglement: local permutation operators $G_A \otimes G_B$ commute with $U_A^{\otimes c} \otimes U_B^{\otimes c}$, making them natural building blocks for entanglement observables. 
        In the two-copy setting $c=2$, this leads to the Hermitian ansatz $\mathcal{G} =\{\Id, \S_{A_0 A_1}, \S_{B_0 B_1}, \S_{A_0 A_1}\S_{B_0 B_1}\}$, which we studied both analytically and numerically for pure and isotropic states using squared negativity as an entanglement measure (for additional results concerning $c=2$ copies, see Appendix~\ref{app:est_theor}). 
        The analytical coefficients are obtained via Haar integration and subsequent optimization; our numerical method recovers the same coefficients to within $10^{-4}$, as demonstrated in the appendices, where relevant.

To improve the prediction performance, we extended the construction to more copies $c$. 
As shown in Table~\ref{tab:c_counts}, the number of permutation operators grows rapidly with $c$. 
We address this by first restricting to linearly independent operators and then grouping them into classes corresponding to distinct functional forms. 
For $c = 4$, we reduced 100 Hermitian operators to 10 classes capturing the essential structure of the problem. 
Additional 16 Hermitized non-Hermitian operators (see Table~\ref{tab:c_counts}) reduce to six further classes, whose contribution is negligible---and can even degrade performance on structured datasets as we show in Appendix~\ref{app:sec:class_extention}. 
We therefore adopted the 10-class Hermitian ansatz as our final choice.
For completeness, we similarly address the case of $c=3$ copies in  Appendix~\ref{app:sec:3_copies}, where we identified four Hermitian classes and one non-Hermitian class; the inclusion of the latter through Hermitization (corresponding to measuring $\Tr[\rho^3]$) significantly improves performance. 

    Furthermore, we introduced \textit{easy} states (pure and isotropic states) and demonstrated that training the 10-class ansatz on these structured datasets yields fairly accurate predictions for random mixed states (Sec.~\ref{sec:easy_states}). 

    We emphasize the optimality of the 10-class Hermitian ansatz, among all subsets of the 576 operators at $c = 4$, in terms of the polynomials they generate. 
    In Appendix~\ref{app:sec:trainability_models}, we benchmark it against several alternative models.
    For instance, we consider the polynomials of $\rho$ and its partial states up to a certain degree, or removing less effective terms from the 10 functions generated by the 10-class ansatz. 
    Among the tested models, the 10-class Hermitian ansatz presented in Section~\ref{sec:herm} is found to be the optimal choice, performing well on either easy or mixed states.

Compared to Pauli-string (Fig.~\ref{fig:Pauli_entanglement}) and, e.g., hardware-efficient variational ans\"atze used in \cite{kard2025}, our method achieves significantly higher accuracy with fewer training samples (as few as 20--100 states) and substantially reduced computational cost. Training the 10-class Hermitian ansatz with variance regularization takes only seconds on a desktop PC, whereas Pauli-based and hardware-efficient approaches can require orders of magnitude more time and are more prone to overfitting on limited or structured datasets.

Finally, we note that our ansatz can closely approximate other standard quantities. The PT moments derived from cyclic permutations on subsystems $A$ and $B$~\cite{gray2018machine},
\begin{equation}
    \label{eq:pt_moments}
    p_n = \Tr\big[\forwardPi_A \backwardPi_B\, \rho^{\otimes n}\big],
\end{equation}
involve operators $\forwardPi_A$ and $\backwardPi_B$ that, while present in the full 576 operator set at $c = 4$, are not directly among our 10 classes. Nevertheless, as we show in Appendix~\ref{app:sec:class_extention}, they can be accurately approximated by linear combinations of our 10 class ansatz.
    
    \subsection{Future directions}
    Our work proposes several directions for a future investigation. First, while we numerically observe that increasing the number of copies to $c=4$ improves prediction performance, it remains an open question how this improvement scales for larger $c$ and whether a fundamental limit exists beyond which no substantial gains are achieved. In particular, an analytical understanding of the scaling of the number of independent operators in Table~\ref{tab:c_counts} is still lacking. Second, it would be interesting to explore more structured and expressive training datasets beyond the easy states considered here, such as X-states~\cite{mendoncca2014entanglement} and stabilizer states~\cite{fattal2004entanglement}, to further improve generalization to random mixed states. Third, a natural extension of our entanglement quantification framework would be to incorporate the witness expansion tools of Ref.~\cite{tang2026witness}, which potentially enhance the entanglement detection through moments up to the fourth order.
    Fourth, although not emphasized in this work, our numerical observations suggest that variance regularization not only reduces the variance but can also moderately improve prediction accuracy, particularly when training on structured datasets and testing on random mixed states, which can be studied in the future. 
    Fifth, while our results indicate that the 10-class Hermitian ansatz is optimal among the models (for more details, see Appendix~\ref{app:sec:trainability_models}), a rigorous theoretical justification of this optimality remains an important open problem. Finally, the general framework introduced in Sec.~\ref{sec:methods} can be extended to a wide range of other quantum estimation and learning tasks.

\section{Conclusion}

    In this paper, we introduced a regression-based framework with variance regularization that exploits symmetry-inspired ans\"atze. The key advantage of our scheme is that the input quantum states need to be measured only once, after which the optimization can be completed classically. The main limitation is that one must identify an ansatz respecting the symmetries of the problem. We applied our method to two tasks: (a) predicting the relative strength of an external field in the transverse-field Ising Hamiltonian, and (b) predicting the entanglement negativity in bipartite qubit systems. In (a), we identified an ansatz respecting spin-flip and time-reversal symmetries that achieved high accuracy with relatively low variance. In (b), we found an ansatz consisting of 10 Hermitian operators that performs well even when trained on easy states, whereas Pauli-based or generic ans\"atze overfit or underperform when the dataset is small or limited to structured states such as easy (pure and isotropic) states.

\section{Code and data availability}

    The Python code that generates the data and produces the numerical results is available on GitHub at \cite{github_repo}.

\newpage

\bibliography{bibliography}
\bibliographystyle{unsrt}

\onecolumngrid
\appendix

\section{Proof of the positive semi-definiteness of the Hessian Matrix}
\label{app:sec:hessian}

In this section, we show that $S_{\Sigma}$ in Eq.~\eqref{eq:cost_2der} is positive semidefinite, which ensures that the second derivative $\partial_{\btheta}(\partial_{\btheta} f(\btheta))$ is non-negative. 
Consider a matrix with elements
\begin{equation}
    M_{mn} = \Tr\!\big[(G_m G_n + G_n G_m)\rho\big].
\end{equation}
We show that $\langle v|M|v\rangle \ge 0$ for any vector $\ket{v}$. Expanding $\ket{v} = \sum_m v_m \ket{m}$ in the computational basis, we obtain

\begin{equation}
    \label{eq:vMv}
    \langle v|M|v\rangle 
    = \sum_{m,n} \Tr\!\big(v_m^* v_n G_m G_n \rho + v_m^* v_n G_n G_m \rho\big).
\end{equation}
Defining $B = \sum_m v_m G_m$, we have $B^\dagger = \sum_m v_m^* G_m$. Using linearity and the cyclicity of the trace, Eq.~\eqref{eq:vMv} becomes
\begin{equation}
    \langle v|M|v\rangle = \Tr\!\big(B^\dagger \rho B + B \rho B^\dagger\big).
\end{equation}
Since $\rho$ is positive semidefinite, both terms $\Tr(B^\dagger \rho B)$ and $\Tr(B \rho B^\dagger)$ are non-negative. Therefore, $\langle v|M|v\rangle \ge 0$, and hence $M$ is positive semidefinite.


\section{Theoretical optimization of variance regularization problem with respect to the parameters}
\label{app:sec:theory_Ab}
We consider the integral form of Eq. (\ref{eq:regression-H}) into which we put the parametrized observable in Eq. (\ref{eq:ham-par}),
    \begin{align}
        \label{eq:min_cost_theta}
            H^*&=\min_{\{\theta_j\}_j}\:  w_\mathrm{ls} \int_{a}^{b} \left( \sum_{j=1}^M\theta_j \Tr (\rho_\alpha G_j)  - \alpha\right)^2 \,d \alpha  +w_\mathrm{var} \int_{a}^{b}\left(2\sum_{j,k=1}^{M}\theta_j\theta_k\Tr (\rho_\alpha \frac{1}{2}(G_l G_j+G_j G_l))-(\sum_{j=1}^k\theta_j\Tr\rho_\alpha G_j)^2\right) \,d \alpha.
    \end{align}
For an ansatz $\mathcal{G}=\{G_j\}_{j=1}^M$, we can optimize $H$ via parameters $\theta_j$ through differentiation as it was done in Eq. (\ref{eq:cost_1der}). Differentiating with respect to $\theta_l$
\begin{align}
    2&\wls \int_{a}^{b}\left(\sum_{j=1}^M\theta_j \Tr (\rho_\alpha G_j)-\alpha\right) \Tr(\rho_\alpha G_l) \nonumber \\
    + &2\wvar \int_{a}^{b} \left( \sum_{j=1}^M\theta_j \Tr(\rho_\alpha \frac{1}{2}(G_l G_j+G_j G_l)) \right)\nonumber \\
    - &2\wvar \int_{a}^{b}\left(\sum_{j=1}^M \theta_j \Tr(\rho_\alpha G_j)\Tr(\rho_\alpha G_l) \right)=0
\end{align}
Let us introduce the following  (sorted) arrays:
\begin{align}
    &\boldsymbol{\theta}= [\theta_j]_{j=1}^M,\nonumber \\
    &\boldsymbol{v}(\alpha)= [\Tr(\rho_\alpha G_j)]_{j=1}^M, \nonumber \\
    &S(\alpha)=[\Tr(\rho_\alpha \frac{1}{2}(G_l G_j+G_j G_l) )]_{j,l=1}^{M,M}.
\end{align}
This allows to write
\begin{align}
    \wls \int_a^b(\boldsymbol{v}\boldsymbol{v}^T\boldsymbol{\theta}-\alpha \boldsymbol{v}) \mathrm{d}\alpha+\wvar \int_a^bS\boldsymbol{\theta}\mathrm{d}\alpha
    -\wvar \int_a^b \boldsymbol{v}\boldsymbol{v}^T\boldsymbol{\theta}\mathrm{d}\alpha = 0.
\end{align}
Denoting $k=\wls/\wvar$, we obtain
\begin{align}
    A=(k-1) \int_a^b\boldsymbol{v}\boldsymbol{v}^T\mathrm{d}\alpha+\int_a^bS\mathrm{d}\alpha \boldsymbol{b}= k\int_a^b\alpha \boldsymbol{v} \mathrm{d}\alpha.
\end{align}
Thus, we arrive at $A\boldsymbol{\theta}=\boldsymbol{b}$, in which we get $\boldsymbol{\theta}=A^+\boldsymbol{b}$.


\section{Combination of datasets}
\label{app:sec:comb_datasets}
Let us consider two datasets $\mathcal{T}_1=\{(\rho_{i_1}, \alpha_{i_1})\}_{i_1=1}^{T_1}$ and $\mathcal{T}_2=\{(\rho_{i_2}, \alpha_{i_2})\}_{i_2=1}^{T_2}$. From the structure of $L$ in Eq.~\ref{eq:L} and $S$ in Eq.~\ref{eq:S}, one could consider the following relations:
\begin{align}
    L_1^TL_1=[\sum_{i_1=1}^{T_1}\Tr \rho_{i_1}G_k\Tr \rho_{i_1}G_j]_{k,j}^{M,M},\\
    L_2^TL_2=[\sum_{i_2=2}^{T_2}\Tr \rho_{i_2}G_k\Tr \rho_{i_2}G_j]_{k,j}^{M,M}.
\end{align}
When the two datasets $\mathcal{T}_1$ and $\mathcal{T}_2$ are combined, one can derive the following expression
\begin{align*}
    L_1^TL_1+L_2^TL_2= \big[\sum_{i_1=1}^{T_1}\Tr \rho_{i_1}G_k\Tr \rho_{i_1}G_j +
    \sum_{i_2=2}^{T_2}\Tr \rho_{i_2}G_k\Tr \rho_{i_2}G_j\big]_{k,j}^{M,M}= \big[\sum_{i=1}^{T_1+T_2}\Tr \rho_{i}G_k\Tr \rho_{i_1}G_j\big]_{k,j}^{M,M}.
\end{align*}
In a similar logic,
\begin{align*}
    &S_{\Sigma_1}+S_{\Sigma_2}=\frac{1}{2}\big[\Tr\sum_{i=1}^{T_1+T_2}\rho_i (G_jG_k+G_kG_j) \big],\\
    &b_1+b_2=kL_1^T\boldsymbol{\alpha}_1+kL_2^T\boldsymbol{\alpha}_2 = 
    \big[\sum_{i_1=1}^{T_1}\Tr \rho_{i_1}G_j \big]_j^M + \big[\sum_{i_2=1}^{T_2}\Tr \rho_{i_2}G_j \big]_j^M=
    \big[\sum_{i=1}^{T_1+T_2}\Tr \rho_{i}G_j \big]_j^M = b.
\end{align*}

Therefore, using the ansatz $\{G_j\}_{j=1}^M$ one could construct $A_1$ and $b_1$ using the dataset $\mathcal{T}_1$ and $A_2$ and $b_2$ using the dataset $\mathcal{T}_2$, and calculate the matrix $A$ and the vector $b$ as

\begin{align}
    A&=A_1+A_2,\nonumber\\
    b&=b_1+b_2\label{eq:additive_Ab}
\end{align}
which can be used to obtain the solution for the combined dataset $\mathcal{T}=\mathcal{T}_1\cup \mathcal{T}_2$ as $\btheta^*=A^+b$. This way one could add up different datasets at different sizes as well, and proof can be similarly extended to multi-dataset.


\section{On entanglement symmetry}
\label{app:sec:ent_sym_intu}

    In this section, we provide additional intuition for choosing permutation operators as our ansatz. A key property of entanglement is its invariance under local unitary transformations. That is, for any valid entanglement measure $\mathcal{E}$ of a bipartite state $\rho \equiv \rho_{AB}$,
    \begin{equation}
        \mathcal{E}(U_A \otimes U_B \, \rho \, U_A^\dagger \otimes U_B^\dagger) = \mathcal{E}(\rho).
    \end{equation}
    
    We first argue that a single copy of $\rho$ is insufficient to construct a model $H$ that accurately predicts entanglement. Suppose, by contradiction, that such a model exists and respects the symmetry, i.e.,
    \[
    \Tr(\rho H) = \Tr(U_A \otimes U_B \, \rho \, U_A^\dagger \otimes U_B^\dagger H)
    \]
    for all local unitaries $U_A, U_B$. Averaging over the Haar measure, we obtain
    \begin{align*}
        &\int dU_A dU_B \, \Tr(U_A \otimes U_B \, \rho \, U_A^\dagger \otimes U_B^\dagger H)
        = \Tr\!\left( \int dU_A dU_B \, U_A \otimes U_B \, \rho \, U_A^\dagger \otimes U_B^\dagger H \right) \\
        &= \Tr\!\left( \int dU_A \, U_A \left( \int dU_B \, U_B \rho U_B^\dagger \right) U_A^\dagger H \right)= \Tr\!\left( \int dU_A \, U_A \frac{\Tr_B(\rho)}{d_B} U_A^\dagger H \right) = \frac{\Tr(H)}{d_A d_B}.
    \end{align*}
    This expression is independent of $\rho$, contradicting the assumption that $\Tr(\rho H)$ encodes entanglement information. Therefore, no observable $H$ acting on a single copy of $\rho$ can be trained to predict entanglement for arbitrary states.
    
    Now, we consider $k$ copies $\rho^{\otimes k}$. With the same logic, 
    \begin{align}
        &\int_{U_A,U_B} d_{U_A}d_{U_B}\Tr(U_A^{\otimes k}\otimes U_B^{\otimes 2}\rho U^{\dagger \otimes k}_A\otimes U^{\dagger \otimes k}_B H) =\Tr\left(\int_{U_A,U_B} d_{U_A}d_{U_B}U_A^{\otimes k}\otimes U_B^{\otimes k}\rho U^{\dagger \otimes k}_A\otimes U^{\dagger \otimes k}_B H\right) \nonumber\\
        & =\Tr\left(\rho\int_{U_A,U_B} d_{U_A}d_{U_B} U_A^{\otimes k}\otimes U_B^{\otimes k} H U^{\dagger \otimes k}_A\otimes U^{\dagger \otimes k}_B \right) = \Tr\left(\rho\mathcal{M}_{A\otimes B} (H)\right). \label{app:eq:trace_H_proof}
    \end{align}
    where in the second equality, we applied the cyclicality of trace operator and in the third equality, we defined the moment operator $\mathcal{M}(\cdot)$. Further, we define the permutation operators and commutants as in \cite{mele2024introduction}. First, we consider a unitary group $U(d)$, symmetric group $S_k$ and define in \eqref{app:eq:perm_def} the permutation operators, in \eqref{app:eq:comm_def} the commutants, and the Schur-Weyl duality in \eqref{app:eq:Shur-Weyl}
    \begin{align}
        &\mathbb{P}_d (\ket{\psi_0}\otimes\ket{\psi_1} \dots\otimes\ket{\psi_k}) = \ket{\psi_{\pi^{-1}(0)}}\otimes\ket{\psi_{\pi^{-1}(1)}} \dots\otimes\ket{\psi_{\pi^{-1}(k)}}, \label{app:eq:perm_def}\\
        &\operatorname{Comm} (U(d), k) =  \{A\in \mathcal{L}((\mathbb{C}^d)^{\otimes k}) :\, [A, U(d)^{\otimes k}]=0\,\, \text{for all }\, U \in U(d) \}, \label{app:eq:comm_def}\\
        &\operatorname{Comm} (U(d), k) = \operatorname{span} (\mathbb{P}_d(\pi) \text{ for all } \pi \in S_k) .\label{app:eq:Shur-Weyl}
    \end{align}
    According to the properties of commutants, $\mathcal{M}_{A\otimes B} (H)\in \operatorname{Comm} (U(d_A), k)\otimes \operatorname{Comm} (U(d_B), k)$, and if $H \in \operatorname{Comm} (U(d_A), k)\otimes \operatorname{Comm} (U(d_B), k)$, then $\mathcal{M}_{A\otimes B} (H) = H$. Thus, one can simply prepare a model $H$ that is already in $\operatorname{Comm} (U(d_A), k)\otimes \operatorname{Comm} (U(d_B), k)$. Then Eq.~\eqref{app:eq:trace_H_proof} concludes as $\Tr\left(\rho\mathcal{M}_{A\otimes B} (H)\right) = \Tr\left(\rho H\right)$. One way to prepare $H$ is to pick the ansatz as a linear combination of permutation operators with trainable coefficients $\theta \in \mathbb{R}$
    \begin{equation}
        H = \sum_{\pi, \sigma \in S_k}\theta_{\pi\sigma}\mathbb{P}_{d_A}(\pi)\otimes \mathbb{P}_{d_B}(\sigma).
    \end{equation}
    This observable can be used in Eq.~\eqref{eq:ham-par} to implement our algorithm described in Section~\ref{sec:method_param}. However, in Section~\ref{sec:herm} we propose more refined and efficient observable for quantifying the entanglement of bipartite states.


\section{Entanglement quantification for random states. Theory, examples, and comparison to numerics}
\label{app:est_theor}
    
    In this section, we address the problem introduced in Section~\ref{sec:anz_choosing}. Specifically, we solve the task defined by Eq.~\eqref{eq:opt_obs_reg} for bipartite qubit and qudit systems, employing two entanglement quantification measures: negativity and tangle. First, we start with random pure and isotropic states, and continue with the combination of these two states.
    
    \subsection{Haar-integration framework for regression with variance regularization for bipartite pure qudit states}
    \label{app:pure_qudits}
        
        We consider a bipartite pure qudit system $\ket{\psi}_{\scriptscriptstyle \alpha,U_A, U_B} \in \mathcal{H_A}\otimes \mathcal{H_B}$ of dimension $d\times d$, 
        \begin{equation}
            \label{eq:pure_qudit2}
            \ket{\psi}_{\scriptscriptstyle \alpha,U_A, U_B}  = \left(U_A\otimes U_B\right)\sum_{i=1}^d\sqrt{\lambda_i}\ket{i_Ai_B},
        \end{equation}
        where $\sum_i \lambda_i=1$.
        Further we consider two copies of $ \ket{\psi}_{\scriptscriptstyle \alpha,U_A, U_B}$ in a density matrix form of $\rho_{\scriptscriptstyle \alpha,U_AU_B}=\ketbra{\psi}_{\scriptscriptstyle \alpha,U_A, U_B}$
        \begin{equation}
            \label{eq:density_qudit}
            \rho_{\scriptscriptstyle \alpha,U_A, U_B}^{\otimes2} \equiv \left(U_{A_1}\otimes U_{B_1}\otimes U_{A_2}\otimes U_{B_2}\right)\sum_{i,j, k, l=1}^d\sqrt{\lambda_i \lambda_j\lambda_{k} \lambda_{l}}\ketbra{i_{A_1}i_{B_1}k_{A_2}k_{B_2}}{j_{A_1}j_{B_1}l_{A_2}l_{B_2}}\left(U_{A_1}^\dagger\otimes U_{B_1}^\dagger\otimes U_{A_2}^\dagger\otimes U_{B_2}^\dagger\right).
        \end{equation}
        We perform Haar integration in Eq. (\ref{eq:rho_global}) initially over the subsystem $A$ and then continue with the subsystem $B$. Integration through subsystem $A$ 
        \begin{align}\label{eq:int_A}
            \operatorname{\mathbb{E}}_{U_A(d)}\left[\rho_{\scriptscriptstyle \alpha,U_A,U_B}^{\otimes2}\right] = &\int_{U(d)}\,dU_A\rho_{\scriptscriptstyle \alpha,U_A,U_B}^{\otimes2}=\int_{U(d)}\,dU_A{U_A}^{\otimes2}\rho_{\scriptscriptstyle \alpha,U_B}^{\otimes2}{U_A}^{\dagger\otimes2}\\
            =&\frac{\operatorname{Tr}_{A_0A_1}(\rho_{\scriptscriptstyle \alpha,U_B}^{\otimes2})-d^{-1}\operatorname{Tr}_{A_0A_1}(\S_{A_0A_1}\rho_{\scriptscriptstyle \alpha,U_B}^{\otimes2})}{d^2-1}\Id_{A_0A_1}+\frac{\operatorname{Tr}_{A_0A_1}(\rho_{\scriptscriptstyle \alpha,U_B}^{\otimes2}\S_{A_0A_1})-d^{-1}\operatorname{Tr}_{A_0A_1}(\rho_{\scriptscriptstyle \alpha,U_B}^{\otimes2})}{d^2-1}\S_{A_0A_1}.\nonumber
        \end{align}
        The second equality implies that we are acting with the moment operator of the second order on two copies of $\rho_{\scriptscriptstyle AB}^{\otimes 2}$ as in \cite{mele2024introduction}. In the third equality, we integrate the Haar measure over the second-order moment operator for subsystem $A$, leading to the final equality. Taking into account that,
        \begin{equation}
            \rho_{\scriptscriptstyle \alpha,U_B}^{\otimes2} \equiv \left(U_{B_1}\otimes U_{B_2}\right)\sum_{i,j,k,l=1}^d\sqrt{\lambda_i \lambda_j\lambda_{k} \lambda_{l}}\ketbra{i_{A_1}i_{B_1}k_{A_2}k_{B_2}}{j_{A_1}j_{B_1}l_{A_2}l_{B_2}}\left(U_{B_1}^\dagger\otimes U_{B_2}^\dagger\right)
        \end{equation}
        each term in the Eq. (\ref{eq:int_A}) can be computed as follows:
        \begin{equation}
        \label{int_A_Id}
            \operatorname{Tr}_{A_0A_1}(\rho_{\scriptscriptstyle \alpha,U_B}^{\otimes2}) = \left(U_{B_0}\otimes U_{B_1}\right)\sum_{i,k=1}^d\lambda_i \lambda_{k}\ketbra{i_{B_0}k_{B_1}}{i_{B_0}k_{B_1}}\left(U_{B_0}^\dagger\otimes U_{B_1}^\dagger\right),
        \end{equation}
        \begin{multline}
            \operatorname{Tr}_{A_0A_1}(\S_{A_0A_1}\rho_{\scriptscriptstyle \alpha,U_B}^{\otimes2})
            =\left(U_{B_0}\otimes U_{B_1}\right)\sum_{i,j=1}^d\sqrt{\lambda_i \lambda_j\lambda_{k} \lambda_{l}}\operatorname{Tr}_{A_0A_1}\left(\S_{A_0A_1}\ketbra{i_{A_0}i_{B_0}k_{A_1}k_{B_1}}{j_{A_0}j_{B_0}l_{A_1}l_{B_1}}\right)\left(U_{B_0}^\dagger\otimes U_{B_1}^\dagger\right)\\
            =\left(U_{B_0}\otimes U_{B_1}\right)\sum_{i,j,k, l=1}^d\sqrt{\lambda_i \lambda_j\lambda_{k} \lambda_{l}}\operatorname{Tr}_{A_0A_1}\left(\ketbra{k_{A_1}i_{B_0}i_{A_0}k_{B_1}}{j_{A_0}j_{B_0}l_{A_1}l_{B_1}}\right)\left(U_{B_0}^\dagger\otimes U_{B_1}^\dagger\right)\\
            =\left(U_{B_0}\otimes U_{B_1}\right)\sum_{i,k=1}^d\lambda_i \lambda_{k}\ketbra{i_{B_0}k_{B_1}}{k_{B_1}i_{B_0}}\left(U_{B_0}^\dagger\otimes U_{B_1}^\dagger\right),
        \end{multline}
        \begin{equation}
            \operatorname{Tr}_{A_0A_1}(\rho_{\scriptscriptstyle \alpha,U_B}^{\otimes2}\S_{A_0A_1}) 
            =\operatorname{Tr}_{A_0A_1}(\S_{A_0A_1}\rho_{\scriptscriptstyle \alpha,U_B}^{\otimes2}) =
            \left(U_{B_0}\otimes U_{B_1}\right)\sum_{i,k=1}^d\lambda_i \lambda_{k}\ketbra{i_{B_0}k_{B_1}}{k_{B_1}i_{B_0}}\left(U_{B_0}^\dagger\otimes U_{B_1}^\dagger\right).
        \end{equation}

        We now integrate Eq. (\ref{int_A_Id}) through subsystem $B$
        \begin{align}
            \nonumber    \operatorname{\mathbb{E}}_{U_B(d)}\left[\operatorname{Tr}_{A_0A_1}(\rho_{\scriptscriptstyle \alpha,U_B}^{\otimes2})\right] &= \frac{\operatorname{Tr}_{B_0B_1}(\operatorname{Tr}_{A_0A_1}(\rho_{\scriptscriptstyle \alpha}^{\otimes2}))-d^{-1}\operatorname{Tr}_{B_0B_1}(\S_{B_0B_1}\operatorname{Tr}_{A_0A_1}(\rho_{\scriptscriptstyle \alpha}^{\otimes2}))}{d^2-1}\Id_{A_0A_1}\Id_{B_0B_1}\\
            &+\frac{\operatorname{Tr}_{B_0B_1}(\operatorname{Tr}_{A_0A_1}(\rho_{\scriptscriptstyle \alpha}^{\otimes2})\S_{B_0B_1})-d^{-1}\operatorname{Tr}_{B_0B_1}(\S_{B_0B_1}\operatorname{Tr}_{A_0A_1}(\rho_{\scriptscriptstyle \alpha}^{\otimes2}))}{d^2-1}\Id_{A_0A_1}\S_{B_0B_1}.\label{int_BB}
        \end{align}
        Taking the traces gives
        \begin{align}
            &\operatorname{Tr}_{B_0B_1}(\operatorname{Tr}_{A_0A_1}(\rho_{\scriptscriptstyle \alpha}^{\otimes2}))=\operatorname{Tr}_{B_0B_1}\left(\sum_{i,k=1}^d\lambda_i \lambda_{k}\ketbra{i_{B_0}k_{B_1}}{i_{B_0}k_{B_1}}\right)=\sum_{i,k=1}^d\lambda_i \lambda_{k},\nonumber\\
            &\operatorname{Tr}_{B_0B_1}(\S_{B_0B_1}\operatorname{Tr}_{A_0A_1}(\rho_{\scriptscriptstyle \alpha}^{\otimes2}))=\operatorname{Tr}_{B_0B_1}\left(\sum_{i,k=1}^d\lambda_i \lambda_{k}\ketbra{i_{B_0}k_{B_1}}{k_{B_1}i_{B_0}}\right)=\sum_{i=1}^d\lambda_i^2,\nonumber\\
            &\operatorname{Tr}_{B_0B_1}(\operatorname{Tr}_{A_0A_1}(\rho_{\scriptscriptstyle \alpha}^{\otimes2})\S_{B_0B_1})=\operatorname{Tr}_{B_0B_1}\left(\sum_{i,k=1}^d\lambda_i \lambda_{k}\ketbra{i_{B_0}k_{B_1}}{k_{B_1}i_{B_0}}\right)=\sum_{i=1}^d\lambda_i^2.\nonumber
        \end{align}
        Taking into account that $\sum_{i=1}^d\lambda_i=1$, we have that $\sum_{i,k=1}^d\lambda_i \lambda_{k}=1$. We perform the same calculations for the second, third and fourth terms in Eq. (\ref{int_BB}), and after further algebraic manipulations, we arrive at
        \begin{align}
            \Tilde{\rho}_\alpha:&=    \operatorname{\mathbb{E}}_{U_B(d)}\operatorname{\mathbb{E}}_{U_A(d)}\left[\rho_{\scriptscriptstyle \alpha,U_A,U_B}^{\otimes2}\right]\nonumber \\&=
            \frac{1}{(d^2-1)^2}\left[ \left(\frac{d^2+1}{d^2}-\frac{2}{d} P \right)(\Id_{AB}+\S_{A_0A_1}\S_{B_0B_1})-\left(\frac{2}{d}-\frac{d^2+1}{d^2} P \right)(\S_{A_0A_1}+\S_{B_0B_1})\right],\label{eq:Gen_d}
        \end{align}
        where $P=\sum_{i=1}^d\lambda_i^2$.
        We could search for an optimal observable as in the form of Eq. (\ref{eq:haar_sol_form}), but due to the lower rank of pure states in Eq.~\eqref{eq:pure_qudit2}, the $H_0$ reduces to $H_0=(a \Id_{A_1A_2} + b\S_{A_1A_2})\otimes (c \Id_{B_1B_2} + f\S_{B_1B_2})$. Thus, we consider the observable form as
        \begin{equation}
            H_0 = ac\Id_{AB} + bc\S_{A_0A_1}+af\S_{B_0B_1} + bf\S_{A_0A_1}\S_{B_0B_1}
            \label{app:eq:ham_form_pure}
        \end{equation}
        and applying it into Eq. (\ref{eq:Gen_d})
        \begin{align}
            \Tilde{\rho}_\alpha H_0 &= \frac{1}{(d^2-1)^2}\left[ \left(\frac{d^2+1}{d^2}-\frac{2}{d} P \right)(\Id_{AB}+\S_{A_0A_1}\S_{B_0B_1})-\left(\frac{2}{d}-\frac{d^2+1}{d^2} P \right)(\S_{A_0A_1}+\S_{B_0B_1})\right] H_0\nonumber\\
            &=\frac{1}{(d^2-1)^2}\left[ \left(\frac{d^2+1}{d^2}-\frac{2}{d} P \right)\left((ac+bf)(\Id_{AB}+\S_{A_0A_1}\S_{B_0B_1})+(bc+af)(\S_{A_0A_1}+\S_{B_0B_1})\right)\right.\nonumber\\ 
            & \quad- \left.\left(\frac{2}{d}-\frac{d^2+1}{d^2} P \right)\left((ac+bf)(\S_{A_0A_1}+\S_{B_0B_1})+(bc+af)(\Id+\S_{A_0A_1}\S_{B_0B_1})\right)\right]\nonumber\\
            &= \frac{1}{(d^2-1)^2}\left[(ac+bf)(\frac{d^2+1}{d^2}-\frac{2}{d}P)+(af+bc)(\frac{d^2+1}{d^2}P-\frac{2}{d})\right](\Id_{AB}+\S_{A_0A_1}\S_{B_0B_1})\nonumber\\
            &\quad + \frac{1}{(d^2-1)^2}\left[(ac+bf)(\frac{d^2+1}{d^2}P-\frac{2}{d})+(af+bc)(\frac{1+d^2}{d^2}-\frac{2}{d}P)\right](\S_{A_0A_1}+\S_{B_0B_1}).\label{eq:Gen_d_H0}
        \end{align}

        We conclude by deriving the Haar-integrated formulation of the regression problem with variance regularization for entanglement quantification of pure states. In the following sections, we apply this framework to negativity and tangle as entanglement measures and present the prediction performance of our scheme.

\subsection{Entanglement negativity quantification}
\label{app:sec:entagnlement_q}

    In this section, we present theoretical and numerical results for the quantification of entanglement in both qubit-qubit and qudit-qudit systems, specifically focusing on negativity or tangle as a measure for pure and mixed states. Following the methodology outlined in Section~\ref{sec:methods}, we solve the regression problem using variance regularization and employ a problem-specific ansatz during training, supplemented by theoretical analysis.

    \subsubsection{Projectors onto symmetric and antisymmetric subspaces and their relations to our ansatz}
    \label{app:sec:sym-asym-proj}

        Consider the following observable acting on two copies of a bipartite system:
        \[
        H = g_1 \, \Id_{A_0A_1} \Id_{B_0B_1} + g_2 \, \S_{A_0A_1} \Id_{B_0B_1} + g_3 \, \Id_{A_0A_1} \S_{B_0B_1} + g_4 \, \S_{A_0A_1} \S_{B_0B_1},
        \]
        where $\S_{A_1A_2}$ and $\S_{B_1B_2}$ are the swap operators acting on the two copies of subsystem $A$ and $B$, respectively, and $\Id$ denotes the identity operator. The swap operators can be expressed in terms of their symmetric and antisymmetric projectors:
        \begin{align*}
        \S_{A_0A_1} &= P_{A_0A_1}^{\mathrm{sym}} - P_{A_0A_1}^{\mathrm{asym}}, \\
        \S_{B_0B_1} &= P_{B_0B_1}^{\mathrm{sym}} - P_{B_0B_1}^{\mathrm{asym}}, \\
        \Id_{A_0A_1} &= P_{A_0A_1}^{\mathrm{sym}} + P_{A_0A_1}^{\mathrm{asym}}, \\
        \Id_{B_0B_1} &= P_{B_0B_1}^{\mathrm{sym}} + P_{B_0B_1}^{\mathrm{asym}},
        \end{align*}
        with 
        \[
        P^{\mathrm{sym}} = \frac{1}{2}(\Id + \S), \qquad P^{\mathrm{asym}} = \frac{1}{2}(\Id - \S),
        \]
        for each subsystem. Substituting these expressions into $H$ yields a diagonal form in the joint eigenbasis of the swap operators:
        \begin{equation}
            \label{eq:H_eigendecomp_sym_asym}
            H = \lambda_1 \, P_{A_0A_1}^{\mathrm{sym}} P_{B_0B_1}^{\mathrm{sym}} 
            + \lambda_2 \, P_{A_0A_1}^{\mathrm{sym}} P_{B_0B_1}^{\mathrm{asym}}
            + \lambda_3 \, P_{A_0A_1}^{\mathrm{asym}} P_{B_0B_1}^{\mathrm{sym}}
            + \lambda_4 \, P_{A_0A_1}^{\mathrm{asym}} P_{B_0B_1}^{\mathrm{asym}},
        \end{equation}
        where the coefficients $\lambda_i$ are linear combinations of the $g_i$:
        \begin{align*}
        \lambda_1 &= g_1 + g_2 + g_3 + g_4, \\
        \lambda_2 &= g_1 + g_2 - g_3 - g_4, \\
        \lambda_3 &= g_1 - g_2 + g_3 - g_4, \\
        \lambda_4 &= g_1 - g_2 - g_3 + g_4.
        \end{align*}
        
        Given a bipartite state $\rho = \rho_{AB}$, we denote its reduced states by $\rho_A = \operatorname{Tr}_B[\rho_{AB}]$ and $\rho_B = \operatorname{Tr}_A[\rho_{AB}]$. The expectation values of the joint projectors on two copies $\rho^{\otimes 2}$ are directly related to the purities of these states:
        \begin{align}
            \label{eq:proj_traces}
            \begin{split}
            \operatorname{Tr}\!\big[ P_{A_0A_1}^{\mathrm{sym}} P_{B_0B_1}^{\mathrm{sym}} \, \rho^{\otimes 2} \big] 
                &= \frac{1}{4}\Big( 1 + \operatorname{Tr}[\rho_A^2] + \operatorname{Tr}[\rho_B^2] + \operatorname{Tr}[\rho^2] \Big),  \\
            \operatorname{Tr}\!\big[ P_{A_0A_1}^{\mathrm{sym}} P_{B_0B_1}^{\mathrm{asym}} \, \rho^{\otimes 2} \big] 
                &= \frac{1}{4}\Big( 1 + \operatorname{Tr}[\rho_A^2] - \operatorname{Tr}[\rho_B^2] - \operatorname{Tr}[\rho^2] \Big), \\
            \operatorname{Tr}\!\big[ P_{A_0A_1}^{\mathrm{asym}} P_{B_0B_1}^{\mathrm{sym}} \, \rho^{\otimes 2} \big] 
                &= \frac{1}{4}\Big( 1 - \operatorname{Tr}[\rho_A^2] + \operatorname{Tr}[\rho_B^2] - \operatorname{Tr}[\rho^2] \Big), \\
            \operatorname{Tr}\!\big[ P_{A_0A_1}^{\mathrm{asym}} P_{B_0B_1}^{\mathrm{asym}} \, \rho^{\otimes 2} \big] 
                &= \frac{1}{4}\Big( 1 - \operatorname{Tr}[\rho_A^2] - \operatorname{Tr}[\rho_B^2] + \operatorname{Tr}[\rho^2] \Big).
            \end{split}
        \end{align}
        These identities follow from the relations $\operatorname{Tr}[S_{A_0A_1} \rho^{\otimes 2}] = \operatorname{Tr}[\rho_A^2]$, $\operatorname{Tr}[S_{B_0B_1} \rho^{\otimes 2}] = \operatorname{Tr}[\rho_B^2]$, and $\operatorname{Tr}[S_{A_0A_1} S_{B_0B_1} \rho^{\otimes 2}] = \operatorname{Tr}[\rho^2]$.
        They will be used later for computing cFI for the ansatz $H$ when it is used for predicting the negativity of states.

    \subsubsection{Qubit-qubit pure states negativity}
    \label{sec:results_neg_pure}

        We consider a bipartite pure state $\ket{\psi}_{AB} \in \mathcal{H}_A \otimes \mathcal{H}_B$ of two qudits, each of dimension $d$, expressed in Schmidt decomposition form. The state is transformed by local random unitary operators $U_A$ and $U_B$ acting on subsystems $A$ and $B$, respectively:
        \begin{equation}
            \label{eq:pure_qudit}
            \ket{\psi}_{\scriptscriptstyle U_A, U_B}  = \left(U_A\otimes U_B\right)\sum_{i=1}^d\sqrt{\lambda_i}\ket{i_Ai_B},
        \end{equation}
        where $\{\lambda_i\}$ are the Schmidt coefficients satisfying $\lambda_i \ge 0$ and $\sum_{i=1}^{d} \lambda_i = 1$. For a pure state~\eqref{eq:pure_qudit}, the negativity can be expressed directly in terms of the Schmidt coefficients as
        \begin{equation}
            \label{eq:neg_schmidt}
            N(\rho) = \left(\sum_{i=1}^d\sqrt{\lambda_i}\right)^2 - 1.
        \end{equation}
        For the special case $d=2$ (two qubits), the state~\eqref{eq:pure_qudit} simplifies to
        \begin{equation}
            \label{eq:bell_neg}
            \ket{\psi}_{N^2} = \frac{1}{\sqrt{2}}\left(U_A\otimes U_B\right)\big(c_1\ket{00} + c_2 \ket{11}\big),
        \end{equation}
        where $c_{1,2} = \sqrt{1 \pm \sqrt{1 - N^2}}$.
        
        A solution to Eq.~\eqref{eq:opt_obs_reg} can be sought in the general form of Eq.~(\ref{eq:haar_sol_form}). However, for pure qudit states, a more specialized ansatz is available, where the observable factorizes as a tensor product of local operators acting on the two copies of each subsystem: $H = H_{A_0A_1} \otimes H_{B_0B_1}$. 
        Specifically, we consider
        \begin{equation}
            H_0=(a \Id_{A_0A_1} + b\S_{A_0A_1})\otimes (c \Id_{B_0B_1} + f\S_{B_0B_1}).
        \end{equation}
        This leads to the coefficients
        \begin{align}
            &a=2\frac{5c+ck+fk}{(c^2-f^2)(20+k)}, \\
            &b=-2\frac{5f+ck+fk}{(c^2-f^2)(20+k)},
        \end{align}
        where $c, f$ are free variables. Setting $c=0,\, f=1$ (or equivalently $a=1,\, b=0$) it is possible to quantify the entanglement of pure states swapping only one of the subsystems $A$ or $B$. For further details, see Appendix \ref{app:pure_qudits}. With these coefficients, the expected value of the observable and its reduced variance can be expressed as
        \begin{align}
            \langle H \rangle_{\rho_{N^2}} &= \frac{10+kN^2}{20+k},\label{eq:negSq_qubit_est}\\
            \operatorname{rVar}(H)&= \frac{\Delta_{\rho_{N^2}}^2 H} {\big| \partial_{N^2} \langle H \rangle_{\rho_{N^2}} \big|^2}=N^2(4-N^2).
            \label{eq:negSq_qubit_var}
        \end{align}

        Let us now compute cFI. 
        The purities of the reduced states of \eqref{eq:bell_neg} are
        \[
        \operatorname{Tr}[\rho_A^2] = \operatorname{Tr}[\rho_B^2] = 1 - \frac{N^2}{2}, \qquad \operatorname{Tr}[\rho^2] = 1.
        \]
        Substituting into Eqs.~\eqref{eq:proj_traces} gives
        \begin{align*}
        \operatorname{Tr}\!\big[ P_{A_0A_1}^{\mathrm{sym}} P_{B_0B_1}^{\mathrm{sym}} \, \rho^{\otimes 2} \big] &= 1 - \frac{N^2}{4}, \\
        \operatorname{Tr}\!\big[ P_{A_0A_1}^{\mathrm{sym}} P_{B_0B_1}^{\mathrm{asym}} \, \rho^{\otimes 2} \big] &= 0, \\
        \operatorname{Tr}\!\big[ P_{A_0A_1}^{\mathrm{asym}} P_{B_0B_1}^{\mathrm{sym}} \, \rho^{\otimes 2} \big] &= 0, \\
        \operatorname{Tr}\!\big[ P_{A_0A_1}^{\mathrm{asym}} P_{B_0B_1}^{\mathrm{asym}} \, \rho^{\otimes 2} \big] &= \frac{N^2}{4}.
        \end{align*}
        Thus, only two outcomes are nonzero: the fully symmetric and fully antisymmetric projections. 
        Using Eq.~\eqref{eq:fisher_classical}, we can now calculate the classical Fisher information:
        \begin{equation}
            I_c^{-1}(\boldsymbol{\Pi}, \rho_{N^2}) = N^2(4-N^2) = \operatorname{rVar}(H),
        \end{equation}
        where $\boldsymbol{\Pi}$ are the eigenprojectors of $H$ as in \eqref{eq:H_eigendecomp_sym_asym}.
        That is, the observable saturates the classical Cramer-Rao bound \eqref{eq:var-qcrb}.
        
        These results were used in the main text, where
        Fig.~\ref{fig:pre_neg_sq_pure} compares the performance of our numerical approach with the analytical results given in Eqs.~(\ref{eq:negSq_qubit_est}) and (\ref{eq:negSq_qubit_var}) for different variance weights $\wvar = \{10^{-4}, 10^{-2}, 10^{-1}\}$, in which the lower the weight for variance regularization, the higher we have precision. 
        For the cCRB analysis, we employed the eigenprojectors $\boldsymbol{\Pi}$ onto symmetric and antisymmetric subspaces as described in Appendix~\ref{app:sec:sym-asym-proj}. 
        These projectors match both our numerical results and theoretical predictions, though they do not saturate the qCRB.

    \subsubsection{Qubit-qubit isotropic states}
    \label{sec:results_neg_iso}

        Consider the two-qubit isotropic state
        \begin{equation}
            \label{eq:iso}
            \rho_q = q\ketbra{\Phi} + \frac{1-q}{4}\Id,
        \end{equation}
        where $\ket{\Phi} = \frac{1}{\sqrt{2}}(\ket{00}+\ket{11})$ is the Bell state. 
        Isotropic states serve as a standard noise model in quantum information and possess the distinctive property of being invariant under local unitary transformations of the form $U \otimes \overline{U}$, where $\overline{U}$ denotes the complex conjugate of $U$. This symmetry makes them particularly useful as intermediate states in entanglement distillation protocols~\cite{Hordist99}.
        
        The negativity of the isotropic state is given by
        \begin{equation}
            N(\rho_q) = 
            \begin{cases}
                0, & q \leqslant \frac{1}{3}, \\
                \frac{3q - 1}{2}, & q > \frac{1}{3}.
            \end{cases}
        \end{equation}
        For the entangled regime ($q > 1/3$), we can invert this relationship and express \eqref{eq:iso} in terms of the negativity as
        \begin{equation}
            \label{eq:iso_neg}
            \rho_N = \frac{2N + 1}{3}\ketbra{\Phi} + \frac{1 - N}{6}\Id.
        \end{equation}
        
        We employ the parameterization in \eqref{eq:iso_neg} to solve the optimization problem in \eqref{eq:opt_obs_reg} for predicting the squared negativity $N^2$ of isotropic states. Using the two-copy ansatz from \eqref{eq:ansatz_2copies}, we consider an observable of the form $H = g_1 \Id_{AB} + g_2 \S_{A_0A_1} + g_3 \S_{B_0B_1} + g_4 \S_{A_0A_1}\S_{B_0B_1}$ with trainable parameters $\{g_i\}_{i=1}^4$. The parameters $g_i$ can be obtained either by solving Eq. \eqref{eq:opt_obs_reg} analytically (see Appendix~\ref{app:sec:theory_Ab}) or via numerical minimization using the approach described in Section~\ref{sec:methods}. 
        
        The analytical solution yields the following parameters:
        \begin{equation}
            \label{eq:g-coeffs-iso}
            g_1 = \frac{14847 - 1091 k}{29694 + 2156 k}, \quad\quad
            g_2 = g_3 = -\frac{405 k}{98(303 + 22k)},\quad\quad g_4 = \frac{1782 k}{49(303 + 22k)}.    
        \end{equation}
        Notably, the symmetry $g_2 = g_3$ suggests that the ansatz can be reduced to $\mathcal{G} = \{\Id_{AB}, \S_{A_0A_1} + \S_{B_0B_1}, \S_{A_0A_1}\S_{B_0B_1}\}$ for efficient training and prediction of entanglement negativity in isotropic states. 
        In addition, analytical calculations can also lead to the data matrix $A$ and the data label vector $b$ as in Eqs.~\eqref{eq:matrix_A_num}, \eqref{eq:vector_b_num}, and one can solve to retrieve the parameters as $\boldsymbol{g}=A^+\boldsymbol{b}$, where $A^+$ stands for the pseudoinverse of $A$.

        Using the coefficients \eqref{eq:g-coeffs-iso}, one can find th expectation and the reduced variance of $H$: 
        \begin{align}
            \langle H \rangle_{\rho_N} &= \frac{14847+44k (-7+27N(1+N))}{98(303+22k)}\label{eq:negSq_iso_qubit_ave},\\
            \operatorname{rVar}(H)&=-\frac{N^2(-13223+2N(1+N)(1201+968N(1+N))))}{484(1+2N)^2}.
            \label{eq:negSq_iso_qubit_var}
        \end{align}
        We can also compute the classical Fisher information. 
        For this, we find the purities of the reduced states of 
        \[
            \operatorname{Tr}[\rho_A^2] = \operatorname{Tr}[\rho_B^2] = \frac{1}{2}, \qquad \operatorname{Tr}[\rho^2] = \frac{N^2 + N + 1}{3}.
        \]
        Using Eqs.~\eqref{eq:proj_traces}, we obtain
        \begin{align*}
            \operatorname{Tr}\!\big[ P_{A_0A_1}^{\mathrm{sym}} P_{B_0B_1}^{\mathrm{sym}} \, \rho^{\otimes 2} \big] &= \frac{N^2 + N + 7}{12}, \\
            \operatorname{Tr}\!\big[ P_{A_0A_1}^{\mathrm{sym}} P_{B_0B_1}^{\mathrm{asym}} \, \rho^{\otimes 2} \big] &= \frac{2 - N^2 - N}{12}, \\
            \operatorname{Tr}\!\big[ P_{A_0A_1}^{\mathrm{asym}} P_{B_0B_1}^{\mathrm{sym}} \, \rho^{\otimes 2} \big] &= \frac{2 - N^2 - N}{12}, \\
            \operatorname{Tr}\!\big[ P_{A_0A_1}^{\mathrm{asym}} P_{B_0B_1}^{\mathrm{asym}} \, \rho^{\otimes 2} \big] &= \frac{N^2 + N + 1}{12}.
        \end{align*}
        There are four distinct outcomes, two of which are degenerate. 
        Using Eq.~\eqref{eq:fisher_classical}, taking the eigenprojectors $\boldsymbol{\Pi}$ from \eqref{eq:H_eigendecomp_sym_asym}, we calculate:
        \begin{equation}
            I_c(\boldsymbol{\Pi}, \rho_N) = \frac{(2N+1)^2 (2N^2+2N+5)}{8 N^2 (N^2+N+7)(2-N^2-N)(N^2+N+1)}.
        \end{equation}

        The results obtained in this section were shown in the main text, where Figure~\ref{fig:pre_neg_sq_iso} presents both theoretical and numerical prediction performance for various regularization weights $\wvar = \{10^{-4}, 10^{-2}, 10^{-1}\}$, comparing precision and variance across different regularization strengths.
        This time, the observable $H$ does not allow to achieve the classical Cramer-Rao bound in \eqref{eq:var-qcrb}.
        

        
    \subsubsection{Qubit-qubit random mixed states}
    
        Consider the training set $\mathcal{T} = \left\{ \rho_{N_i}^{\otimes 2},\, N_i^2 \right\}_{i=1}^{1000}$, where $\rho_{N_i}$ are random mixed states with negativities $N_i$. The goal is thus to learn a predictor for the \emph{squared} negativity.
        Motivated by our intuition from negativity prediction for pure and isotropic states, we choose the following ansatz for $H_{\boldsymbol{\theta}}$ in Eq.~\eqref{eq:ham-par}:
        \begin{equation}
            \label{eq:neg_sq-ansatz}
            \mathcal{G} = \left\{ \Id_{AB},\; \S_{A_0A_1},\; \S_{B_0B_1},\; \S_{A_0A_1}\S_{B_0B_1} \right\}.
        \end{equation}
        This choice results in only $|\boldsymbol{\theta}| = 4$ variational parameters.
        
\begin{figure*}[tbh]
            \centering
            \includegraphics[width=.495\textwidth]{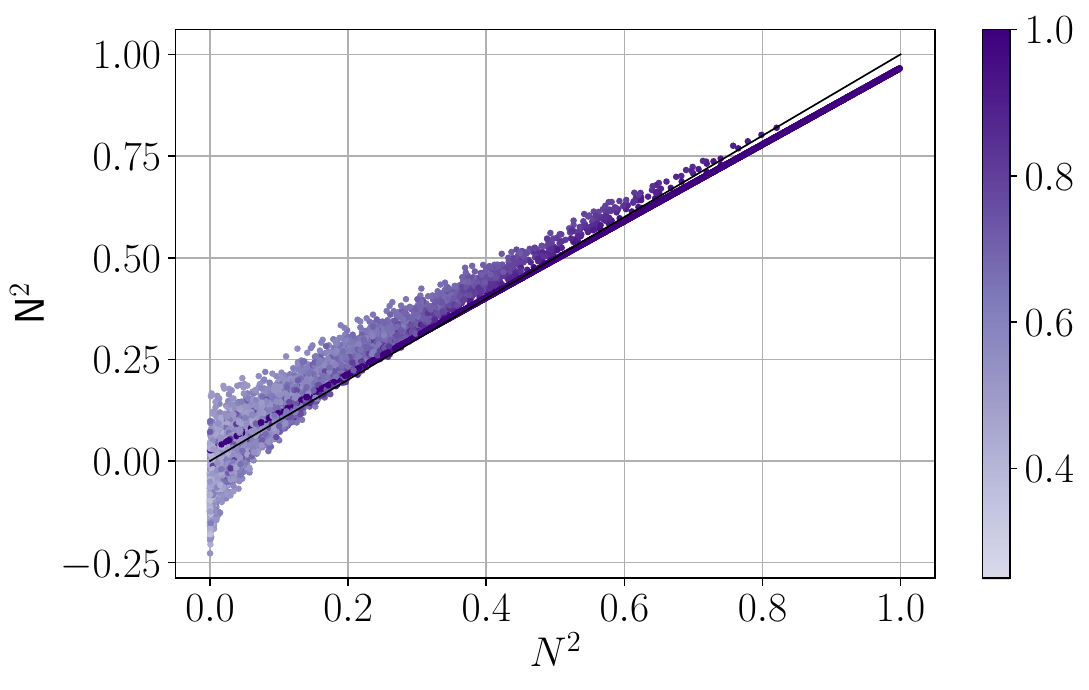}
            \includegraphics[width=.495\textwidth]{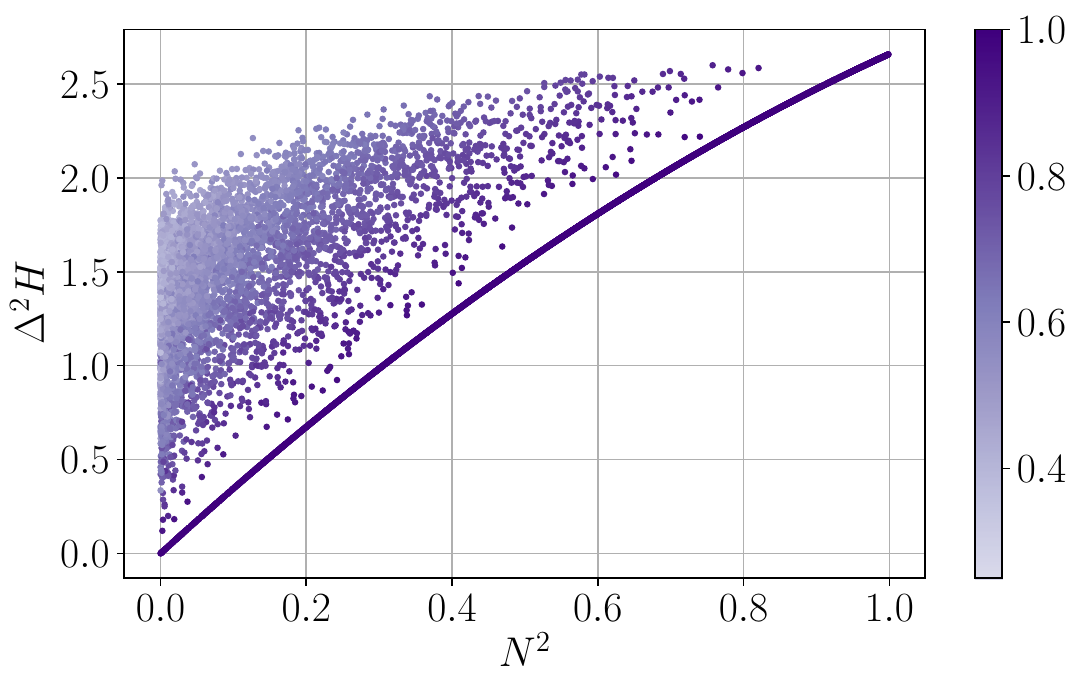}
            \caption{
                Left: Predicted squared negativity $\mathsf{N}^2$ of $10^4$ random mixed states vs. the true squared negativity $N^2$ for $c=2$ copies. The black line connecting $(0,0)$ and $(1,1)$ is the ground truth. 
                Right: Variance of the trained observable $H$.
                The color of the points indicates the purity of the corresponding states.
                The model was trained on an ansatz in Eq.~\ref{eq:neg_sq-ansatz} and on a set $\mathcal{T} = \big\{(\rho_j^{\otimes 2}, N_j )\big\}_{j=1}^{1000}$ with random $\rho_j$ and $N_j^2$ evenly distributed on $[0, 1]$.
            }
            \label{fig:pre-neg_sq}
        \end{figure*}

        The negativity prediction results for this ansatz are shown in Fig.~\ref{fig:pre-neg_sq}. Despite its simplicity, this ansatz achieves better prediction accuracy than a variational quantum circuit with $l=2$ layers acting on all $m=4$ qubits \cite{kard2025}. Moreover, the training and inference require only a few seconds, highlighting the efficiency of the proposed approach.

\section{Train on easy and test on mixed states with $c=2$ copies}
\label{app:sec:easy_2copies}

\begin{figure*}[tbh]
        \centering
        \includegraphics[width=.495\textwidth]{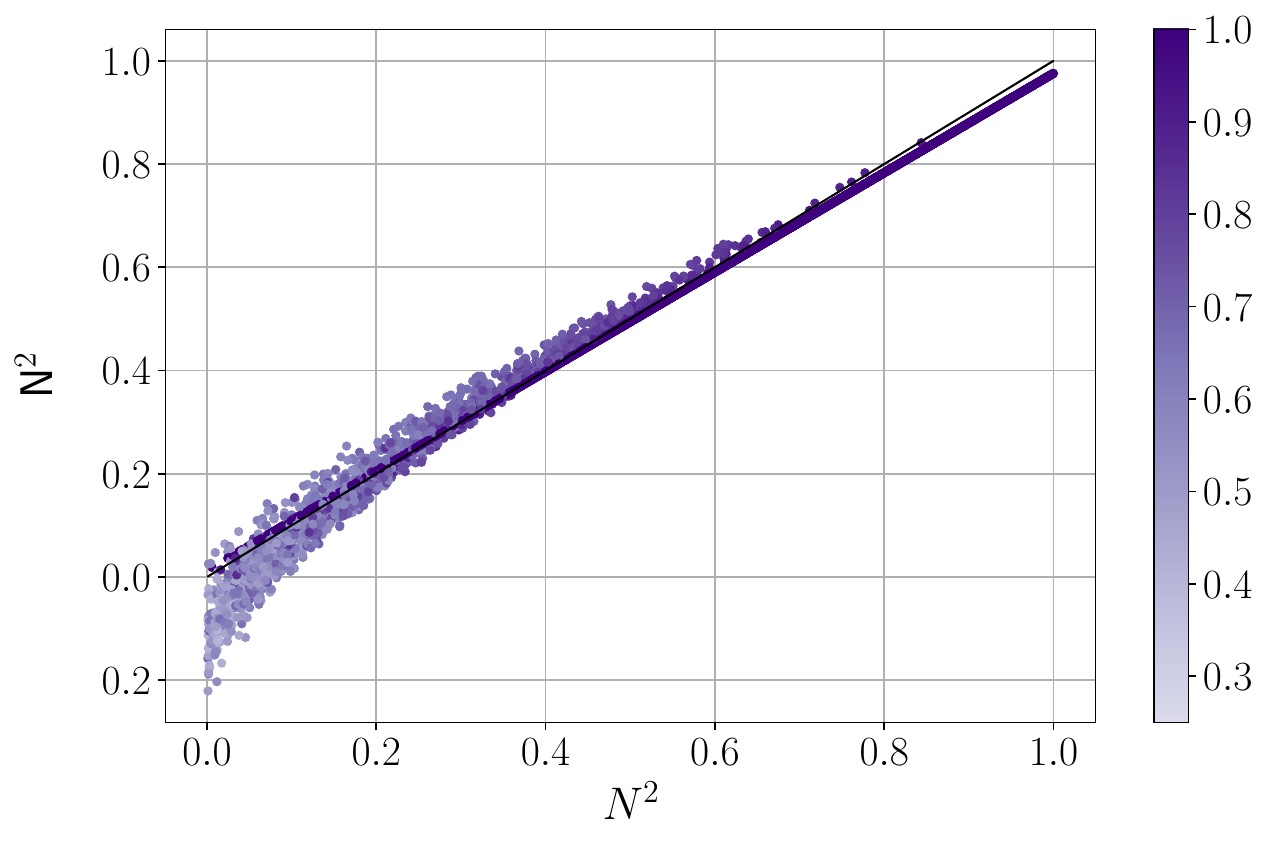}
        \includegraphics[width=.495\textwidth]{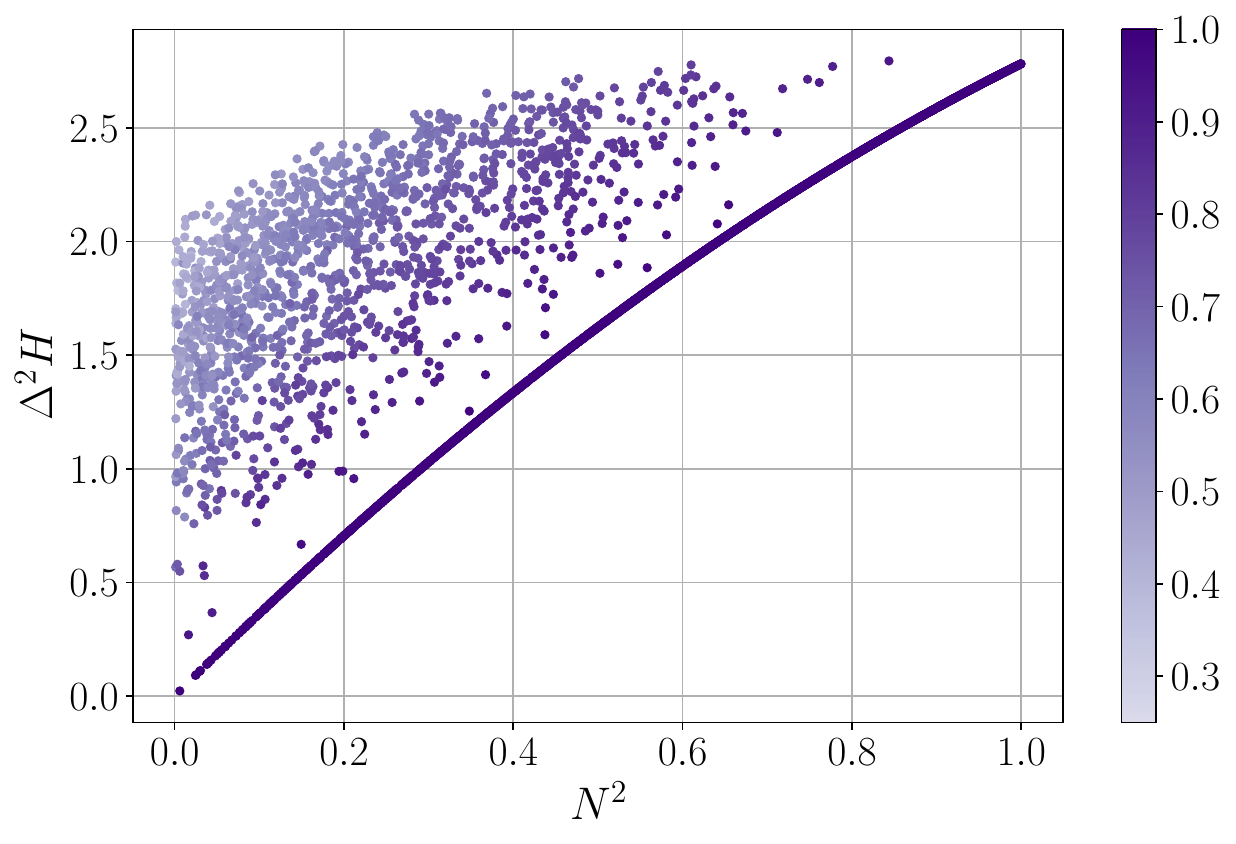}
        \caption{
             Left: Predicted negativity $\mathsf{N}^2$ of $5000$ random mixed states vs. the true negativity $N^2$ for $c=2$ copies. The black line connecting $(0,0)$ and $(1,1)$ is the ground truth. 
             Right: Variance of the trained observable $H$.
             The color of the points indicates the purity of the corresponding states. The model is trained on an ansatz in Eq.~\ref{eq:ansatz_2copies} and on a dataset consisting of $50$ pure and $50$ isotropic states as \textit{easy} states.
        }
        \label{fig:neg_square_2copies_isopure}
    \end{figure*}

In Sec.~\ref{sec:methods} and in Appendix \ref{app:sec:comb_datasets}, we have shown that various datasets can be measured at different times and their corresponding $L$, $S$, and $b$ values are calculated once and stored. In addition, we also separately consider pure states in Sec. \ref{sec:results_neg_pure} and isotropic states in Sec. \ref{sec:results_neg_iso}. In this section, for the overall optimal parameter $\theta^*$ calculation for combined dataset of pure and isotropic states, we only need to calculate the matrix $A$ and data label vector $b$ for each of them and calculate the total values of $A$ and $b$ for the combined datasets as
\begin{align*}
    A_{\mathrm{easy}} =& A_{\mathrm{pure}}+A_{\mathrm{isotropic}},\nonumber\\
    b_\mathrm{{easy}} =& b_{\mathrm{pure}}+b_{\mathrm{isotropic}}\label{eq:isopure_Ab2cop}
\end{align*}
and obtain the trained values as $\theta^*=A_{\mathrm{easy}}^+b_{\mathrm{easy}}$. 
On the other hand, if the mathematical expression of the states with respect to their labels are known, then one can estimate the parameters $\btheta^*$ theoretically as it is demonstrated in Appendix~\ref{app:sec:theory_Ab}, in which one can calculate their corresponding matrix $A$ and label vector $b$. As in Sec. \ref{sec:results_neg_pure} and Sec. \ref{sec:results_neg_iso}, we have computed the parameters for both numerical $\btheta^*$ and theoretical $\boldsymbol{g}^*$ approaches
\begin{align*}
    \boldsymbol{g}^*=(0.24227896, -0.964737  , -0.96474408,  1.69966592),\\
    \btheta^*=(0.24276018, -0.9646805 , -0.9646805 ,  1.69907277).
\end{align*}
As you can see, both parameters ($\boldsymbol{g}^*$ and $\btheta^*$) pretty close to each other. In Fig. \ref{fig:neg_square_2copies_isopure} we present the prediction quality in terms of accuracy and variance for two copies.

\section{Training with $c=3$ copies}
\label{app:sec:3_copies}
In Table~\ref{tab:c_counts} of the main text, we summarized the number of linearly independent permutation operators that can appear for $c = 2, 3, 4$ copies. In this section, we provide the detailed form of these operators for $c = 3$. Table~\ref{tab:SwapDetails_3copies} lists five classes of operators: Classes~1--4 correspond to linearly independent Hermitian operators, while Class~5 represents a cyclic operation. After hermitization, as discussed in Section~\ref{sec:herm}, these five classes collectively yield linearly independent permutation operators for $c = 3$.

\begin{figure*}[tbh]
            \centering
            \includegraphics[width=.495\textwidth]{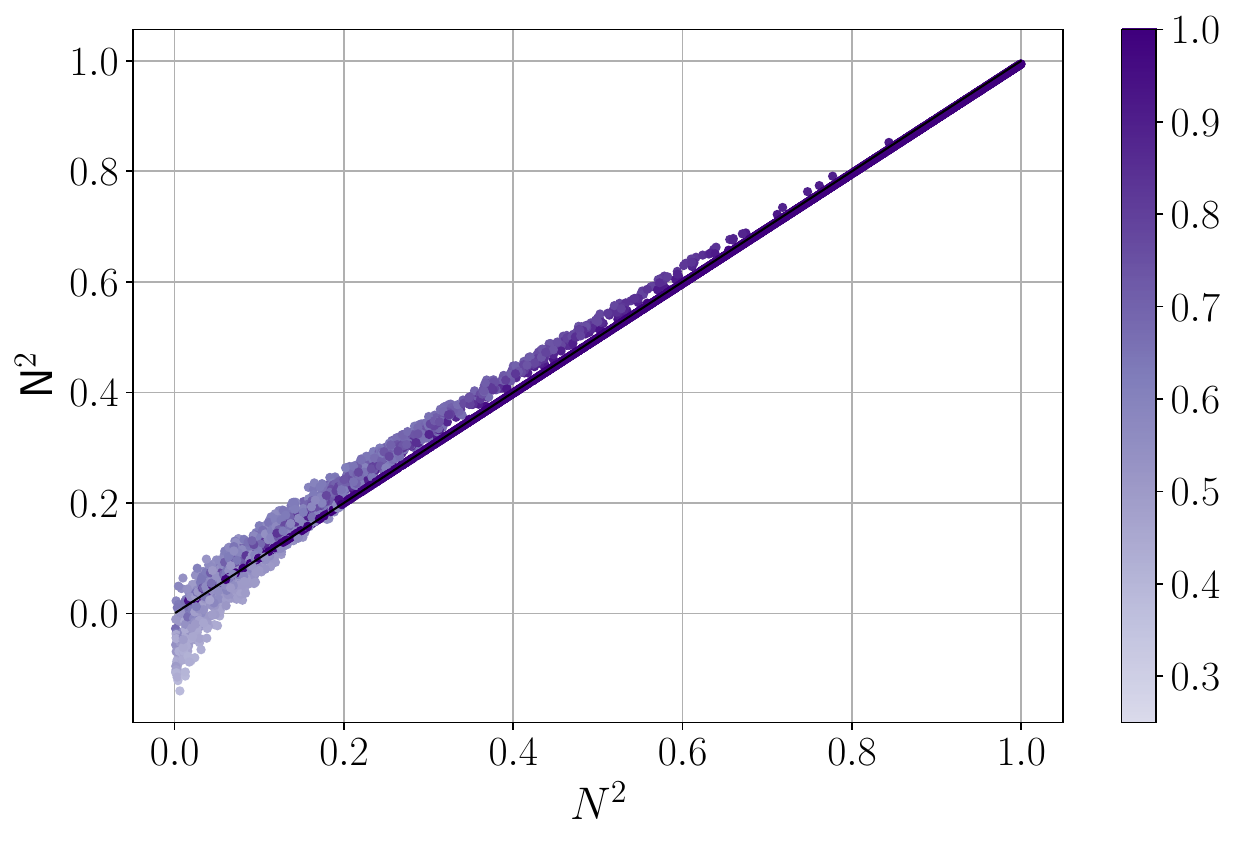}
            \includegraphics[width=.495\textwidth]{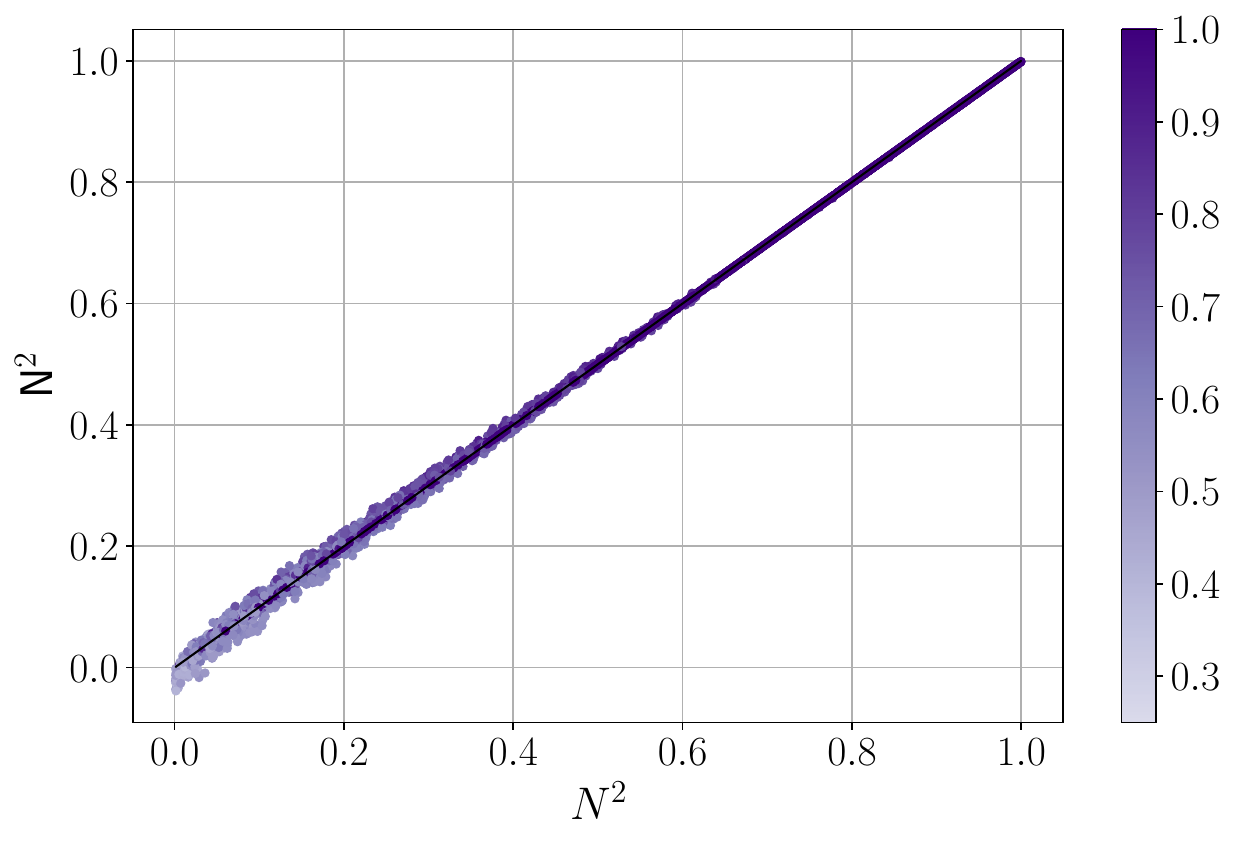}
            \caption{
                Predicted squared negativity $\mathsf{N}^2$ of $5000$ random mixed states vs. the true squared negativity $N^2$ for $c=3$ copies. As an ansatz, either first four Hermitian classes (left) or all five (including the non-Hermitian) classes (right) in Table~\ref{tab:SwapDetails_3copies} were chosen.
                The model was trained on a dataset $\mathcal{T} = \big\{(\rho_j^{\otimes 3}, N_j )\big\}_{j=1}^{100}$ with random $\rho_j$ and $N_j^2$ evenly distributed on $[0, 1]$. The black lines connecting $(0,0)$ and $(1,1)$ are the ground truth.
            }
            \label{fig:pre-neg_sq_3copies}
        \end{figure*}

We construct an ansatz following Eqs.~\eqref{eq:Hj_ansatz} and \eqref{eq:H10_ansatz} using the first four classes from Table~\ref{tab:SwapDetails_3copies}. Additionally, we incorporate Class~5 via hermitization, represented explicitly as
\begin{equation}
\frac{1}{2}\bigl(\S_{A_0A_2}\S_{A_1A_2}\S_{B_0B_2}\S_{B_1B_2} + \S_{A_1A_2}\S_{A_0A_2}\S_{B_1B_2}\S_{B_0B_2}\bigr),
\end{equation}
and extend the ansatz accordingly. Both models are trained on a dataset $\mathcal{T} = \big\{(\rho_j^{\otimes 3}, N_j )\big\}_{j=1}^{100}$ of random mixed states. In Fig.~\ref{fig:pre-neg_sq_3copies}, the left panel shows predictions obtained from training on the first four classes only, while the right panel includes all five classes. The comparison clearly demonstrates that incorporating the cyclic (non-Hermitian) operator through hermitization substantially improves prediction quality.

\begin{table*}[!t]
\centering
\caption{Permutation operators of 5 classes expressed in terms of Swap operators for $c=3$}
\label{tab:SwapDetails_3copies}
\begin{tabular}{|l|c|c|c|c|}
\hline
Class & Swaps over $A$ & Swaps over $B$& Formula & parameter $\btheta^*$\\
\hline 
1&&&1&-0.8164\\
\hline
2&$\S_{A_0A_1}$&$\S_{B_0B_1}$&$\Tr(\rho^2)$&-0.5355\\
&$\S_{A_1A_2}$&$\S_{B_1B_2}$& &\\
&$\S_{A_0A_2}$&$\S_{B_0B_2}$& &\\
\hline 
3&$\S_{A_0A_1}$&$\S_{B_0B_2}$&$\Tr \rho(\rho_A\otimes \rho_B)$&-0.6655\\
&$\S_{A_1A_2}$&$\S_{B_0B_2}$& &\\
&$\S_{A_1A_2}$&$\S_{B_0B_1}$& &\\
&$\S_{A_0A_1}$&$\S_{B_1B_2}$& &\\
&$\S_{A_0A_2}$&$\S_{B_1B_2}$& &\\
&$\S_{A_0A_2}$&$\S_{B_0B_1}$& &\\
\hline 
4&$\S_{A_0A_1}$&&$\Tr(\rho_A^2)$&0.6578\\
&$\S_{A_0A_2}$&& &\\
&$\S_{A_1A_2}$&& &\\
&&$\S_{B_0B_1}$& $\Tr(\rho_B^2)$&0.6728\\
&&$\S_{B_0B_2}$& &\\
&&$\S_{B_1B_2}$& &\\
\hline 
5&$\S_{A_0A_2}\S_{A_1A_2}$&$\S_{B_0B_2}\S_{B_1B_2}$&$\Tr(\rho^3)$&2.4241\\
\hline
\end{tabular}
\end{table*}

\section{Derivation of mathematical expressions for Tables~\ref{tab:10-class_short} and \ref{tab:SwapDetails_3copies}}
\label{sec:app:10-class_derivations}

\begin{figure}
    \centering
    \includegraphics[width=0.5\linewidth]{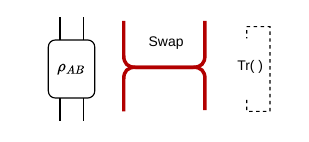}
    \caption{Designations: density matrix $\rho_{AB}$, swap operator $\S$, and trace operation $\Tr(\cdot)$ (shown as wire-binding in dashed lines).}
    \label{fig:diag_descr}
\end{figure}

In this section, we formulate the permutation operators for each class in terms of elementary swap operators, as systematically categorized in Table~\ref{tab:SwapDetails_3copies} and \ref{tab:SwapDetails_1-6}. We present detailed diagrammatic solutions for several key operators, while noting that the remaining classes can be constructed using analogous methods. Table~\ref{tab:SwapDetails_1-6} demonstrates the classes in decreasing order with respect to the optimized parameters $\btheta^*$ trained on easy states. In our diagrammatic representation, we denote density matrices $\rho$, swap operators $\S$, and the trace operation (represented by dashed-line wire connections) as illustrated in Fig.~\ref{fig:diag_descr}. In all the diagrams, we assume that the operations flow either from left to right, or from top to bottom.

\begin{figure}
    \centering
    \includegraphics[width=0.6\linewidth]{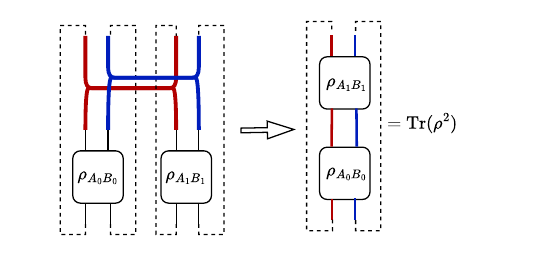}
    \caption{Diagrammatic representation for $\Tr(\S_{A_0A_1}\S_{B_0B_1}\rho^{\otimes 4})$ corresponding to Class 2 in Table~\ref{tab:10-class_short}}
    \label{fig:class2}
\end{figure}

 We begin with Class 2 in Tables \ref{tab:10-class_short} and \ref{tab:SwapDetails_3copies}, which implement the purity measurement $\Tr(\rho^2)$ through the action of the operator $\S_{A_0A_1}\S_{B_0B_1}$. As in Fig~\ref{fig:class2}, we started acting with $\S_{A_0A_1}\S_{B_0B_1}$ on density states $\rho_{A_0A_1}\otimes \rho_{B_0B_1}$ and taking the trace operator via dashed lines, in other words, the action of $\Tr(\S_{A_0A_1}\S_{B_0B_1} \rho_{A_0B_0}\otimes \rho_{A_1B_1})$. However, following the trajectory of the dashed lines, we can see that the subsystem $A_0$ connects to the subsystem $A_1$, and similarly subsystem $B_0$ connects to $B_1$, yielding $\Tr(\rho_{A_0A_1}\otimes \rho_{B_0B_1})$. Since the states $\rho_{A_0B_0}$ and $\rho_{A_1B_1}$ are identical, we conclude with the formula $\Tr \rho^2$.

\begin{figure}
    \centering
    \includegraphics[width=1\linewidth]{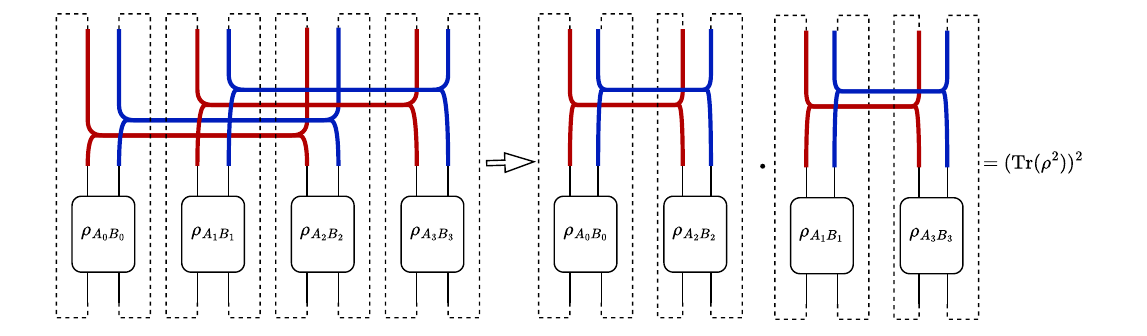}
    \caption{Diagrammatic representation for $\Tr(\S_{A_0A_2}\S_{A_1A_3}\S_{B_0B_2}\S_{B_1B_3}\rho^{\otimes 4})$ corresponding to Class 1 in Table~\ref{tab:10-class_short}}
    \label{fig:class1}
\end{figure}

Class 1 in Table~\ref{tab:10-class_short} can be derived from Class 2 by considering the measurement of the observable $\S_{A_0A_2}\S_{A_1A_3}\S_{B_0B_2}\S_{B_1B_3}$ on the state $\rho^{\otimes 4}$, which corresponds to evaluating $\Tr(\S_{A_0A_2}\S_{A_1A_3}\S_{B_0B_2}\S_{B_1B_3} \rho^{\otimes 4})$. As shown in Fig~\ref{fig:class1}, this measurement decomposes into two independent measurements: $\Tr(\S_{A_0A_2}\S_{B_0B_2} \rho^{\otimes 2})$ and $\Tr(\S_{A_1A_3}\S_{B_1B_3} \rho^{\otimes 2})$. Since each of these measurements corresponds to the Class 2 observable that yields $\Tr(\rho^2)$, and the measurements are performed on independent copies, the overall result is $(\Tr(\rho^2))^2$. 

\begin{figure}
    \centering
    \includegraphics[width=0.4\linewidth]{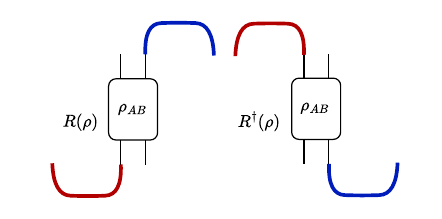}
    \caption{A diagrammatic description of the realignment operator $R(\cdot)$}
    \label{fig:class3r}
\end{figure}

\begin{figure}
    \centering
    \includegraphics[width=0.9\linewidth]{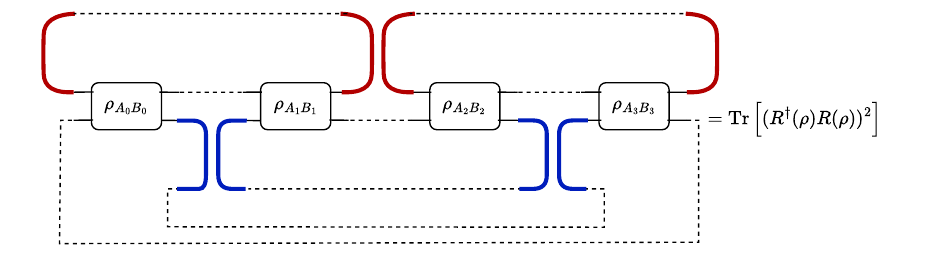}
    \caption{Diagrammatic representation for $\Tr(\S_{A_0A_1}\S_{A_2A_3}\S_{B_0B_3}\S_{B_1B_2}\rho^{\otimes 4})$ corresponding to Class 3 in Table~\ref{tab:10-class_short}}
    \label{fig:class3}
\end{figure}

In Class~3, we consider the realignment operator $R(\cdot)$, defined by its matrix elements as $R(\rho)_{i,j;k,l} = \rho_{i,k;j,l}$. This operation underlies the realignment criterion---also known as the Computable Cross-Norm or Realignment (CCNR) criterion--- for entanglement detection~\cite{chen2002matrix, wang2024moments}: a state $\rho_{AB}$ is certified entangled if $\|R(\rho_{AB})\|_1 > 1$. Following Ref.~\cite{wang2024moments}, $\Tr[R^\dagger R]$ is the first realignment moment, which coincides with the purity considered in Class~2, and $\Tr[(R^\dagger R)^2]$ is the second realignment moment, corresponding to Class~5. Diagrammatically, we represent this operation in Fig.~\ref{fig:class3r}, where the ket of subsystem $B$ becomes a bra and the bra of subsystem $A$ becomes a ket. The adjoint operator $R^\dagger(\cdot)$ reverses this realignment, converting the bra of $B$ back to a ket and the ket of $A$ back to a bra. 
Class 3 encompasses permutations where if two states are connected via the realignment operator through either subsystem $A$ or $B$ and closed with tracing operator, then the complementary subsystem must close with other copies. This constraint yields exactly six possible combinations, as detailed in Table~\ref{tab:SwapDetails_1-6}. 
Fig.~\ref{fig:class3} illustrates the measurement of the operator $\S_{A_0A_1}\S_{A_2A_3}\S_{B_0B_3}\S_{B_1B_2}$. Interpreting the upper indices as subsystem $A$ and bottom indices as $B$, this measurement yields $\Tr\left[R^\dagger(\rho_{AB}) R(\rho_{AB}) R^\dagger(\rho_{AB}) R(\rho_{AB})\right]$. Applying this methodology to the remaining five cases results in the consolidated expression $\Tr\left[\left(R^\dagger(\rho) R(\rho)\right)^2\right]$.

We now evaluate the quantity $\Tr\big[(R(\rho)^\dagger R(\rho))^2\big]$ analytically. Specifically, we show that
\begin{equation}
    \mathcal{E}:=\Tr\big[(\S_{A_0A_1}\S_{A_2A_3}\S_{B_0B_3}\S_{B_1B_2})\,\rho^{\otimes 4}\big]=\Tr\big[(R(\rho)^\dagger R(\rho))^2\big].
\end{equation}
To this end, we express $\rho$ in the operator-Schmidt form $\rho=\sum_{\alpha} s_\alpha\, P_\alpha\otimes Q_\alpha$, with $\Tr(P^\dagger_\alpha P_\beta)=\delta_{\alpha\beta}$ and $\Tr(Q^\dagger_\alpha Q_\beta)=\delta_{\alpha\beta}$, in which the coefficients $s_\alpha$ are the singular values of the realignment matrix $R(\rho)$. That the operator-Schmidt coefficients of $\rho$ coincide with the singular values of $R(\rho)$ is a standard result~\cite{ChenWu2003, Rudolph2003}; for completeness, we give an explicit derivation in the canonical operator basis.

Consider the canonical bases $\{E_{aa'}\}$ and $\{E_{bb'}\}$ of operators on $\mathcal{H}_A$ and $\mathcal{H}_B$, defined by $E_{aa'}=\ketbra{a}{a'}$ and $E_{bb'}=\ketbra{b}{b'}$ for $a,a'=1,\dots,d_A$ and $b,b'=1,\dots,d_B$; that is, $E_{aa'}$ has a single nonzero entry, equal to $1$, in position $(a,a')$, and likewise for $E_{bb'}$. Then, 
\begin{equation}
    \label{eq:app:rho_E_aa'}
    \rho_{AB}=\sum_{a,a'}\sum_{b,b'} \rho_{a,b;a',b'} E_{aa'}\otimes E_{bb'},
\end{equation}
where the coefficients $\rho_{a,b;a',b'}$ is matrix elements of $\rho_{AB}$. The coefficients $\rho_{a,b;a',b'}$ can also be defined through the realignment matrix 
 \begin{equation}
    \label{eq:app:R_aa'}
     R(\rho)_{a,a';b,b'}=\rho_{a,b;a',b'}.
 \end{equation}
We redefine the density state matrix elements for a bipartite state $\rho_{AB}$ as $\rho_{i,j}$ for $i=1,... , d_A^2$,\ $j=1,... , d_B^2$. Then for all $E_i\in\Big\{\ketbra{a}{a'}\Big\}_{a,a'}$ and $E_j\in\Big\{\ketbra{b}{b'}\Big\}_{b,b'}$, we express
\eqref{eq:app:rho_E_aa'} in terms of \eqref{eq:app:R_aa'} as
\begin{equation}
    \label{eq:app:R_ij'}
    \rho = \sum_{i,j} R(\rho)_{i,j} E_i\otimes E_j.
\end{equation}
Now we consider the SVD of the matrix $R(\rho)$ as $R(\rho) = USV^\dagger = \sum_\alpha s_\alpha U_{(:,\alpha)} V_{(\alpha, :)}^\dagger$. Then 
\begin{equation}
    \bigg[R(\rho)\bigg]_{i,j}
    =\bigg[\sum_\alpha s_\alpha\, U_{(:,\alpha)}\, (V^\dagger)_{(\alpha,:)}\bigg]_{i,j}
    =\sum_\alpha s_\alpha\, U_{(i,\alpha)}\, (V^\dagger)_{(\alpha,j)}
    =\sum_\alpha s_\alpha\, U_{(i,\alpha)}\, V_{(j,\alpha)}^*.
\end{equation}
Substituting the last expression for $\bigg[R(\rho)\bigg]_{i,j}$ into \eqref{eq:app:R_ij'},
\begin{equation}
    \rho = \sum_{i,j} \bigg(\sum_\alpha s_\alpha U_{(i,\alpha)} V_{(j,\alpha)}^*\bigg) \big(E_i \otimes E_j\big) = \sum_\alpha s_\alpha \bigg(\sum_i U_{(i,\alpha)} E_i\bigg) \otimes \bigg(\sum_j V_{(j,\alpha)}^* E_j\bigg).
\end{equation}
Defining $P_\alpha=\sum_i U_{(i,\alpha)} E_i$, $Q_\alpha = \sum_j V_{(j,\alpha)}^* E_j$, we could obtain the operator-Schmidt form for a bipartite density state 
\begin{equation}
    \rho_{AB}=\sum_\alpha s_\alpha P_\alpha\otimes Q_\alpha
    \label{eq:app:op_schmidt}
\end{equation}
with $\Tr(P^\dagger_\alpha P_\beta)=\delta_{\alpha\beta}$ and $\Tr(Q^\dagger_\alpha Q_\beta)=\delta_{\alpha\beta}$, because
\begin{align*}
    \Tr(P^\dagger_\alpha P_\beta)
    =\sum_{k,l}U^*_{(k,\alpha)}U_{(l,\beta)}\,\Tr(E_k^\dagger E_l)=\sum_{k,l}U^*_{(k,\alpha)}U_{(l,\beta)}\,\delta_{kl}=\sum_{k}U^*_{(k,\alpha)}U_{(k,\beta)}=\delta_{\alpha\beta}.
\end{align*}
and similarly for $Q_\alpha$. The last equality is due to $U^\dagger U = I$.

Consider $\rho^{\otimes 4}$ using the expansion from Eq.~\eqref{eq:app:op_schmidt}
\begin{equation}
    \rho^{\otimes 4} = \sum_{\alpha \beta \gamma\delta} s_\alpha s_\beta s_\gamma s_\delta (P_\alpha\otimes Q_\alpha)\otimes (P_\beta\otimes Q_\beta)\otimes (P_\gamma\otimes Q_\gamma)\otimes (P_\delta\otimes Q_\delta).
\end{equation}

\begin{figure}
    \centering
    \includegraphics[width=0.9\linewidth]{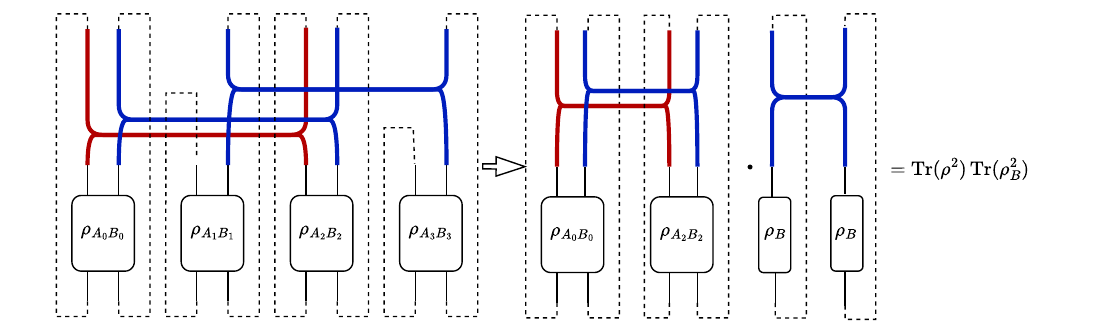}
    \caption{Diagrammatic representation for $\Tr(\S_{A_0A_2}\S_{B_0B_2}\S_{B_1B_3}\rho^{\otimes 4})$ corresponding to Class 4 in Table~\ref{tab:10-class_short}}
    \label{fig:class4}
\end{figure}

Then
\begin{align*}
    \mathcal{E}:&=\Tr\bigg[(\S_{A_0A_1}\S_{A_2A_3}\S_{B_0B_3}\S_{B_1B_2})\rho^{\otimes 4}\bigg] \\
    &=  \sum_{\alpha \beta \gamma\delta} s_\alpha s_\beta s_\gamma s_\delta \Tr\bigg[\S_{A_0A_1}\S_{A_2A_3} (P_\alpha\otimes P_\beta \otimes P_\gamma\otimes P_\delta) \bigg] \Tr\bigg[\S_{B_0A_3}\S_{B_1B_2} (Q_\alpha\otimes Q_\beta \otimes Q_\gamma\otimes Q_\delta) \bigg]\\
    &=\sum_{\alpha \beta \gamma\delta} s_\alpha s_\beta s_\gamma s_\delta \Tr\bigg[\S_{A_0A_1} (P_\alpha\otimes P_\beta)\bigg]  \Tr\bigg[\S_{A_2A_3}(P_\gamma\otimes P_\delta) \bigg] \Tr\bigg[\S_{B_0B_3} (Q_\alpha\otimes Q_\delta)\bigg] \Tr\bigg[\S_{B_1B_2} (Q_\beta\otimes Q_\gamma)\bigg]\\
    &=\sum_{\alpha \beta \gamma\delta} s_\alpha s_\beta s_\gamma s_\delta \Tr\bigg[P_\alpha P_\beta\bigg] \Tr\bigg[P_\gamma P_\delta\bigg] \Tr\bigg[Q_\alpha Q_\delta\bigg] \Tr\bigg[Q_\beta Q_\gamma \bigg]\\
    &=\sum_{\alpha \beta \gamma\delta} s_\alpha s_\beta s_\gamma s_\delta \delta_{\alpha\beta} \delta_{\gamma\delta} \delta_{\alpha\gamma} \delta_{\beta\gamma}
    = \sum_{\alpha} s_\alpha^4.
\end{align*}
On the other hand, since $R(\rho) = USV^\dagger$ then $\Tr\big[ (R^\dagger(\rho) R(\rho))^2 \big]=\Tr\big[S^4\big]=\sum_{\alpha} s_\alpha^4$, where $S$ is a diagonal matrix containing the elements $s_\alpha$. Thus, we have shown that making measurements using the operator $\S_{A_0A_1}\S_{A_2A_3}\S_{B_0B_3}\S_{B_1B_2}$ produces outcomes equivalent to $\Tr\big[ (R^\dagger(\rho) R(\rho))^2 \big]$. It is worth mentioning that the expression $\Tr\big[ R^\dagger(\rho) R(\rho) \big]=\sum_{\alpha} s_\alpha^2$ is equivalent to purity test that is involved in Class 2 in Table~\ref{tab:SwapDetails_1-6}. Wang et. in Ref.~\cite{wang2024moments} considers $\Tr\big[ R^\dagger(\rho) R(\rho) \big]$ as a first realignment moment and $\Tr\big[ (R^\dagger(\rho) R(\rho))^2 \big]$ as a second realignment moment in their entanglement detection problem.
In Class 4 of Table~\ref{tab:10-class_short}, 
we consider the measurement operator $\S_{A_0A_2}\S_{B_0B_2}\S_{B_1B_3}$ acting on $\rho^{\otimes 4}$. First, we note that this overall measurement can be split into two independent parts $\Tr(\S_{A_0A_2}\S_{B_0B_2} \rho^{\otimes 2})$ and $\Tr(\S_{B_1B_3} \rho^{\otimes 2})$. The diagram corresponding to $\Tr(\S_{A_0A_2}\S_{B_0B_2} \rho^{\otimes 2})$ is equivalent to the diagram in Fig.~\ref{fig:class2}, thus $\Tr(\S_{A_0A_2}\S_{B_0B_2} \rho^{\otimes 2})=\Tr(\rho^2)$. $\Tr(\S_{B_1B_3} \rho^{\otimes 2})$ is reduced first tracing out subsystems $A_1$ and $A_3$ since they are not bound with other subsystems through swap operators. Thus, as shown in Fig~\ref{fig:class4}, $\Tr(\S_{B_1B_3} \rho^{\otimes 2}) = \Tr(\S_{B_1B_3}\rho_{B_1}\otimes \rho_{B_3}) = \Tr(\rho_B^2)$. Finally, $\Tr(\S_{A_0A_2}\S_{B_0B_2}\S_{B_1B_3}\rho^{\otimes 4})=\Tr(\rho^2)\Tr(\rho_B^2)$. 

\begin{figure}
    \centering
    \includegraphics[width=0.9\linewidth]{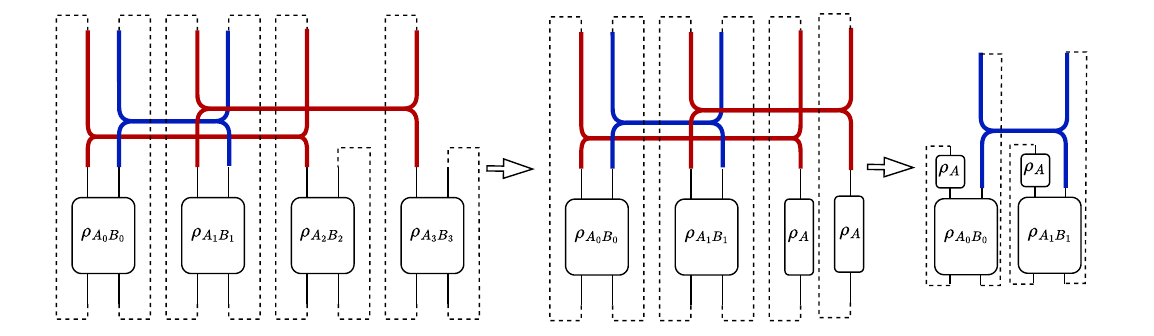}
    \caption{Diagrammatic representation (first half) for $\Tr(\S_{A_0A_2}\S_{A_1A_3}\S_{B_0B_1}\rho^{\otimes 4})$ corresponding to Class 5 in Table~\ref{tab:10-class_short}}
    \label{fig:class5a}
\end{figure}

\begin{figure}
    \centering
    \includegraphics[width=0.75\linewidth]{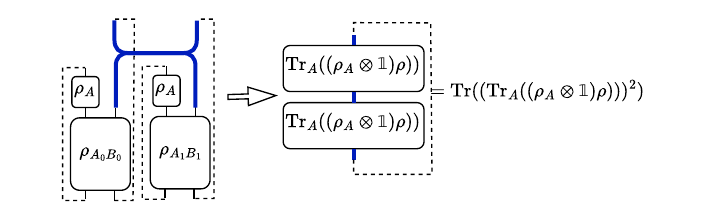}
    \caption{Diagrammatic representation (second half) for $\Tr(\S_{A_0A_2}\S_{A_1A_3}\S_{B_0B_1}\rho^{\otimes 4})$ corresponding to Class 5 in Table~\ref{tab:10-class_short}}
    \label{fig:class5b}
\end{figure}

In Class 5 of Table~\ref{tab:10-class_short}, we consider the measurement operator $\S_{A_0A_2}\S_{A_1A_3}\S_{B_0B_1}$ acting on $\rho^{\otimes 4}$. From Fig.~\ref{fig:class5a}, we can see that subsystem $B$ of $\rho_{A_2B_2}$ and $\rho_{A_3B_3}$ traces out resulting in $\rho_A$ in both cases. Thus, we can translate these two $\rho_A$ states along the swap wires on top of $\rho_{A_0B_0}$ and $\rho_{A_1B_1}$ respectively. Further, as in Fig.~\ref{fig:class5b}, we can reformulate previously obtained result as $\Tr\Big[\S_{B_0B_1}\big(\left((\rho_A\otimes\Id)\rho_{A_0B_0}\right)\otimes \left((\rho_A\otimes\Id)\rho_{A_0B_0}\right)\big)\Big]$. Since $\rho_{A_iB_i}$ are equivalent states, we can consider the latest result as a swap test as in Class 2 in Table~\ref{tab:10-class_short} and the final outcome becomes $\Tr\big[\S_{A_0A_2}\S_{A_1A_3}\S_{B_0B_1}\rho^{\otimes 4}\Big]=\Tr\big[\left(\Tr_A\big[(\rho_A\otimes\Id)\rho_{A_0B_0}\big]\right)^2\big]$.


\begin{figure}
    \centering
    \includegraphics[width=0.6\linewidth]{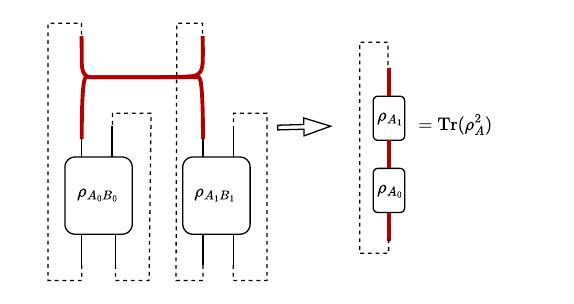}
    \caption{Diagrammatic representation for $\Tr(\S_{A_0A_1}\rho^{\otimes 4})$ corresponding to Class 6 in Table~\ref{tab:10-class_short} and Class 4 of Table~\ref{tab:SwapDetails_3copies}}
    \label{fig:class6}
\end{figure}

In Class 6 of Table~\ref{tab:10-class_short} and in Class 4 of Table~\ref{tab:SwapDetails_3copies}, we consider the measurement operator $S_{A_0A_1}$ acting on four copies of the state, i.e., on $\rho^{\otimes 4}$. Since $S_{A_0A_1}$ only acts on the first two copies, the last two copies are traced out. More specifically, as illustrated in Fig.~\ref{fig:class6}, we first trace over the $B_0$ and $B_1$ subsystems, which reduces the state to $\rho_{A_0} \otimes \rho_{A_1}$. The expectation value then reduces to a swap test between these two reduced density matrices $\Tr\bigl( \S_{A_0A_1} (\rho_{A_0} \otimes \rho_{A_1}) \bigr) = \Tr(\rho_A^2)$.

\begin{figure}
    \centering
    \includegraphics[width=0.7\linewidth]{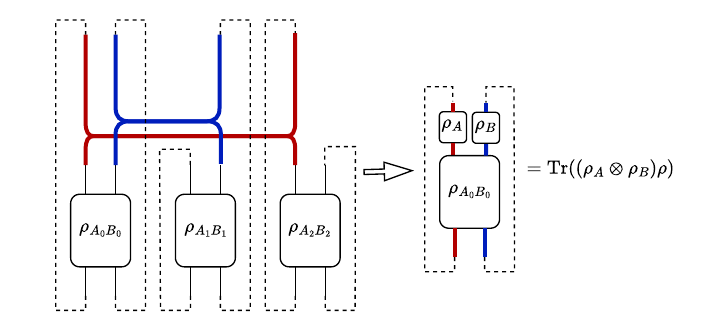}
    \caption{Diagrammatic representation for $\Tr(\S_{A_0A_2}\S_{B_0B_1}\rho^{\otimes 4})$ corresponding to Class 7 in Table~\ref{tab:10-class_short} and Class 3 of Table~\ref{tab:SwapDetails_3copies}}
    \label{fig:class7}
\end{figure}

In Class 7 of Table~\ref{tab:10-class_short} and in Class 3 of Table~\ref{tab:SwapDetails_3copies}, we consider the measurement operator $\S_{A_0A_2}\S_{B_0B_1}$ acting on four copies of the state, i.e., on $\rho^{\otimes 4}$ as in Fig.~\ref{fig:class7}. Since the operator is acting on first three copies, fourth copy traces out. Tracing over subsystems $A_1$ and $B_2$ leaves only $\rho_{B_1}$ and $\rho_{A_2}$. The remaining two states are connected to the state $\rho_{A_0B_0}$ via the measurement operator $\S_{A_0A_2}\S_{B_0B_1}$. Following the swap and trace wires, we can conclude that $\Tr(\S_{A_0A_2}\S_{B_0B_1}\rho^{\otimes 4}) = \Tr((\rho_A\otimes\rho_B)\rho)$.

\begin{figure}
    \centering
    \includegraphics[width=0.75\linewidth]{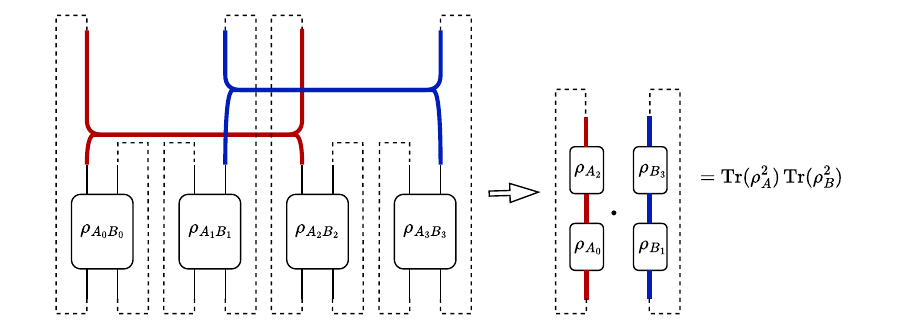}
    \caption{Diagrammatic representation for $\Tr(\S_{A_0A_2}\S_{B_1B_3}\rho^{\otimes 4})$ corresponding to Class 8 in Table~\ref{tab:10-class_short}}
    \label{fig:class8}
\end{figure}

In Class 8 of Table~\ref{tab:10-class_short}, we consider the measurement operator $\S_{A_0A_2}\S_{B_1B_3}$ acting on four copies of the state, i.e., on $\rho^{\otimes 4}$ as in Fig.~\ref{fig:class8}. Since the operators $\S_{A_0A_2}$ and $\S_{B_1B_3}$ are disconnected, we can separately consider them within their copies. Each operator results to a formulation described in Class 6 of Table~\ref{tab:10-class_short}. Thus, we conclude with the solution $\Tr(\S_{A_0A_2}\S_{B_1B_3}\rho^{\otimes 4}) = \Tr(\rho_A^2)\Tr(\rho_B^2)$

\begin{figure}
    \centering
    \includegraphics[width=0.7\linewidth]{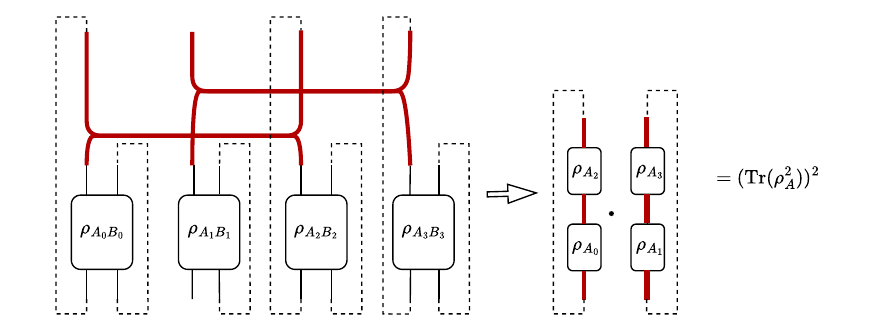}
    \caption{Diagrammatic representation for $\Tr(\S_{A_0A_2}\S_{A_1A_3}\rho^{\otimes 4})$ corresponding to Class 9 in Table~\ref{tab:10-class_short}}
    \label{fig:class9}
\end{figure}

In Class 9 of of Table~\ref{tab:10-class_short}, we consider the measurement operator $\S_{A_0A_2}\S_{A_1A_3}$ acting on four copies of the state, i.e., on $\rho^{\otimes 4}$ as in Fig.~\ref{fig:class9}. Since the operators $\S_{A_0A_2}$ and $\S_{A_1A_3}$ are disconnected, we can separately consider them within their copies. Each of these two swap measurements reveals us $\Tr(\rho_A^2)$ as in Fig.~\ref{fig:class9} and as a result $\Tr(\S_{A_0A_2}\S_{A_1A_3}\rho^{\otimes 4}) = (\Tr(\rho_A^2))^2$

\begin{figure}
    \centering
    \includegraphics[width=0.7\linewidth]{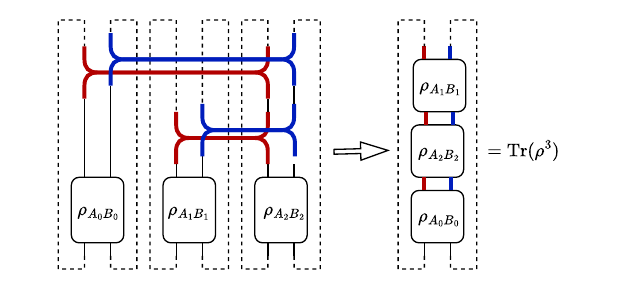}
    \caption{Diagrammatic representation for $\Tr(S_{A_0A_2}S_{A_1A_2}S_{B_0B_2}S_{B_1B_2}\rho^{\otimes 3})$ corresponding to Class 5 in Table~\ref{tab:SwapDetails_3copies}}
    \label{fig:class5_3copies}
\end{figure}

Finally, in Class 5 of Table~\ref{tab:SwapDetails_3copies}, we consider the measurement operator $S_{A_0A_2}S_{A_1A_2}S_{B_0B_2}S_{B_1B_2}$ acting on three copies of the state, i.e., on $\rho^{\otimes 3}$, as illustrated in Fig.~\ref{fig:class5_3copies}. Tracing through the swap network, we observe that subsystem $A_0$ connects directly to $A_2$, and similarly $B_0$ connects to $B_2$. Following the same logic, we find that $\rho_{A_2B_2}$ is connected to $\rho_{A_1B_1}$. Tracing the wires of $A_1$ and $B_1$, we first encounter the swap operators $S_{A_1A_2}S_{B_1B_2}$, followed by $S_{A_0A_2}S_{B_0B_2}$, ultimately connecting to the state $\rho_{A_0B_0}$. As depicted on the right side of Fig.~\ref{fig:class5_3copies}, this configuration corresponds to measuring three consecutive identical copies of $\rho$, yielding the expectation value $\operatorname{Tr}(\rho^{\otimes 3})$.

\section{Extension of Hermitian operators in Table~\ref{tab:10-class_short} with non-Hermitian operators for 4-copies}
\label{app:sec:class_extention}

\begin{table*}[!t]
\centering
\caption{Permutation operators of 6 non-Hermitian classes expressed in terms of swap operators for $c=4$ and their expected values in the state $\rho^{\otimes 4}.$}
\label{tab:SwapDetails_4copies_nonH}
\begin{tabular}{|l|c|c|c|}
\hline
Class & Swaps over $A$ & Swaps over $B$& Formula \\
\hline 
1&$\S_{A_0A_2}\S_{A_1A_2}$&$\S_{B_1B_3}\S_{B_2B_3}$&\\
&$\S_{A_1A_3}\S_{A_2A_3}$&$\S_{B_0B_2}\S_{B_1B_2}$&$\Tr(\rho^2(\rho_A\otimes\rho_B))$\\
\hline 
2&$\S_{A_0A_3}\S_{A_1A_3}\S_{A_2A_3}$&$\S_{B_0B_3}\S_{B_1B_3}\S_{B_2B_3}$&$\Tr (\rho^4)$\\
&$\S_{A_0A_2}\S_{A_1A_2}\S_{A_2A_3}$&$\S_{B_0B_2}\S_{B_1B_2}\S_{B_2B_3}$&\\
\hline 
3&$\S_{A_0A_3}\S_{A_1A_3}\S_{A_2A_3}$&$\S_{B_0B_2}\S_{B_1B_2}\S_{B_2B_3}$&\\
&$\S_{A_0A_2}\S_{A_1A_2}\S_{A_2A_3}$&$\S_{B_0B_3}\S_{B_1B_3}\S_{B_2B_3}$&$\Tr \left((\rho^{T_B})^2(\rho^2)^{T_B}\right)$\\
\hline 
4&$\S_{A_0A_3}\S_{A_1A_3}\S_{A_2A_3}$&$\S_{B_1B_3}\S_{B_2B_3}$&$\Tr(\Tr_B(\rho^3)\rho_A)$\\
&$\S_{A_0A_3}\S_{A_1A_3}\S_{A_2A_3}$&$\S_{B_0B_2}\S_{B_1B_2}$& \\
&$\S_{A_0A_2}\S_{A_1A_2}\S_{A_2A_3}$&$\S_{B_0B_2}\S_{B_1B_2}$&\\
&$\S_{A_1A_3}\S_{A_2A_3}$&$\S_{B_0B_3}\S_{B_1B_3}\S_{B_2B_3}$&$\Tr(\Tr_A(\rho^3)\rho_B)$\\
&$\S_{A_0A_2}\S_{A_1A_2}$&$\S_{B_0B_3}\S_{B_1B_3}\S_{B_2B_3}$&\\
&$\S_{A_0A_2}\S_{A_1A_2}$&$\S_{B_0B_2}\S_{B_1B_2}\S_{B_2B_3}$&\\
\hline 
5&$\S_{A_0A_2}\S_{A_1A_2}\S_{A_2A_3}$&$\S_{B_1B_3}\S_{B_2B_3}$&$\Tr(\Tr_B((\rho^{T_B})^3)\rho_A)$\\
&$\S_{A_1A_3}\S_{A_2A_3}$&$\S_{B_0B_2}\S_{B_1B_2}\S_{B_2B_3}$&$\Tr(\Tr_A((\rho^{T_A})^3)\rho_B)$\\
\hline
6&$\S_{A_0A_2}\S_{A_1A_2}$&$\S_{B_0B_2}\S_{B_1B_2}$&\\
&$\S_{A_0A_2}\S_{A_1A_2}$&$\S_{B_0B_2}\S_{B_1B_2}$&$\Tr(\rho^3)$\\
\hline
\end{tabular}
\end{table*}

In this section, we examine whether prediction performance can be improved by including all 116 Hermitized permutation operators for $c=4$, consisting of 100 Hermitian operators and 16 additional Hermitized operators. The corresponding mathematical expressions for the latter are presented in Table~\ref{tab:SwapDetails_4copies_nonH}, where they can be grouped into six classes. Since using all individual operators is inefficient for both numerical optimization and practical implementation, we follow the construction of the ten Hermitian class operators in Eq.~\eqref{eq:H10_ansatz} and group these 16 operators into six additional Hermitian classes. Together with the original ten classes, this yields a total of 16 classes derived from 576 permutation operators for four copies.

\begin{figure}[h]
            \centering
            \includegraphics[width=.495\textwidth]{10classSwaps4copiesMixedTrainMSE.pdf}
            \includegraphics[width=.495\textwidth]{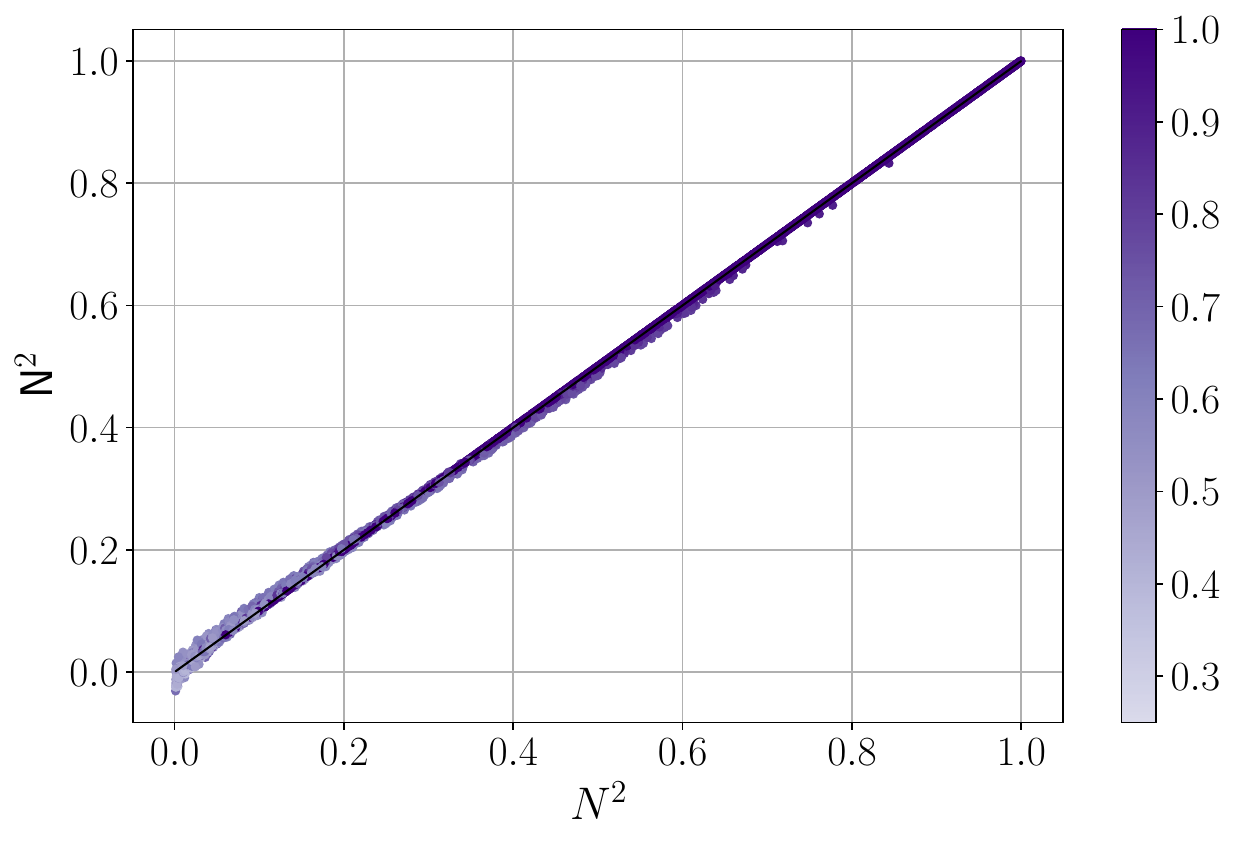}
            \caption{
                Training with \textit{mixed} states. Comparison of prediction performance for the 10-class (left) and 16-class (right) ansätze. Predicted squared negativity $\mathsf{N}^2$ versus true values $\mathsf{N}^2$, where the models are trained on 100 uniformly distributed random mixed states and tested on random mixed states. The colorbar represents the purity of predicted states. The black lines connecting $(0,0)$ and $(1,1)$ are the ground truth.
            }
            \label{fig:10-16mixed_training}
        \end{figure}

\begin{figure}[h]
            \centering
            \includegraphics[width=.495\textwidth]{10classSwaps4copiesEasyTrainMSE.pdf}
            \includegraphics[width=.495\textwidth]{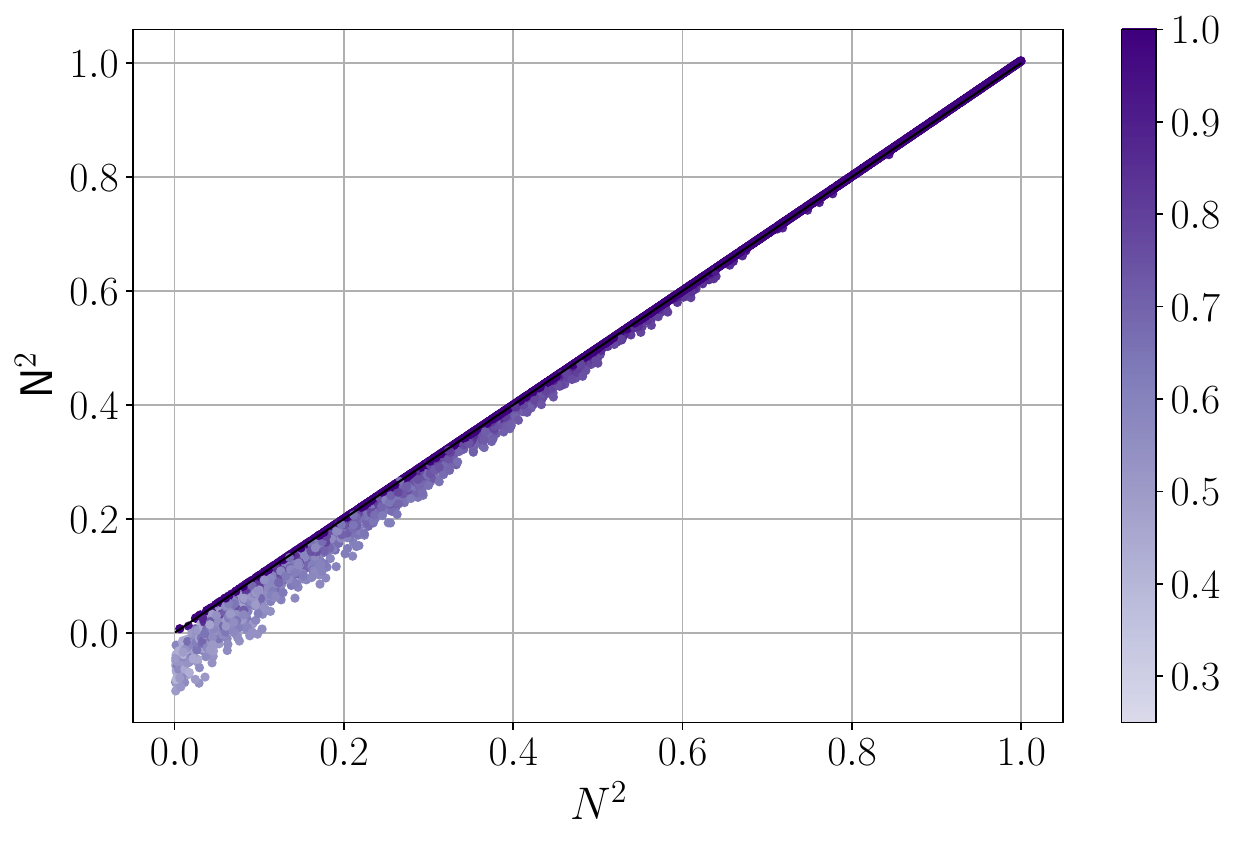}
            \caption{
                Training with \textit{easy} states. Comparison of prediction performance for the 10-class (left) and 16-class (right) ansätze. Predicted squared negativity $\mathsf{N}^2$ versus true values $\mathsf{N}^2$, where the models are trained on 100 uniformly distributed easy states and tested on random mixed states. The colorbar represents the purity of predicted states. The black lines connecting $(0,0)$ and $(1,1)$ are the ground truth.
            }
            \label{fig:10-16easy_training}
        \end{figure}

Optimizing the parameters of these 16 classes yields prediction performance nearly identical to that obtained with the original ten classes (see Fig.~\ref{fig:10-16mixed_training}). In particular, for training on random mixed states, the mean squared errors are comparable: $\mathrm{MSE}_{16\text{ class}} = 1.969 \times 10^{-5}$ versus $MSE_{10\text{ class}} = 2.165 \times 10^{-5}$. However, when training on easy states and testing on random mixed states, the performance significantly deteriorates for the 16-class model, with $\mathrm{MSE}_{16\text{ class}} = 3.351 \times 10^{-4}$ compared to $\mathrm{MSE}_{10\text{ class}} = 9.721 \times 10^{-5}$. This behavior is also reflected in Fig.~\ref{fig:10-16easy_training}, which shows the degradation in prediction quality. While the variance $\Delta^2 H$ remains comparable for mixed-state training, it slightly worsens for easy-state training (see Table~\ref{tab:SwapDetails_16classes}). These results indicate that the additional operators are redundant and can be safely omitted from the ansatz.

In addition, we numerically observe that the expectation values of these non-Hermitian operators with respect to a random state $\rho$ —i.e., the expressions in Table~\ref{tab:SwapDetails_4copies_nonH}— can be highly accurately estimated using our 10 Hermitian class Hermitian ansatz. This observation suggests a potential explanation for the redundancy of Classes~11–16 in Table~\ref{tab:SwapDetails_16classes}, although a formal proof remains an open question. Moreover, since the 116 operators are linearly independent, we can argue that the expectation value of any non-Hermitian permutation operator arising from the full set of 576 operators (after hermitization) can be accurately expressed as a linear combination of expectation values of the 10 Hermitian class operators for any state $\rho$. In other words,
\begin{equation*}
\frac{1}{2}\Tr((\mathbb{P}+\mathbb{P}^\dagger)\rho_{AB}^{\otimes 4}) = \sum_{j=1}^{10} \theta_j \Tr(H_j \rho_{AB}^{\otimes 4}),
\end{equation*}
where $\mathbb{P}$ is a non-Hermitian permutation operator from the set of 576 operators and $\rho$ is a bipartite quantum state.

\begin{figure}[h]
            \centering
            \includegraphics[width=.495\textwidth]{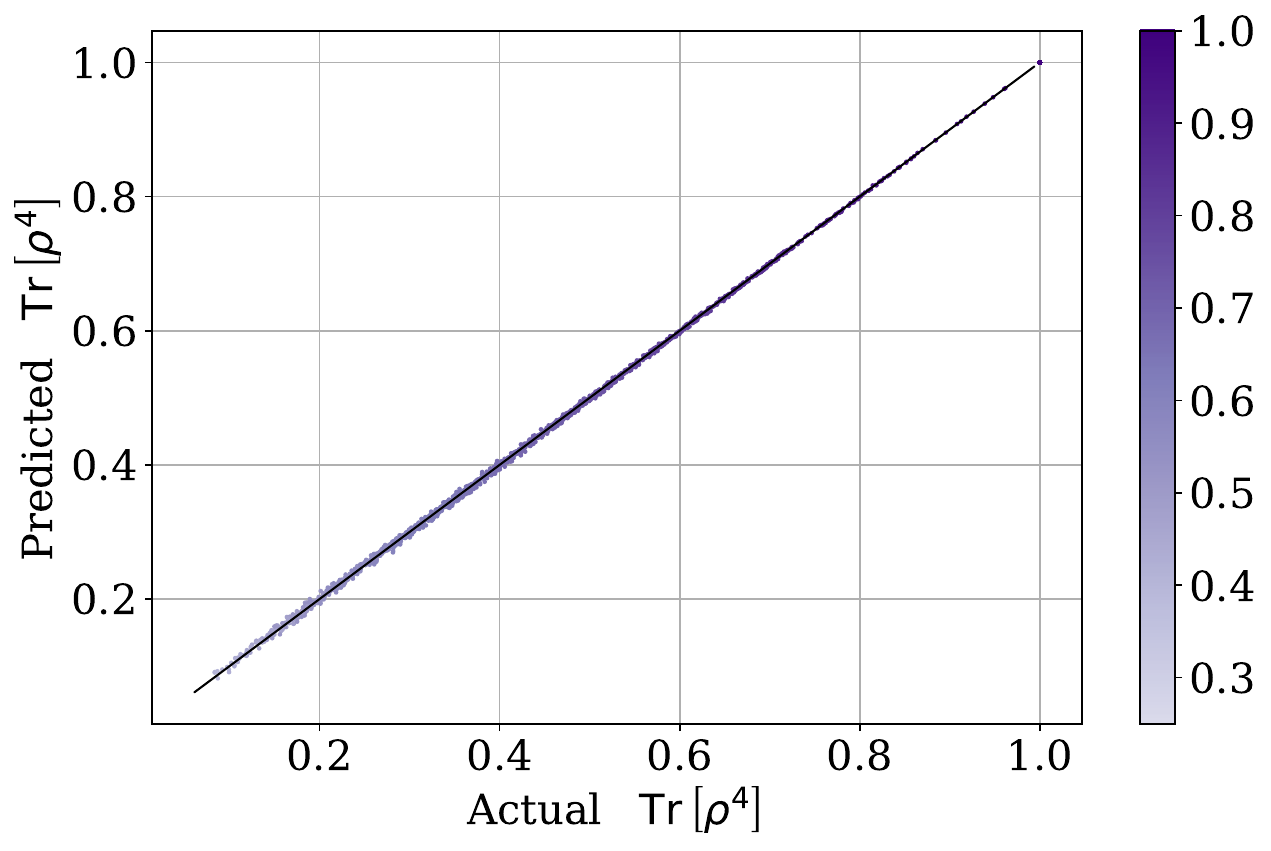}
            \includegraphics[width=.495\textwidth]{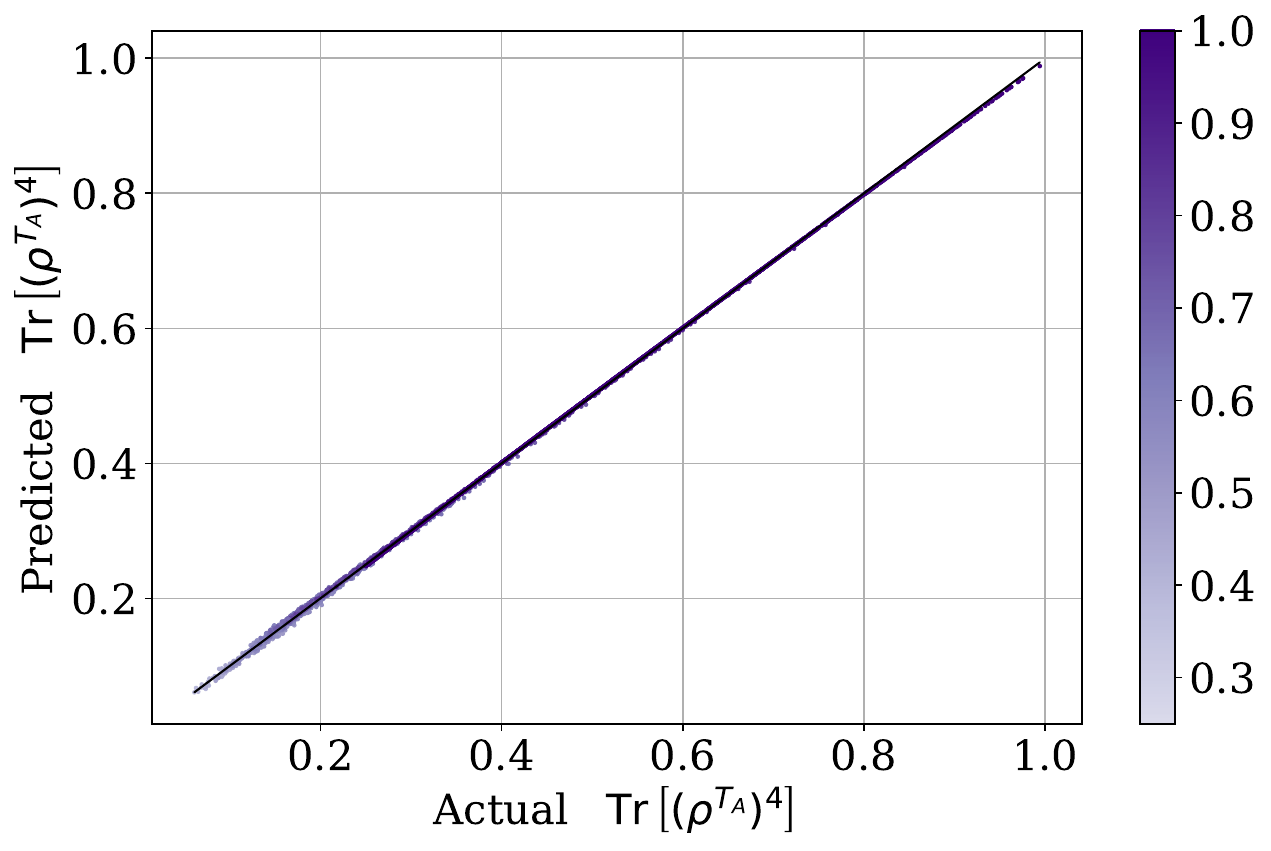}
            \caption{
                Left: Predicted values of $\Tr(\rho^{\otimes 4})$ vs true values of $\Tr(\rho^{\otimes 4})$. Right: Predicted values of $p_4$ vs true values of $p_4$. The models (10 class Hermitian ansatz) is trained on 100 random mixed states $\rho_j^{\otimes 4}$ and true values of $p_4$. The colorbar represents the purity of predicted states. The black lines connecting $(0,0)$ and $(1,1)$ are the ground truth.
            }
            \label{fig:p4_pt4}
        \end{figure}

As an illustrative example, we consider the well-known fourth-order partial-transposition moment
\begin{equation}
\label{eq:p4}
p_4 = \Tr\!\left[(\rho^{T_A})^4\right]=\tfrac{1}{2}\Tr\!\left[(\rho^{T_A})^4 + (\rho^{T_B})^4\right]
     = \tfrac{1}{2}\Tr\!\left[(\forwardPi_A \backwardPi_B + \backwardPi_A \forwardPi_B)\rho^{\otimes 4}\right],
\end{equation}
where $\forwardPi$ denotes a four-cycle permutation acting on one subsystem. In fact,
\begin{equation}
    \Tr[(\forwardPi_A+\backwardPi_A)\otimes (\forwardPi_B+\backwardPi_B)\rho^{\otimes 4}] = \Tr\left[(\forwardPi_A\forwardPi_B + \backwardPi_A\backwardPi_B)\rho^{\otimes 4}\right] + \Tr\left[(\forwardPi_A\backwardPi_B + \backwardPi_A\forwardPi_B)\rho^{\otimes 4}\right] = 2p_4+ 2\Tr(\rho^{4}).
\end{equation}
On the other hand, if we set $\forwardPi_A= \S_{A_0A_1}\S_{A_1A_2}\S_{A_2A_3}$ and $\forwardPi_B= \S_{B_0B_1}\S_{B_1B_2}\S_{B_2B_3}$, then for \textit{two-qubit} states
\begin{align}
    \nonumber
    \forwardPi_A+\backwardPi_A = \S_{A_0A_1}\S_{A_2A_3}+\S_{A_0A_3}\S_{A_1A_2}-\S_{A_0A_2}\S_{A_1A_3}+\S_{A_0A_2}+\S_{A_1A_3} - \Id,
    \\
    \label{eq:fbPi}
    \forwardPi_B+\backwardPi_B = \S_{B_0B_1}\S_{B_2B_3}+\S_{B_0B_3}\S_{B_1B_2}-\S_{B_0B_2}\S_{B_1B_3}+\S_{B_0B_2}+\S_{B_1B_3} - \Id.
\end{align}
The Eq.~\eqref{eq:fbPi} gives a hint that $(\forwardPi_A+\backwardPi_A)\otimes (\forwardPi_B+\backwardPi_B)$ can be expressed exactly in terms of Hermitian permutation operators that are listed in Tables~\ref{tab:SwapDetails_1-6} and \ref{tab:SwapDetails_7-10}. This implies that we can explicitly express $\Tr[(\forwardPi_A+\backwardPi_A)\otimes (\forwardPi_B+\backwardPi_B)\rho^{\otimes 4}]$ as a linear combination of expectation values of 10 class Hermitian operators for any random bipartite qubit state $\rho$. Since we also have highly accurate estimation of $\Tr(\rho^{\otimes 4})$ in terms of 10 Class Hermitian operators, we can indeed have highly accurate estimation for $p_4$ in Eq.~\eqref{eq:p4}.
In Fig.~\ref{fig:p4_pt4}, we demonstrate the approximation accuracy of $\Tr(\rho^{\otimes 4})$ (left) and $p_4$ (right) using the 10-class Hermitian ansatz.

\begin{figure}
    \centering
    \includegraphics[width=0.9\linewidth]{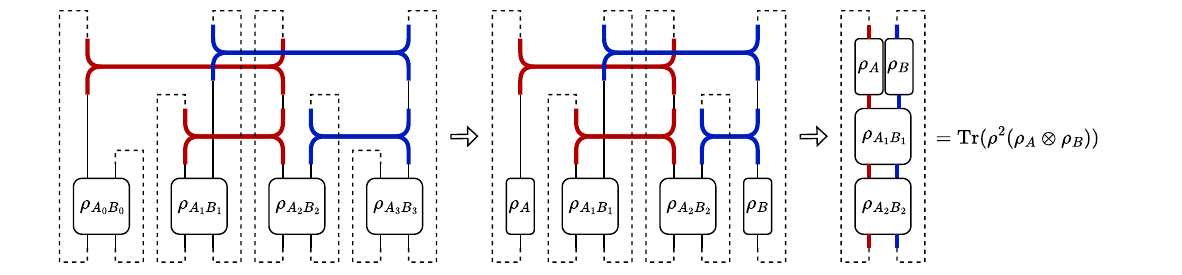}
    \caption{Diagrammatic representation for $\Tr(\S_{A_0A_2}\S_{A_1A_2}\S_{B_1B_3}\S_{B_2B_3}\rho^{\otimes 4})$ corresponding to Class 1 in Table~\ref{tab:SwapDetails_4copies_nonH}}
    \label{fig:class1nonHerm}
\end{figure}

\begin{figure}
    \centering
    \includegraphics[width=0.9\linewidth]{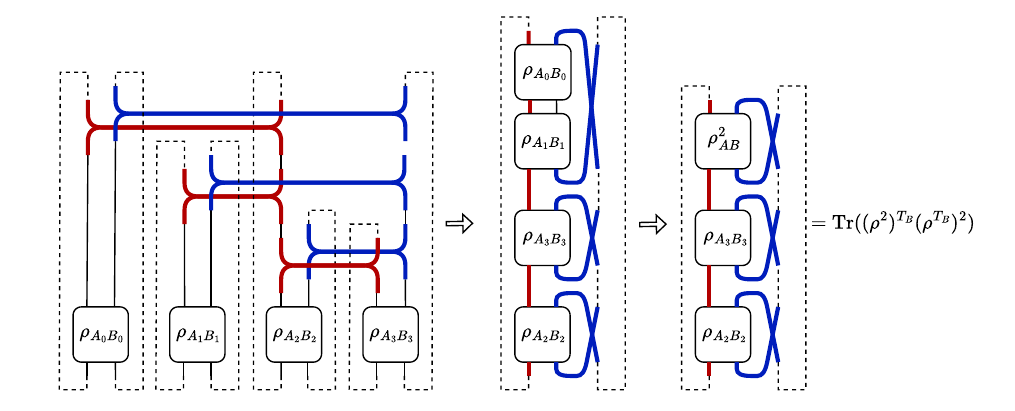}
    \caption{Diagrammatic solution for $\Tr(\S_{A_0A_2}\S_{A_1A_2}\S_{A_2A_3}\S_{B_0B_3}\S_{B_1B_3}\S_{B_2B_3}\rho^{\otimes 4})$ corresponding to Class 3 in Table~\ref{tab:SwapDetails_4copies_nonH}}
    \label{fig:class3nonHerm}
\end{figure}

\begin{figure}
    \centering
    \includegraphics[width=0.8\linewidth]{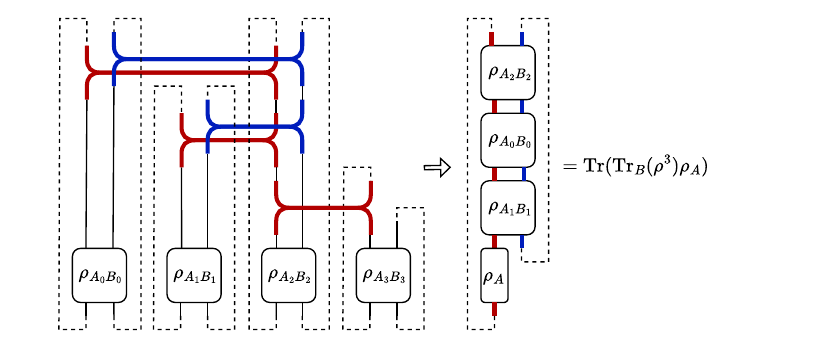}
    \caption{Diagrammatic solution for $\Tr(\S_{A_0A_2}\S_{A_1A_2}\S_{A_2A_3}\S_{B_0B_2}\S_{B_1B_2}\rho^{\otimes 4})$ corresponding to Class 4 in Table~\ref{tab:SwapDetails_4copies_nonH}}
    \label{fig:class4nonHerm}
\end{figure}

\begin{figure}
    \centering
    \includegraphics[width=0.9\linewidth]{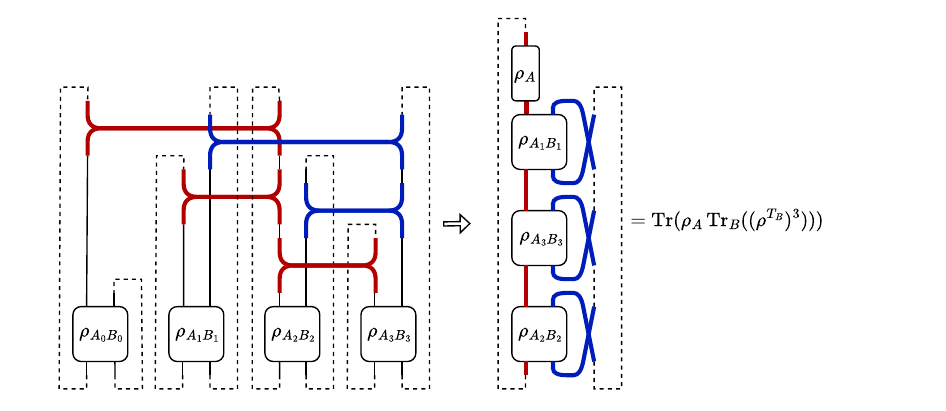}
    \caption{Diagrammatic solution for $\Tr(\S_{A_0A_2}\S_{A_1A_2}\S_{A_2A_3}\S_{B_1B_3}\S_{B_2B_3}\rho^{\otimes 4})$ corresponding to Class 5 in Table~\ref{tab:SwapDetails_4copies_nonH}}
    \label{fig:class5nonHerm}
\end{figure}

Further, in Figs.~\ref{fig:class1nonHerm}, \ref{fig:class3nonHerm},\ref{fig:class4nonHerm},\ref{fig:class5nonHerm} we present the diagrammatic solutions for the Classes 1,3,4 and 5 listed in Table~\ref{tab:SwapDetails_4copies_nonH} respectively. In these diagrams, following the swap wires, we can simplify the operations and deduce the final mathematical formulations. The Classes 2 and 6 can be analogously obtained as in the diagram in Fig.~\ref{fig:class5_3copies}.

\begin{table*}[!t]
\centering
\caption{Parameters trained on random mixed or easy states for 10 Hermitian classes and 6 non-Hermitian classes as an ansatz.}
\label{tab:SwapDetails_16classes}  
\begin{tabular}{|l|c|c|c|c|c|}
\hline
Class & Formula &  $\btheta_{\text{mixed}}^*$&  $\btheta_{\text{easy}}^*$ &$\btheta_{\text{mixed}}^*$&$\btheta_{\text{easy}}^*$\\
\hline 
1 & $(\Tr \big[\rho^2\big])^2$ & -0.0514 & -0.3627&0.6453&0.3895 \\
\hline 
2 & $\Tr \rho^2$  & 0.0082 & -0.1130&0.2725&0.1781 \\
\hline 
3 &  $\Tr\big[(R^\dagger R)^2\big]$ & 0.0767 &  0.1717 &-0.4870&-0.1434\\
\hline 
4 & $\Tr \big[\rho_A^2\big]\Tr \big[\rho^2\big]+\Tr \big[\rho_B^2\big]\Tr \big[\rho^2\big]$ & -0.0155 &0.0636 &-0.3042&-0.1355\\
\hline 
5 & $\Tr\big[\Tr_A\big[((\rho_A\otimes\Id)\rho)^2\big]\big]+\Tr(\Tr_A((\Id\otimes \rho_B)\rho))^2$ & 0.0395 & -0.1044 &0.3531&0.0560\\
\hline 
6 & $\Tr \rho_A^2+\Tr \rho_B^2$ & 0.4851 & 0.1367&0.3980&-0.0390 \\
\hline 
7 &$\Tr \rho(\rho_A\otimes \rho_B)$ & -0.2332 & -0.0457&-0.3227&-0.0378\\
\hline 
8 & $\Tr(\rho_A^2) \Tr(\rho_B^2)$ &  0.0938 & -0.0444&0.1910&0.0371 \\
\hline 
9 & $\Tr(\rho_A^2)^2+\Tr(\rho_B^2)^2$ & -0.2390 & 0.1004&-0.4083&0.0058 \\
\hline 
10 & 1 & -1.5643 & -0.3893 &-1.1998&0.0249\\
\hline 
11 & $\Tr(\rho^2(\rho_A\otimes\rho_B))$ & 0.0013 & -0.0153&-&- \\
\hline 
12 & $\Tr (\rho^4)$ & 0.4342 & 0.7251&-&- \\
\hline 
13 & $\Tr \left((\rho^T)^2(\rho^2)^T\right)$ & -0.0274 & -0.0362&-&- \\
\hline 
14 & $\Tr(\Tr_B(\rho^3)\rho_A)+\Tr(\Tr_A(\rho^3)\rho_B)$ & -0.0031 & -0.0062&-&- \\
\hline 
15 & $\Tr(\Tr_B((\rho^T)^3)\rho_A)+\Tr(\Tr_A((\rho^T)^3)\rho_B)$ & -0.0535 & -0.0680&-&- \\
\hline 
16 & $\Tr(\rho^3)$ & 0.2014 & 0.4074&-&- \\
\hline
& Total MSE &$1.969\times 10^{-5}$&$3.3511\times 10^{-4}$&$2.165\times 10^{-5}$&$9.721\times 10^{-5}$ \\
\hline
& Total Mean Variance &0.4656&0.4879&0.4682&0.4476 \\
\hline
\end{tabular}
\end{table*}
        
\begin{table*}[!htb]
\centering
\caption{Permutation operators of 10 classes (1-6 classes) expressed in terms of swap operators for $c=4$ and their expected values in the state $\rho^{\otimes 4}$.}
\label{tab:SwapDetails_1-6}
\begin{tabular}{|l|c|c|c|c|}
        \hline
        Class & Swaps over $A$ & Swaps over $B$& Formula & $\btheta^*_{easy}$\\

\hline 
1&$\S_{A_0A_2}$$\S_{A_1A_3}$&$\S_{B_0B_2}$$\S_{B_1B_3}$&$(\Tr \rho^2)^2$&0.3685\\
&$\S_{A_0A_1}$$\S_{A_2A_3}$&$\S_{B_0B_1}$$\S_{B_2B_3}$& &\\
&$\S_{A_0A_3}$$\S_{A_1A_2}$&$\S_{B_0B_3}$$\S_{B_1B_2}$& &\\
\hline  
2&$\S_{A_0A_3}$&$\S_{B_0B_3}$&$\Tr \rho^2$&0.1877\\
&$\S_{A_2A_3}$&$\S_{B_2B_3}$& &\\
&$\S_{A_0A_1}$&$\S_{B_0B_1}$& &\\
&$\S_{A_1A_2}$&$\S_{B_1B_2}$& &\\
&$\S_{A_1A_3}$&$\S_{B_1B_3}$& &\\
&$\S_{A_0A_2}$&$\S_{B_0B_2}$& &\\
\hline 
3&$\S_{A_0A_1}$$\S_{A_2A_3}$&$\S_{B_0B_2}$$\S_{B_1B_3}$&$\Tr(R^\dagger R)^2$&-0.1388\\
&$\S_{A_0A_1}$$\S_{A_2A_3}$&$\S_{B_0B_3}$$\S_{B_1B_2}$& &\\
&$\S_{A_0A_2}$$\S_{A_1A_3}$&$\S_{B_0B_3}$$\S_{B_1B_2}$& &\\
&$\S_{A_0A_2}$$\S_{A_1A_3}$&$\S_{B_0B_1}$$\S_{B_2B_3}$& &\\
&$\S_{A_0A_3}$$\S_{A_1A_2}$&$\S_{B_0B_2}$$\S_{B_1B_3}$& &\\
&$\S_{A_0A_3}$$\S_{A_1A_2}$&$\S_{B_0B_1}$$\S_{B_2B_3}$& &\\
\hline 
4&$\S_{A_1A_3}$&$\S_{B_0B_2}$$\S_{B_1B_3}$&$\Tr (\rho_A^2)\Tr (\rho^2)$&-0.1302\\
&$\S_{A_0A_2}$&$\S_{B_0B_2}$$\S_{B_1B_3}$& &\\
&$\S_{A_2A_3}$&$\S_{B_0B_1}$$\S_{B_2B_3}$& &\\
&$\S_{A_0A_1}$&$\S_{B_0B_1}$$\S_{B_2B_3}$& &\\
&$\S_{A_1A_2}$&$\S_{B_0B_3}$$\S_{B_1B_2}$& &\\
&$\S_{A_0A_3}$&$\S_{B_0B_3}$$\S_{B_1B_2}$& &\\
&$\S_{A_0A_3}$$\S_{A_1A_2}$&$\S_{B_0B_3}$& $\Tr (\rho_B^2)\Tr (\rho^2)$&\\
&$\S_{A_0A_3}$$\S_{A_1A_2}$&$\S_{B_1B_2}$& &\\
&$\S_{A_0A_1}$$\S_{A_2A_3}$&$\S_{B_0B_1}$& &\\
&$\S_{A_0A_2}$$\S_{A_1A_3}$&$\S_{B_1B_3}$& &\\
&$\S_{A_0A_2}$$\S_{A_1A_3}$&$\S_{B_0B_2}$& &\\
&$\S_{A_0A_1}$$\S_{A_2A_3}$&$\S_{B_2B_3}$& &\\
\hline 
5&$\S_{A_0A_1}$$\S_{A_2A_3}$&$\S_{B_0B_3}$&$\Tr(\Tr_A((\rho_A\otimes\Id)\rho))^2$&0.0539\\
&$\S_{A_0A_1}$$\S_{A_2A_3}$&$\S_{B_0B_2}$& &\\
&$\S_{A_0A_2}$$\S_{A_1A_3}$&$\S_{B_1B_2}$& &\\
&$\S_{A_0A_2}$$\S_{A_1A_3}$&$\S_{B_0B_3}$& &\\
&$\S_{A_0A_1}$$\S_{A_2A_3}$&$\S_{B_1B_2}$& &\\
&$\S_{A_0A_1}$$\S_{A_2A_3}$&$\S_{B_1B_3}$& &\\
&$\S_{A_0A_3}$$\S_{A_1A_2}$&$\S_{B_1B_3}$& &\\
&$\S_{A_0A_2}$$\S_{A_1A_3}$&$\S_{B_0B_1}$& &\\
&$\S_{A_0A_3}$$\S_{A_1A_2}$&$\S_{B_0B_2}$& &\\
&$\S_{A_0A_2}$$\S_{A_1A_3}$&$\S_{B_2B_3}$& &\\
&$\S_{A_0A_3}$$\S_{A_1A_2}$&$\S_{B_2B_3}$& &\\
&$\S_{A_0A_3}$$\S_{A_1A_2}$&$\S_{B_0B_1}$& &\\
&$\S_{A_2A_3}$&$\S_{B_0B_3}$$\S_{B_1B_2}$& $\Tr(\Tr_A((\Id\otimes \rho_B)\rho))^2$&\\
&$\S_{A_0A_2}$&$\S_{B_0B_1}$$\S_{B_2B_3}$& &\\
&$\S_{A_1A_3}$&$\S_{B_0B_3}$$\S_{B_1B_2}$& &\\
&$\S_{A_2A_3}$&$\S_{B_0B_2}$$\S_{B_1B_3}$& &\\
&$\S_{A_0A_1}$&$\S_{B_0B_2}$$\S_{B_1B_3}$& &\\
&$\S_{A_1A_3}$&$\S_{B_0B_1}$$\S_{B_2B_3}$& &\\
&$\S_{A_0A_2}$&$\S_{B_0B_3}$$\S_{B_1B_2}$& &\\
&$\S_{A_0A_1}$&$\S_{B_0B_3}$$\S_{B_1B_2}$& &\\
&$\S_{A_1A_2}$&$\S_{B_0B_2}$$\S_{B_1B_3}$& &\\
&$\S_{A_0A_3}$&$\S_{B_0B_1}$$\S_{B_2B_3}$& &\\
&$\S_{A_0A_3}$&$\S_{B_0B_2}$$\S_{B_1B_3}$& &\\
&$\S_{A_1A_2}$&$\S_{B_0B_1}$$\S_{B_2B_3}$& &\\
\hline 
6&$\S_{A_0A_2}$&&$\Tr \rho_A^2$&-0.0393\\
&$\S_{A_1A_2}$&& &\\
&$\S_{A_0A_3}$&& &\\
&$\S_{A_1A_3}$&& &\\
&$\S_{A_2A_3}$&& &\\
&$\S_{A_0A_1}$&& &\\
&&$\S_{B_0B_1}$& $\Tr \rho_B^2$&\\
&&$\S_{B_0B_3}$& &\\
&&$\S_{B_1B_3}$& &\\
&&$\S_{B_0B_2}$& &\\
&&$\S_{B_2B_3}$& &\\
&&$\S_{B_1B_2}$& &\\
        \hline
\end{tabular}
\end{table*}

\begin{table*}[!t]
\centering
\caption{Permutation operators of 10 classes (7-10 classes) expressed in terms of swap operators for $c=4$ and their expected values in the state $\rho^{\otimes 4}$.}
\label{tab:SwapDetails_7-10}
\begin{tabular}{|l|c|c|c|c|}
        \hline
        Class & Swaps over $A$ & Swaps over $B$& Formula & parameter $\btheta^*$\\

\hline 
7&$\S_{A_0A_1}$&$\S_{B_0B_2}$&$\Tr \rho(\rho_A\otimes \rho_B)$&-0.0376\\
&$\S_{A_2A_3}$&$\S_{B_1B_2}$& &\\
&$\S_{A_2A_3}$&$\S_{B_0B_2}$& &\\
&$\S_{A_2A_3}$&$\S_{B_0B_3}$& &\\
&$\S_{A_1A_3}$&$\S_{B_1B_2}$& &\\
&$\S_{A_0A_1}$&$\S_{B_0B_3}$& &\\
&$\S_{A_1A_3}$&$\S_{B_0B_3}$& &\\
&$\S_{A_0A_1}$&$\S_{B_1B_3}$& &\\
&$\S_{A_1A_2}$&$\S_{B_1B_3}$& &\\
&$\S_{A_0A_2}$&$\S_{B_1B_2}$& &\\
&$\S_{A_0A_2}$&$\S_{B_2B_3}$& &\\
&$\S_{A_0A_1}$&$\S_{B_1B_2}$& &\\
&$\S_{A_1A_3}$&$\S_{B_2B_3}$& &\\
&$\S_{A_0A_3}$&$\S_{B_2B_3}$& &\\
&$\S_{A_2A_3}$&$\S_{B_1B_3}$& &\\
&$\S_{A_1A_2}$&$\S_{B_0B_2}$& &\\
&$\S_{A_1A_2}$&$\S_{B_2B_3}$& &\\
&$\S_{A_0A_3}$&$\S_{B_1B_3}$& &\\
&$\S_{A_1A_3}$&$\S_{B_0B_1}$& &\\
&$\S_{A_0A_2}$&$\S_{B_0B_1}$& &\\
&$\S_{A_0A_2}$&$\S_{B_0B_3}$& &\\
&$\S_{A_0A_3}$&$\S_{B_0B_1}$& &\\
&$\S_{A_0A_3}$&$\S_{B_0B_2}$& &\\
&$\S_{A_1A_2}$&$\S_{B_0B_1}$& &\\
\hline 
8&$\S_{A_2A_3}$&$\S_{B_0B_1}$&$\Tr(\rho_A^2) \Tr(\rho_B^2)$&0.0349\\
&$\S_{A_0A_2}$&$\S_{B_1B_3}$& &\\
&$\S_{A_1A_3}$&$\S_{B_0B_2}$& &\\
&$\S_{A_1A_2}$&$\S_{B_0B_3}$& &\\
&$\S_{A_0A_1}$&$\S_{B_2B_3}$& &\\
&$\S_{A_0A_3}$&$\S_{B_1B_2}$& &\\
\hline 
9&$\S_{A_0A_3}$$\S_{A_1A_2}$&&$\Tr(\rho_A^2)^2$&0.0060\\
&$\S_{A_0A_2}$$\S_{A_1A_3}$&& &\\
&$\S_{A_0A_1}$$\S_{A_2A_3}$&& &\\
&&$\S_{B_0B_2}$$\S_{B_1B_3}$& $\Tr(\rho_B^2)^2$&\\
&&$\S_{B_0B_1}$$\S_{B_2B_3}$& &\\
&&$\S_{B_0B_3}$$\S_{B_1B_2}$& &\\
\hline 
10&&&1&0.0011\\
        \hline
\end{tabular}
\end{table*}

\section{Performing the measurement of permutation operators and processing the outcomes.}
\label{app:sec:measurements}

    \subsection{Measuring Hermitian permutation operations}

        Let us consider the measurement of $\Tr (\S_{A_0A_1}\S_{A_2A_3}\S_{B_0B_2}\S_{B_1B_3}\rho^{\otimes 4})$. As we have discussed in Appendix \ref{app:sec:sym-asym-proj}, a swap operator $\S$ has eigenvalues $+1$ (symmetric subspace) and $-1$ (antisymmetric subspace). Denoting the corresponding eigenprojectors by $\operatorname{P}^{\mathrm{sym}}$ and $\operatorname{P}^{\mathrm{asym}}$, we have
        \begin{align}
            S \operatorname{P}^{\mathrm{sym}} &=   \operatorname{P}^{\mathrm{sym}},\\
            \S \operatorname{P}^{\mathrm{asym}} &= - \operatorname{P}^{\mathrm{asym}}.
        \end{align}
        On the other hand, $\operatorname{P}^{\mathrm{sym}}$ and $ \operatorname{P}^{\mathrm{asym}}$ can be expressed in terms of Bell-basis projectors,
        \begin{align}
            \operatorname{P}^{\mathrm{sym}} &= \ketbra{\beta_{00}}{\beta_{00}}+\ketbra{\beta_{01}}{\beta_{01}}+\ketbra{\beta_{10}}{\beta_{10}},\\
            \operatorname{P}^{\mathrm{asym}} &= \ketbra{\beta_{11}}{\beta_{11}},
        \end{align}
        in which the Bell-states are
        \begin{equation}
            \ket{\beta_{xy}} = \frac{\ket{0y}+(-1)^{x}\ket{1(1\oplus y)}}{\sqrt{2}}.
        \end{equation}
        One can generate these states through unitary transformations
        \begin{equation}
            \ketbra{\beta_{xy}}{\beta_{xy}} = \operatorname{CNOT}(H\otimes \Id)\ketbra{xy}{xy}(H\otimes \Id)\operatorname{CNOT},
        \end{equation}
        which can be used for swap measurements
        \begin{align*}
            \Tr(\S\rho_{12}) &=\sum_{x,y=0}^1(1-2xy) \Tr(\ketbra{\beta_{xy}}{\beta_{xy}}\rho_{12})) \\
            &= \sum_{x,y=0}^1(1-2xy) \bra{xy}(H\otimes \Id)\operatorname{CNOT} \rho_{12}\operatorname{CNOT}(H\otimes \Id)\ket{xy}.
        \end{align*}
        Thus, to measure a swap operator, e.g. $\S_{A_0A_1}(\rho_{A_0B_0}\otimes \rho_{A_1B_1})$, we can transform $\rho_{A_0B_0}\otimes \rho_{A_1B_1}$ into the Bell-measurement basis with $U_{A_0A_1}=(H\otimes \Id)_{A_0A_1}\operatorname{CNOT}_{A_0A_1}$ and perform measurements in the computational basis
        \begin{equation}
            \Tr\Big(\S_{A_0A_1}(\rho_{A_0B_0}\otimes \rho_{A_1B_1})\Big) = \sum_{x,y=0}^1(1-2xy) \bra{xy}U_{A_0A_1} (\rho_{A_0B_0}\otimes \rho_{A_1B_1})U_{A_0A_1}^\dagger \ket{xy}.
        \end{equation}
        Returning to the example $\Tr\!\left(\S_{A_0A_1}\S_{A_2A_3}\S_{B_0B_2}\S_{B_1B_3}\rho^{\otimes 4}\right)$, one can implement the measurement by applying the corresponding unitary transformations $U_{A_0A_1},\, U_{A_2A_3},\, U_{B_0B_2},\, U_{B_1B_3}$ to $\rho^{\otimes 4}$, followed by simultaneous measurements in the computational basis.
        
        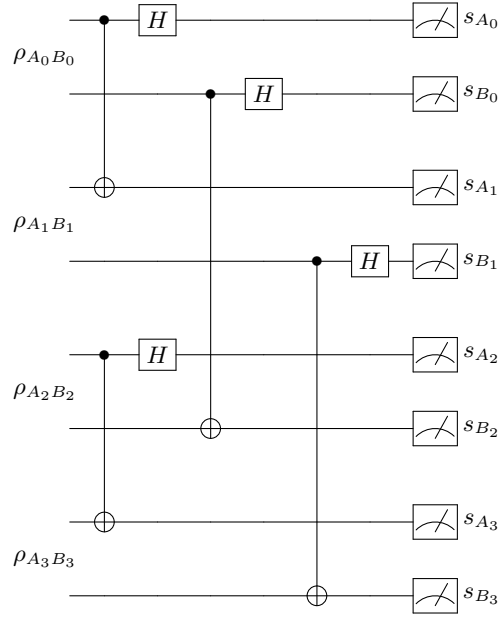
\begin{figure}
            \centering
            \mbox{
        
            \Qcircuit @C=1.0em @R=0.8em {
               && \ctrl{5} & \gate{H}   & \qw       & \qw       & \qw   & \qw &\meter& s_{A_0}\\
               \rho_{A_0B_0}\\
                && \qw      & \qw        & \ctrl{10}  & \gate{H}  & \qw   & \qw&\meter&s_{B_0}\\ 
               \\
               \\
                &&\targ  & \qw        & \qw       & \qw       & \qw   & \qw&\meter&s_{A_1}\\
                \rho_{A_1B_1}\\
                && \qw      & \qw        & \qw       & \qw       &\ctrl{10}  & \gate{H}&\meter&s_{B_1}\\
               \\
               \\
                && \ctrl{5} & \gate{H}   & \qw       & \qw       & \qw   & \qw&\meter&s_{A_2}\\
                \rho_{A_2B_2}\\
                && \qw      & \qw        & \targ  & \qw       & \qw   & \qw&\meter&s_{B_2}\\ 
               \\
               \\
                &&\targ  & \qw        & \qw       & \qw       & \qw   & \qw&\meter&s_{A_3}\\
                \rho_{A_3B_3}\\
                && \qw      & \qw        & \qw       & \qw       & \targ& \qw&\meter&s_{B_3}
            }
            }\caption{Quantum circuit for measuring $\Tr\!\left(\S_{A_0A_1}\S_{A_2A_3}\S_{B_0B_2}\S_{B_1B_3}\rho^{\otimes 4}\right)$.}
            \label{circ:4copies_M}
        \end{figure}
        In the circuit in Fig.~\ref{circ:4copies_M}, measurements in the computational basis yield the bitstrings $(s_{A_0}, s_{B_0}, s_{A_1}, s_{B_1}, s_{A_2}, s_{B_2}, s_{A_3}, s_{B_3})$ in a single shot. 
        We then define the quantities $s_{A_0A_1}$ as
        \begin{equation*}
            s_{A_0A_1} =
                \begin{cases}
                     1 & \text{if } s_{A_0}s_{A_1}\in\{00,01,10\}, \\
                    -1 & \text{if } s_{A_0}s_{A_1}\in\{11\}.
                \end{cases}
        \end{equation*}
        Analogously, we define $s_{A_2A_3}$, $s_{B_0B_2}$, and $s_{B_1B_3}$. In an experiment, one collects the outcomes $\bigcup_{i=1}^{N_{\mathrm{shots}}} \{ s_{A_0A_1}^{(i)}, s_{A_2A_3}^{(i)}, s_{B_0B_2}^{(i)}, s_{B_1B_3}^{(i)} \}$ over $N_{\mathrm{shots}}$ measurement shots.
        To obtain an estimate $\hat{E}$ of $\Tr (\S_{A_0A_1}\S_{A_2A_3}\S_{B_0B_2}\S_{B_1B_3}\rho^{\otimes 4})$ one evaluates
        \begin{equation}
             \hat{E} =\frac{1}{N_{\operatorname{shots}}}\sum_{i=1}^{N_{\operatorname{shots}}} s_{A_0A_1}^{(i)}s_{A_2A_3}^{(i)}s_{B_0B_2}^{(i)}s_{B_1B_3}^{(i)}.
        \end{equation}
        Therefore, by performing simultaneous measurements on 8 qubits over $N_{\mathrm{shots}}$ repetitions, one can estimate $\Tr\!\left(\S_{A_0A_1}\S_{A_2A_3}\S_{B_0B_2}\S_{B_1B_3}\rho^{\otimes 4}\right)$. Moreover, the same measurement outcomes can be reused to estimate simpler Hermitian permutation operators, such as $\Tr\!\left(\S_{A_0A_1}\rho^{\otimes 4}\right)$ and $\Tr\!\left(\S_{A_0A_1}\S_{B_0B_2}\rho^{\otimes 4}\right)$.
        
        This leads to the following observation: by implementing the experimental settings corresponding to Class~1 and Class~3 in Table~\ref{tab:SwapDetails_1-6}, one can reconstruct the expectation values of all permutation operators listed in Tables~\ref{tab:SwapDetails_1-6} and~\ref{tab:SwapDetails_7-10}. Consequently, both in simulations and experiments, it suffices to realize only 9 distinct measurement settings, whose outcomes can be reused to evaluate all Hermitian permutation operators in these tables.

    \subsection{Measuring non-Hermitian permutation operations.}
        
\begin{figure}
        \centering
        \mbox{
            \Qcircuit @C=1em @R=1.em {
                & \lstick{\ket{0}} & \gate{H} & \ctrl{2} & \qw       & \gate{H}      & \meterB{\quad \ketbra{0}{0} \quad}      
                \\
                &&&&&&\\
                &  & \qw      & \multigate{8}{\mathbb{P}}  & \qw& \qw      & \\
                &\rho_{A_0B_0}&&&&&\\
                & &\qw&\ghost{U}&\qw&\qw&\\
                &&&&&&\\
                & \vdots           &          &           &         &   \vdots       &            \\
                &&&&&&\\
                &  & \qw      & \ghost{U}       & \qw      & \qw&      \\
                &\rho_{A_cB_c}&&&&&\\
                &  & \qw      & \ghost{U}  & \qw      & \qw      & \\
            }
        }
        \caption{Quantum circuit for measuring $\Tr\!\left(\mathbb{P}\rho_{AB}^{\otimes c}\right)$.}
        \label{circ:Perm_measure}
    \end{figure}
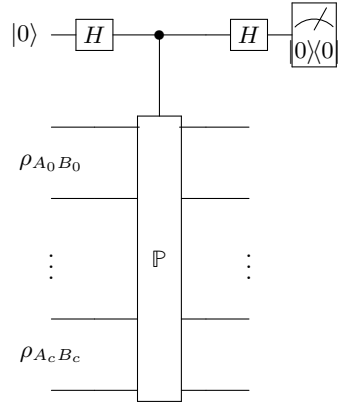

        When the permutation operator $\mathbb{P} = \mathbb{P}_A \otimes \mathbb{P}_B$ acting on $(\rho_{AB})^{\otimes c}$ is non-Hermitian, the quantity $\Tr\!\left(\mathbb{P}\rho_{AB}^{\otimes c}\right)$ cannot be measured directly. Instead, one can estimate its real part via hermitization
        \[
        \operatorname{Re}\!\left(\Tr\!\left(\mathbb{P}\rho_{AB}^{\otimes c}\right)\right)
        = \frac{1}{2}\Tr\!\left((\mathbb{P}+\mathbb{P}^\dagger)\rho_{AB}^{\otimes c}\right).
        \]
        Following the approach of Ref.~\cite{ekert2002direct}, this quantity can be accessed using the circuit shown in Fig.~\ref{circ:Perm_measure}, yielding
        \begin{equation}
        \frac{1}{2}\Tr\!\left((\mathbb{P}+\mathbb{P}^\dagger)\rho_{AB}^{\otimes c}\right)
        = 2\,\Tr\!\left[(\ketbra{0}{0}\otimes \Id_{AB}^{\otimes c})
        (H\otimes \Id_{AB}^{\otimes c})\,\operatorname{C\mathbb{P}}\,(H\otimes \Id_{AB}^{\otimes c})
        (\ketbra{0}{0}\otimes \rho_{AB}^{\otimes c})
        (H\otimes \Id_{AB}^{\otimes c})\,\operatorname{C\mathbb{P}}^\dagger\,(H\otimes \Id_{AB}^{\otimes c})\right] - 1,
        \end{equation}
        where $\Id_{AB}$ denotes the identity operator on the bipartite system and $\operatorname{C\mathbb{P}} = \ketbra{0}{0}\otimes \Id_{AB}^{\otimes c} + \ketbra{1}{1}\otimes \mathbb{P}$.
        
        Although this scheme can, in principle, be applied to Hermitian operators as well, doing so would require up to $100$ distinct measurement settings in our case, compared to only 9 settings for the direct measurement scheme. For this reason, we employ this protocol only for non-Hermitian permutation operators, as listed in Table~\ref{tab:SwapDetails_4copies_nonH}.
\section{Performance of various models}
\label{app:sec:trainability_models}

    In this section, we describe additional models we used for predicting the entanglement of bipartite states.

    \subsection{Expressivity of the functions in Table~\ref{tab:10-class_short}}

\begin{figure*}[tbh]
                \centering
                \includegraphics[width=.495\textwidth]{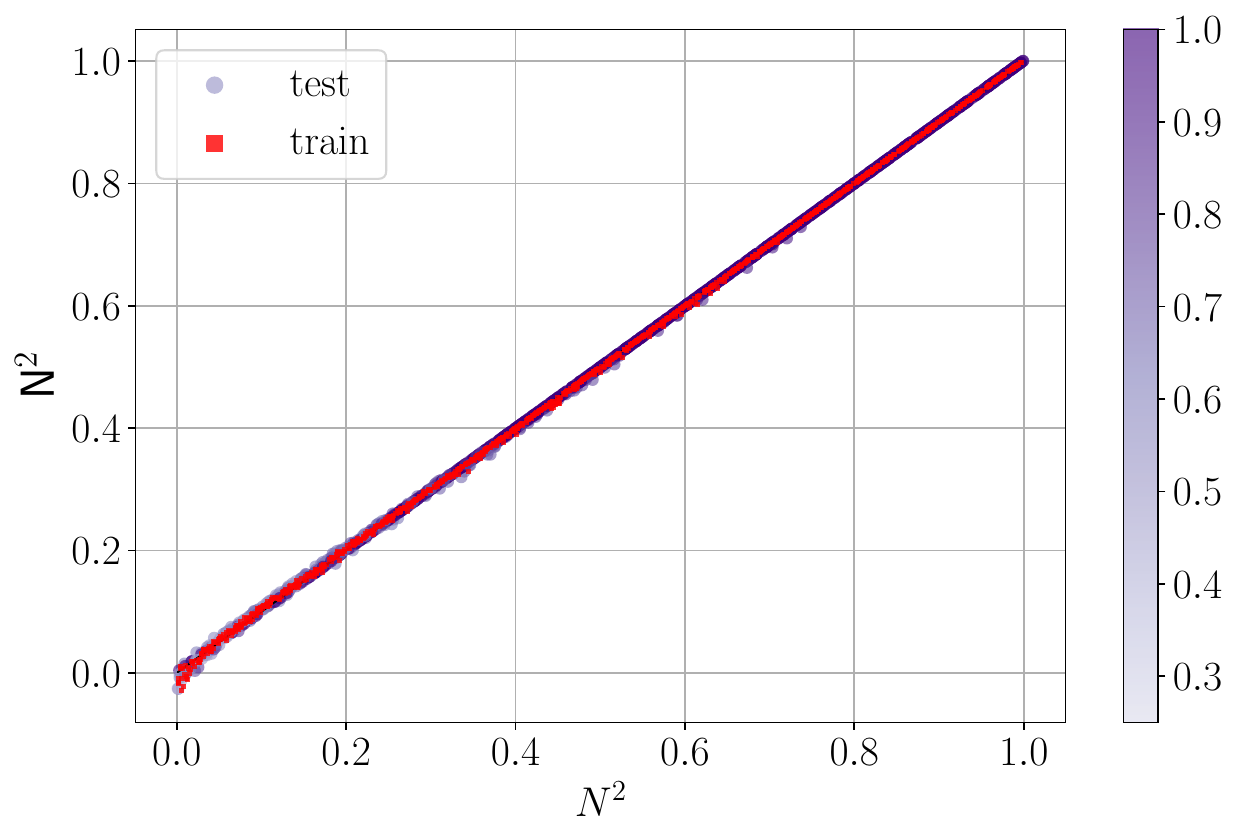}
                \includegraphics[width=.495\textwidth]{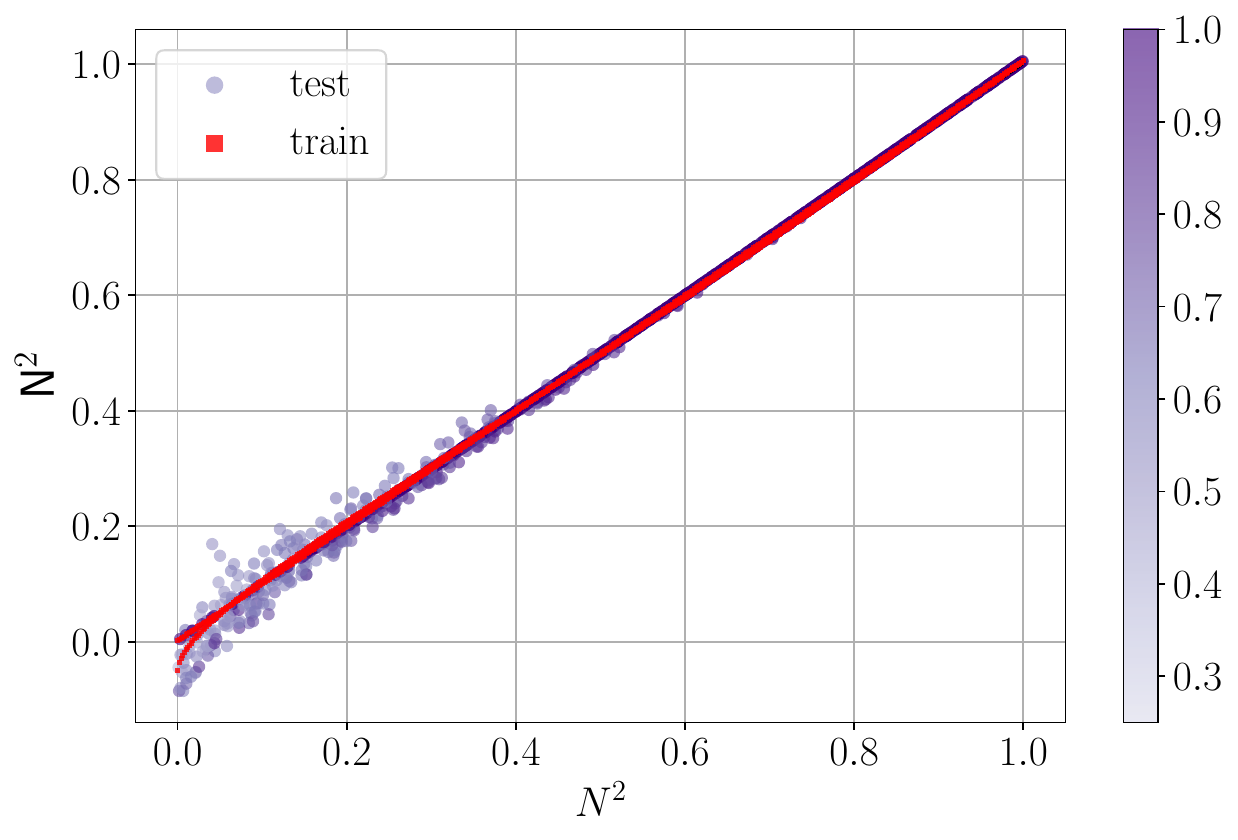}
                \caption{
            Predicted negativity squared $\mathsf{N}^2$ vs true values of negativity squared $N^2$ for 10 functions described in Appendix~\ref{app:sec:trainability_models}. Red points correspond to the predicted training set, while purple points represent predictions on random mixed states. The model is trained either on 1000 random mixed states (left) or on 1000 easy states (right) with their corresponding true labels. The black lines connecting $(0,0)$ and $(1,1)$ are the ground truth.
                }
                \label{fig:f10_easyMix}
            \end{figure*}

        In Fig.~\ref{fig:f10_easyMix}, we present the prediction performance of the ten functions listed in Table~\ref{tab:10-class_short}, trained either on random mixed states (left) or on easy states (right), and tested on random mixed states. Specifically, we consider a linear model $f(\btheta, \rho)=\sum_{j=1}^{10} \theta_j f_j(\rho)$ where $\{f_j\}_{j=1}^{10}$ are the functions defined in Table~\ref{tab:10-class_short}. The prediction quality observed in the left and right panels of Fig.~\ref{fig:f10_easyMix} closely matches the corresponding results in Fig.~\ref{fig:neg_square10mixedXmixed} for training on random mixed states and Fig.~\ref{fig:neg_square10classisopureXmixed} for training on easy states. This agreement is expected, since evaluating the 10-Hermitian-operator ansatz in Eq.~\eqref{eq:H10_ansatz} on a quantum state $\rho$ directly yields these ten functions.
        
    \subsection{Expressivity of $E(x,y,z)$}
        
\begin{figure*}[tbh]
                \centering
                \includegraphics[width=.495\textwidth]{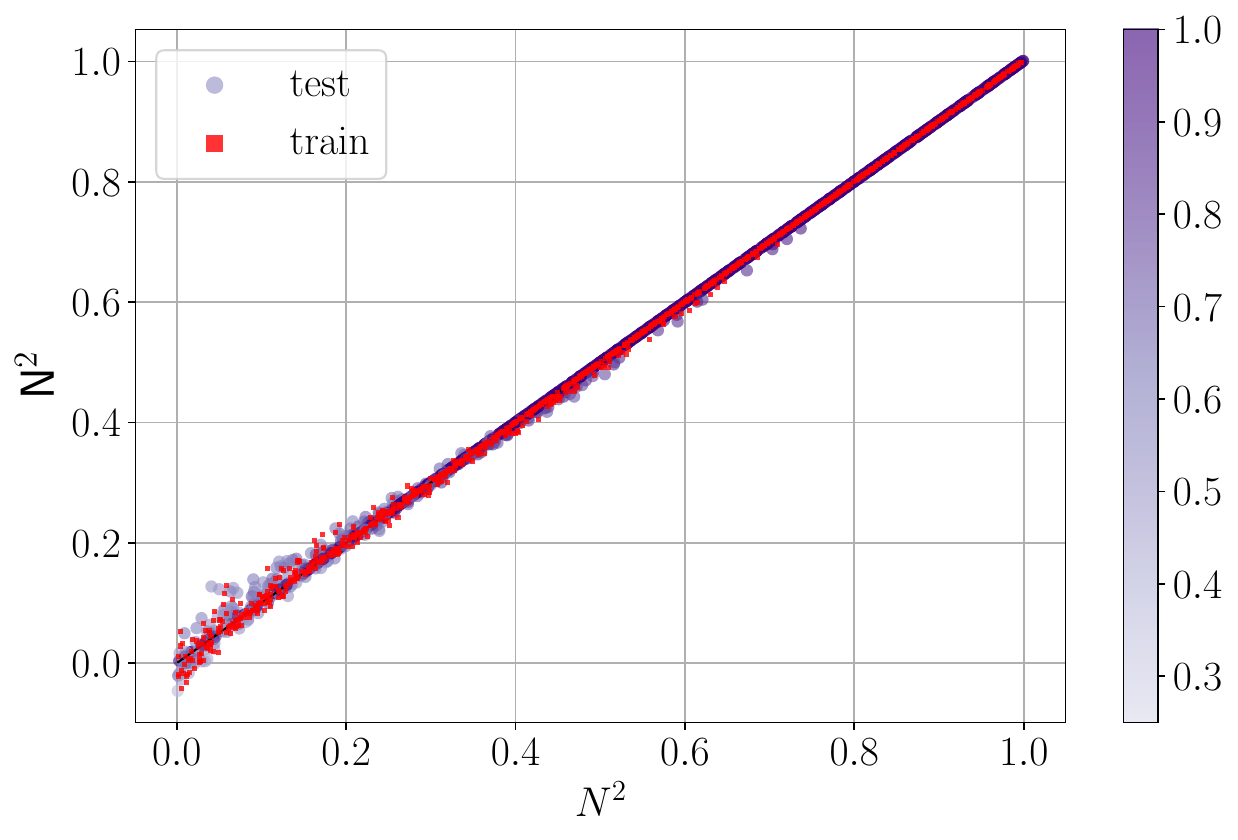}
                \includegraphics[width=.495\textwidth]{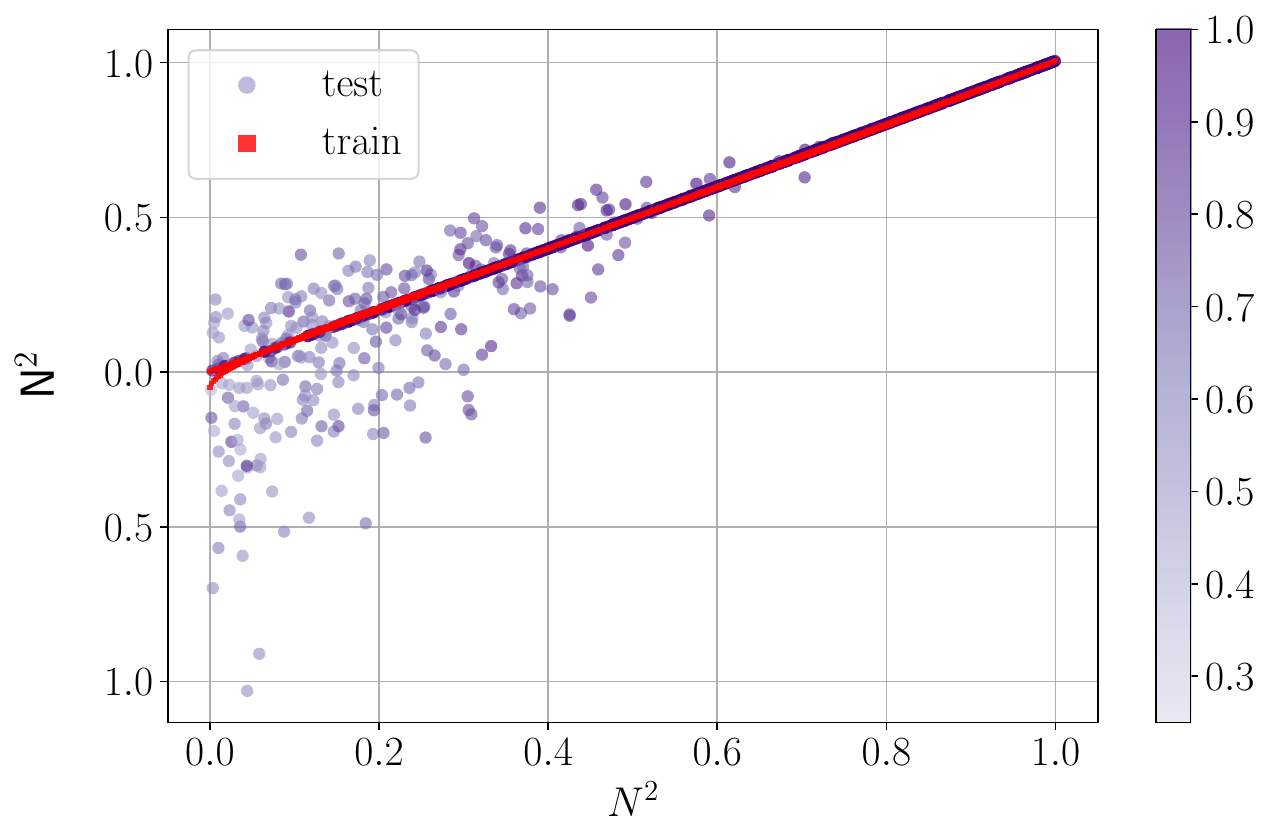}
                \caption{ 
                Predicted negativity squared $\mathsf{N}^2$ vs true values of negativity squared $N^2$ for $E(x,y,z)$ model described in Appendix~\ref{app:sec:trainability_models}. Red points correspond to the predicted training set, while purple points represent predictions on random mixed states. The model is trained either on 1000 random mixed states (left) or on 1000 easy states (right) with their corresponding true labels. The black lines connecting $(0,0)$ and $(1,1)$ are the ground truth.
                }
                \label{fig:f38_easyMix}
            \end{figure*}

        In Fig.~\ref{fig:f38_easyMix} (left), we show the performance of the function 
        \begin{equation}
            \label{eq:E_x}
            E(x_{1,2}, y_{1,2}, z_{1,2}) = \Tr\big[\rho^{x_1} (\rho_A^{y_1} \otimes \rho_B^{z_1})\big] \; \Tr\big[\rho^{x_2} (\rho_A^{y_2} \otimes \rho_B^{z_2})\big],
        \end{equation}
        trained on 1000 random mixed states and evaluate it on another 1000 random mixed states. In the right panel of Fig.~\ref{fig:f38_easyMix}, the same model is trained on a uniformly distributed dataset of easy states consisting of 500 pure and 500 isotropic states and then tested on random mixed states. While training on random mixed states yields highly accurate predictions, training on easy states leads to poor generalization when predicting random mixed states. This behavior is most likely due to the overparameterized nature of the model $E(x_{1,2},y_{1,2},z_{1,2})$, which prevents it from learning features that generalize beyond the restricted easy-state manifold. This example illustrates that entanglement prediction with overparameterized models can be highly sensitive to the choice of training dataset.
    
    \subsection{Expressivity of the functions using PT-moment and linear entropy functions.}
    
\begin{figure*}[tbh]
        \centering
        \includegraphics[width=.495\textwidth]{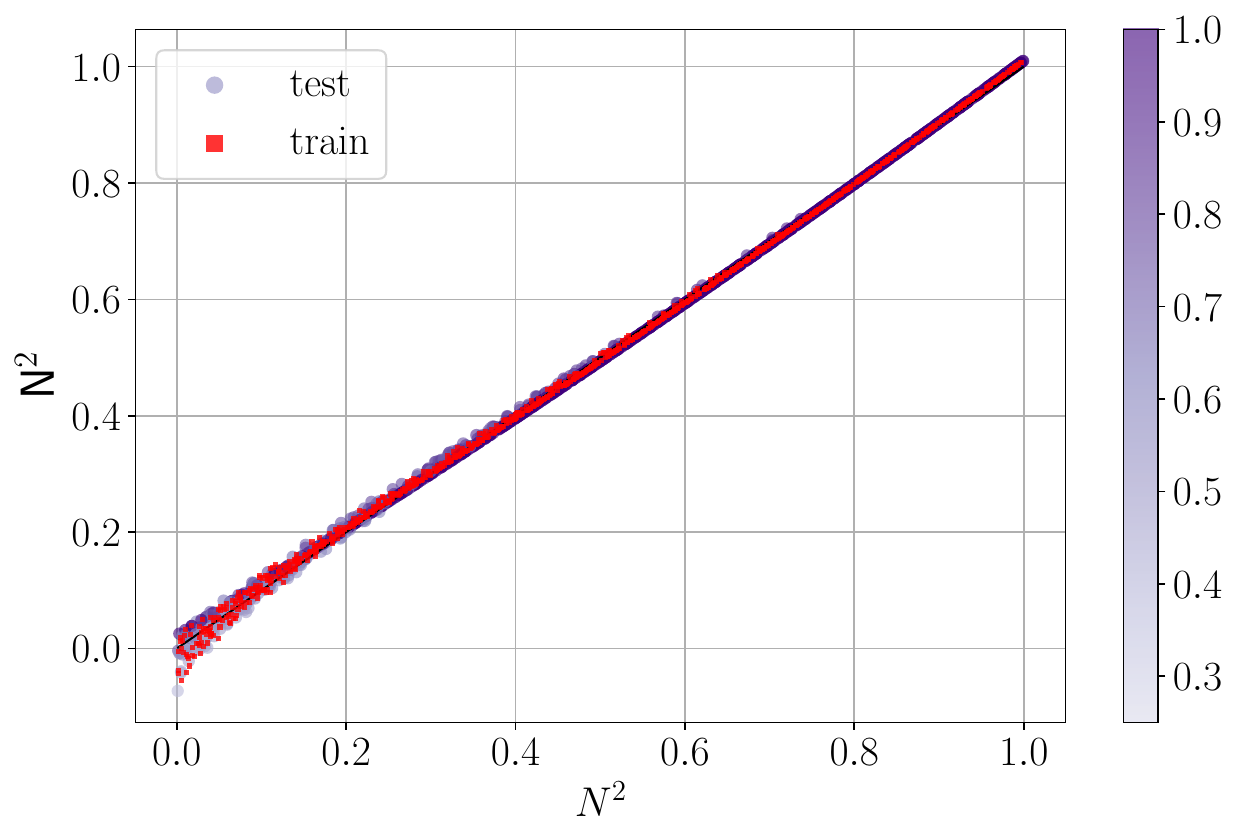}
        \includegraphics[width=.495\textwidth]{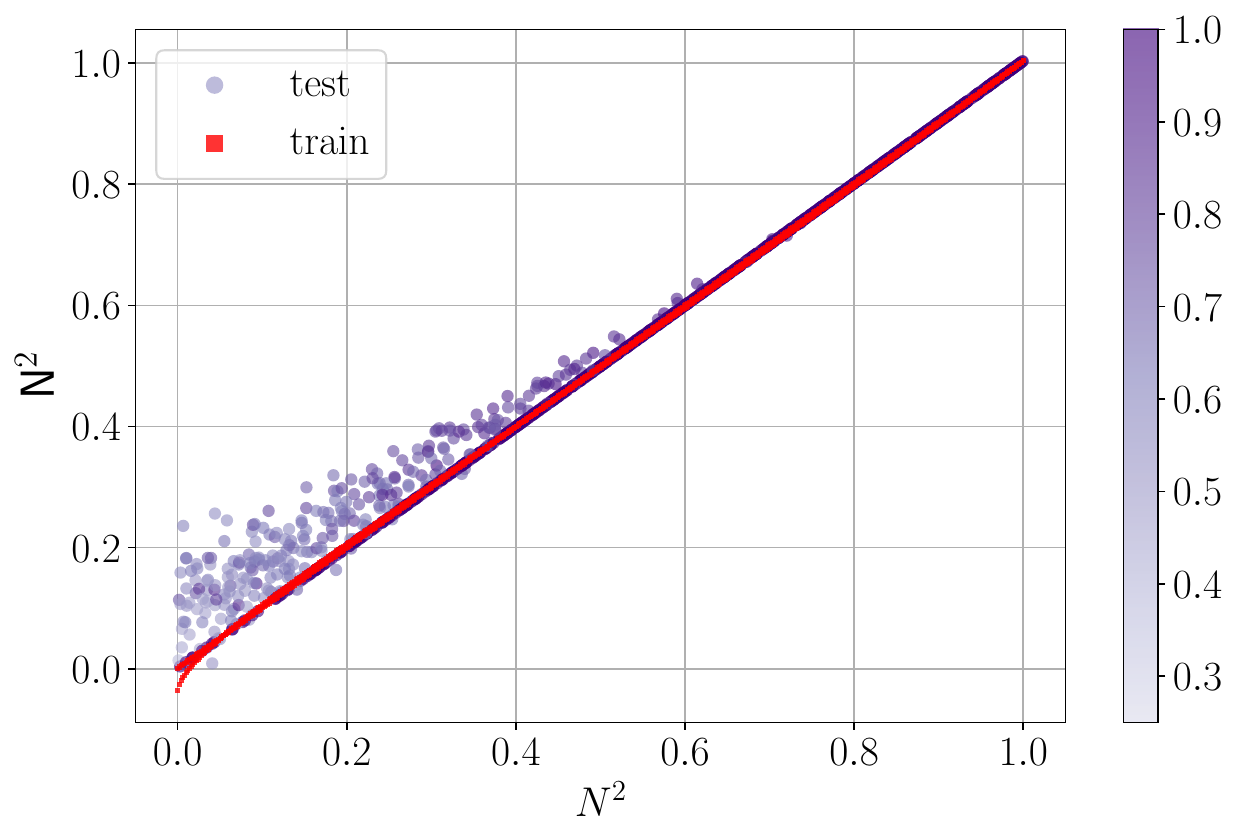}
        \caption{
    Predicted negativity squared $\mathsf{N}^2$ vs true values of negativity squared $N^2$ for PT-based model described in Appendix~\ref{app:sec:trainability_models}. Red points correspond to the predicted training set, while purple points represent predictions on random mixed states. The model is trained either on 1000 random mixed states (left) or on 1000 easy states (right) with their corresponding true labels. The black lines connecting $(0,0)$ and $(1,1)$ are the ground truth.
        }
        \label{fig:pt-moments}
    \end{figure*}

        Now we would like to test the performance of our 10 functions against the set of well known functions such as PT-moments, purity check, and linear entropy entanglement. For that we consider PT-moments described in Eq.~\ref{eq:pt_moments} up to the 4th order. In particular, we are interested in functions $\Big\{1, \quad \Tr\big[\rho_A^2\big]+\Tr\big[\rho_B^2\big], \quad\Tr\big[\rho^2\big], \quad \Tr\big[(\rho^{T_A})^3\big], \quad \Tr\big[(\rho^{T_A})^4\big] \Big\}$. Here, $p_1=1$, $p_2=\Tr[\rho^2]$, $p_3=\Tr[(\rho^{T_A})^3]$, and $p_4=\Tr[(\rho^{T_A})^4]$, while the quantity $\Tr[\rho_A^2]+\Tr[\rho_B^2]$ is directly related to linear-entropy entanglement (or tangle) for bipartite qubits and is equivalent to the sixth-class function listed in Table~\ref{tab:10-class_short}. In Fig.~\ref{fig:pt-moments}, we train this five-function model on random mixed states (left) and on easy states (right), and test both on random mixed states. Training on easy states yields a reasonable prediction quality, although it remains inferior to that achieved with the ten-function model shown in Fig.~\ref{fig:f10_easyMix}. When trained directly on random mixed states, the PT-moment-based model exhibits noticeably poorer accuracy, particularly when compared with the left panel of Fig.~\ref{fig:f10_easyMix}.

So far, we have observed that training directly on the ten functions yields prediction performance comparable to that obtained by training the 10 Hermitian operator ansatz in Eq.~\eqref{eq:H10_ansatz}, for both random mixed states and easy states. However, extending this feature set by including higher-order combinations, such as the expanded model $E(x_{1,2}, y_{1,2}, z_{1,2})$, does not improve the prediction accuracy for random mixed states and significantly degrades performance when trained on easy states, due to overparameterization. 

While one might attempt to incrementally augment the ten-function set with additional features, our numerical results indicate that such extensions do not enhance performance on random mixed states and instead lead to deteriorated generalization when training on easy states. We further compared the ten-function model with selected five-function model that includes linear-entropy entanglement and partial-transpose moments up to fourth order, which consistently yielded inferior prediction quality for both random mixed and easy states training. Altogether, these results suggest that the proposed ten functions constitute a well-balanced and sufficient feature set for accurately predicting entanglement in random mixed states, whether the model is trained on random mixed or easy states.

\begin{figure*}[tbh]
        \centering
        \includegraphics[width=.495\textwidth]{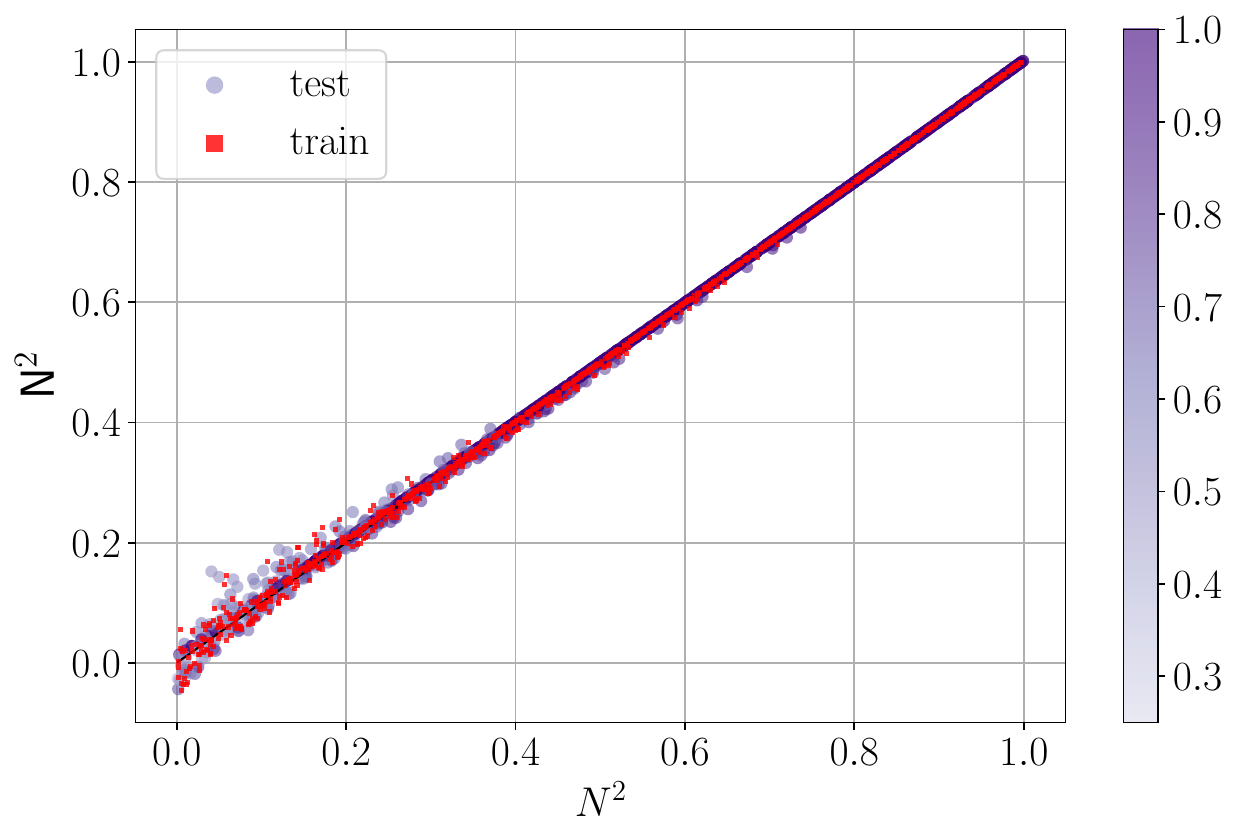}
        \includegraphics[width=.495\textwidth]{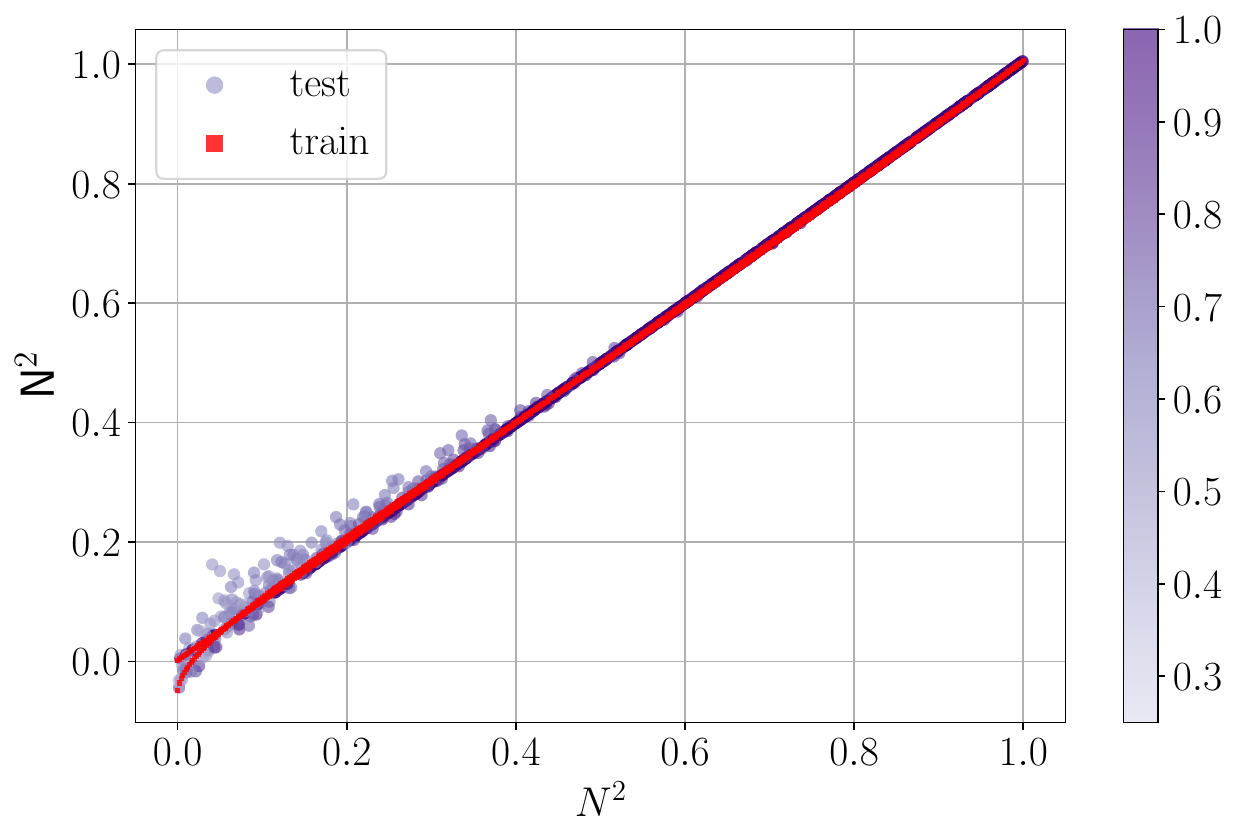}
        \caption{
    Predicted negativity squared $\mathsf{N}^2$ vs true values of negativity squared $N^2$ for reduced \textit{five-selected model} described in Appendix~\ref{app:sec:trainability_models}. Red points correspond to the predicted training set, while purple points represent predictions on random mixed states. The model is trained either on 1000 random mixed states (left) or on 1000 easy states (right) with their corresponding true labels.
        }
        \label{fig:five-selected}
    \end{figure*}

Alternatively, one might consider reducing the number of functions from the ten-function set. However, our numerical results show that removing any function generally increases the mean-squared error (MSE) when training on random mixed states. For easy states, we find that a subset of five functions, class 1 ($\Tr[\rho^2]^2$), class 2 ($\Tr[\rho^2]$), class 6 ($\Tr[\rho_A^2]+\Tr[\rho_B^2]$), class 9 ($\Tr[(\rho_A^2)^2] + \Tr[(\rho_B^2)^2]$), and class 5 ($\Tr\big[\Tr_A((\Id\otimes \rho_B)\rho)^2\big] + \Tr\big[\Tr_A((\rho_A\otimes \Id)\rho)^2\big]$), is sufficient, producing roughly half the MSE of the full ten-function set in our simulations.
However, training on random mixed states with this reduced set significantly degrades performance, as illustrated in Fig.~\ref{fig:five-selected}. In other words, the number of functions can be reduced depending on the complexity of the dataset: for easy states, five carefully chosen functions suffice, whereas for random mixed states, the full ten-function set remains optimal.

\end{document}